\documentclass[structabstract]{aa}  
\usepackage{graphicx}
\usepackage{amsmath,amssymb}
\usepackage{txfonts}
\begin{document}

\def\Arrow{\mathop{\longrightarrow}\limits}
\def\Harpoons{\mathop{\rightleftharpoons}\limits}

   \title{Deuterium fractionation and the degree of ionisation in massive 
clumps within infrared dark clouds\thanks{This publication is based on data 
acquired with the Atacama Pathfinder Experiment (APEX) under programmes 
081.F-9823A, 082.F-9320A, 083.C-0230A, and 085.F-9321A. APEX is a 
collaboration between the MPIfR, the European Southern Observatory, and the 
Onsala Space Observatory.}\fnmsep \thanks{Appendices A and B are only available in electronic form at {\tt http://www.aanda.org}}}

   \author{O. Miettinen\inst{1}, M. Hennemann\inst{2}, \and H. Linz\inst{3}} 

 \offprints{O. Miettinen}

   \institute{Department of Physics, P.O. Box 64, FI-00014 University of Helsinki, Finland\\ \email{oskari.miettinen@helsinki.fi} \and Service d'Astrophysique, Laboratoire AIM, CEA - CNRS - Universit\'e Paris Diderot, CEA-Saclay, 91191 Gif-sur-Yvette cedex, France \and Max-Planck-Institut f\"ur Astronomie, K\"onigstuhl 17, D-69117 Heidelberg, Germany}

   \date{Received ; accepted}

\authorrunning{Miettinen et al.}
\titlerunning{Massive clumps in IRDCs}

  \abstract
   {Massive clumps associated with infrared dark clouds (IRDCs) are 
promising targets to study the earliest stages of high-mass star and star 
cluster formation.}
   {We aim to determine the degrees of CO depletion, deuterium 
fractionation, and ionisation in a sample of seven massive clumps 
associated with IRDCs.}
   {The APEX telescope was used to observe the C$^{17}$O$(2-1)$, 
H$^{13}$CO$^+(3-2)$, DCO$^+(3-2)$, N$_2$H$^+(3-2)$, and N$_2$D$^+(3-2)$ 
transitions towards the clumps. The spectral line data are used in 
conjunction with the previously published and/or archival (sub)millimetre dust 
continuum observations of the sources. The data are used to derive the 
molecular column densities and fractional abundances needed in the analysis 
of deuterium fractionation and ionisation.}
   {The CO molecules do not appear to be significantly depleted in the 
observed clumps. The DCO$^+$/HCO$^+$ and N$_2$D$^+$/N$_2$H$^+$ column density 
ratios are about 0.0002--0.014 and 0.002--0.028, respectively. The former 
ratio is found to decrease as a function of gas kinetic temperature. 
A simple chemical analysis suggests that the lower limit to the ionisation 
degree is in the range $x({\rm e})\sim10^{-8}-10^{-7}$, whereas the estimated 
upper limits range from a few $10^{-6}$ up to $\sim10^{-4}$. 
Lower limits to $x({\rm e})$ imply the cosmic-ray ionisation rate of H$_2$ to 
lie between $\zeta_{\rm H_2}\sim 10^{-17}-10^{-15}$ s$^{-1}$. These are 
the first estimates of $x({\rm e})$ and $\zeta_{\rm H_2}$ towards 
massive IRDCs reported so far. Some additional molecular transitions, mostly 
around 216 and 231 GHz, were detected towards all sources. In particular, 
IRDC 18102-1800 MM1 and IRDC 18151-1208 MM2 show 
relatively line-rich spectra with transitions some of which could be assigned 
to complex organic molecules, though the line blending is hampering the 
identification. The C$^{18}$O$(2-1)$ transition is frequently seen in the 
image band.}
   {The finding that CO is not depleted in the observed sources 
conforms to the fact that they show evidence of star formation 
activity which is believed to release CO from the icy grain mantles back into 
the gas phase. The observed degree of deuteration is lower than in 
low-mass starless cores and protostellar envelopes. Decreasing deuteration 
with increasing temperature is likely to reflect the clump evolution. On the 
other hand, the association with young high-mass stars could enhance 
$\zeta_{\rm H_2}$ and $x({\rm e})$ above the levels usually found in low-mass 
star-forming regions. On the scale probed by our observations, 
ambipolar diffusion cannot be a main driver of clump evolution unless it 
occurs on timescales $\gg 10^6$ yr.}

   \keywords{Astrochemistry - Stars: formation - ISM: abundances - ISM: clouds 
- ISM: molecules - Radio lines: ISM}

   \maketitle
%

\section{Introduction}

Most, if not all, infrared dark clouds (IRDCs; \cite{perault1996}; 
\cite{egan1998}) studied so far are fragmented into clumpy structures. 
Some of the clumps within IRDCs that have been studied with high-resolution 
interferometric observations are found to contain still denser cores, 
indica\-ting that the clumps have further fragmented into smaller pieces
(\cite{rathborne2007}, 2008; \cite{zhang2009}; 
\cite{beuther2009}; \cite{wang2011})\footnote{In this work, we use 
the term ``clump'' to describe a density enhancement within an IRDC with a 
size of about $\sim0.5-1$ pc and a mass of order $10^2-10^3$ M$_{\sun}$. 
The term ``core'' is also often used for these objects.}. 

Massive clumps in IRDCs are promising targets to study the earliest 
stages of high-mass star and star cluster formation 
(e.g., \cite{kauffmann2010} and references therein). 
(Sub)millimetre dust continuum maps of IRDCs have shown that the clumps within 
them have typical radii, masses, beam-averaged H$_2$ column densities, and 
volume-averaged H$_2$ number densities of $\sim0.1-0.5$ pc, 
$\sim10^2-10^3$ M$_{\sun}$, $\gtrsim10^{22}$ cm$^{-2}$, 
and $\sim10^4-10^5$ cm$^{-3}$, respectively (e.g., \cite{rathborne2006}; 
\cite{parsons2009}; \cite{vasyunina2009}; \cite{rygl2010}; 
\cite{miettinenharju2010}). Molecular spectral-line 
observations towards such clumps have shown the typical gas kinetic 
temperature to lie in the range $T_{\rm kin}\approx10-20$ K 
(\cite{carey1998}; \cite{teyssier2002}; \cite{sridharan2005}; 
\cite{pillai2006b}; \cite{sakai2008}, hereafter SSK08; 
\cite{zhang2011}; \cite{devine2011}; \cite{ragan2011}). 
Spectral energy distributions of clumps within IRDCs yield 
dust temperature values of $T_{\rm dust}\approx10-50$ K (e.g., 
\cite{rathborne2010}; \cite{henning2010}). 
The star formation process within some of these sources
mani\-fests itself through embedded infrared emission 
(e.g., \cite{chambers2009}; \cite{ragan2009}; \cite{henning2010}), outflows 
(\cite{beuther2007}; \cite{fallscheer2009}; \cite{sanhueza2010}), and H$_2$O 
and CH$_3$OH maser emission (\cite{pillai2006a}; \cite{ellingsen2006}; 
\cite{wang2006}). Hot cores (\cite{beuther2005a}, 2006; \cite{rathborne2007}, 
2008) and UC H{\scriptsize II} regions (\cite{battersby2010}) 
have also been discovered within IRDCs. These are clear signs of high-mass 
star-formation activity. On the other hand, some of the massive clumps are 
dark at mid-infared (MIR) and do not show any other evidence of ongoing star 
formation [e.g., G024.61-00.33 MM2 (\cite{rygl2010}) and G028.23-00.19 
GLM1 (\cite{battersby2010})]. Such sources are candidates for being the 
high-mass prestellar or precluster objects and are promising targets 
for the studies of the initial conditions of massive star formation. 

In this paper, we aim to further constrain the physical and chemical 
characteristics of massive clumps associated with IRDCs. We study the degrees 
of CO depletion and deuterium fractionation, and the fractional ionisation in 
a sample of seven clumps. So far, only a few studies have been carried out 
to investigate molecular deuteration in IRDCs (\cite{pillai2007}; 
\cite{chen2010}; \cite{fontani2011}; \cite{pillai2011}). 
The level of deute\-ration has the potential to be used as an evolutionary 
indicator (e.g., \cite{crapsi2005}; \cite{emprechtinger2009}; 
\cite{fontani2011}) and as such it is useful to study it in IRDCs. 
The ionisation degree, $x({\rm e})$, and the cosmic-ray ionisation rate of 
H$_2$, $\zeta_{\rm H_2}$, are also essential knowledge towards an understanding 
of the initial conditions of star formation (e.g., \cite{caselli1998}). 
Fractional ionisation determines the coup\-ling between the gas 
and magnetic field, and can thus play a role in the dynamical 
evolution of a star-forming object. In addition, the ionisation 
fraction has an influence on the gas-phase che\-mical reactions, and 
thus assessing the ionisation structure of a source helps to better 
understand its overall chemistry. Because estimates of $x({\rm e})$ and 
$\zeta_{\rm H_2}$ in massive clumps within IRDCs have not been reported so 
far, constraining these parameters through observations seems particularly 
worthwhile. 

This paper is organised as follows. The observations, data-reduction 
procedures, and the source sample are described in Sect.~2. The direct 
observational results are presented in Sect.~3. In Sect.~4, we describe the 
analysis and present the results of the physical and chemical properties of 
the clumps. In Sect.~5, we discuss our results, and in Sect.~6, we summarise 
the main conclusions of this study. 

\section{Source selection, observations, and data reduction}

\subsection{Source selection}

For this study we selected six clumps from the previous 
studies by Beuther et al. (2002a), Sridharan et al. (2005), and Rathborne 
et al. (2006), and one clump, ISOSS J18364-0221 SMM1\footnote{Hereafter, we 
use the appreviated source names, such as J18364 SMM1 etc.}, previously 
studied by Birkmann et al. (2006) and Hennemann et al. (2009).
We selected sources having a high mass inside a typical clump radius 
of $\lesssim0.5$ pc (revised masses from the above reference studies 
are $\sim60-360$ M$_{\sun}$; see Sect.~4.1.3), and which are likely to 
represent different evolutionary stages of (high-mass) star formation: 
two of the clumps are clearly associated with \textit{Spitzer} 8-$\mu$m 
point-sources and five appear dark in the \textit{Spitzer} 8-$\mu$m images. 
Moreover, three of the clumps are associated with H$_2$O and/or CH$_3$OH 
masers. As will be discussed later, the youngest sources in our sample are 
likely to be G015.05 MM1 and G015.31 MM3, whereas I18102 MM1 appears to be 
in the most advanced stage of evolution. In this way, we were aiming to 
investigate if there are any variations or evolutionary trends in the clump 
physical and chemical properties among the sample. 

To gather the source sample, one criterion was that the source distance is 
not very large (revised values are $\sim2.5-3.5$ kpc; see Sect.~4.1.1 
and references therein) in order to achieve a 
reasonable spatial resolution with the single-dish telescope. For example, 
$20\arcsec-30\arcsec$ corresponds to $\sim0.24-0.51$ pc at the cloud 
distances, i.e., the spatial resolution is comparable to the clump sizes. 
We selected sources for which dust continuum data at (sub)mm wavelengths is 
available in order to derive the H$_2$ column density needed in the 
derivation of molecular fractional abundances. 
Also, in order to derive reliable physical parameters of the 
sources, they were supposed to have known gas kinetic temperature or dust 
temperature. The source list is given in Table~\ref{table:sources}, and the 
sources are discussed in more detail in Appendix B. 

\begin{table*}
\caption{Source list.}
\begin{minipage}{2\columnwidth}
\centering
\renewcommand{\footnoterule}{}
\label{table:sources}
\begin{tabular}{c c c c c c c c c}
\hline\hline 
Source & $\alpha_{2000.0}$ & $\delta_{2000.0}$ & $l$ & $b$ & $R_{\rm GC}$ & $d$ & 
\textit{Spitzer} 8 $\mu$m/24 $\mu$m\\
       & [h:m:s] & [$\degr$:$\arcmin$:$\arcsec$] & [$\degr$] & [$\degr$] & [kpc] & [kpc] & \\
\hline
IRDC 18102-1800 MM1 & 18 13 11.0 & -17 59 59 & 12.624 & -0.016 & 5.8 & $2.7^{+0.5}_{-0.6}$ & point/point\\
G015.05+00.07 MM1 & 18 17 50.4 & -15 53 38 & 15.006 & 0.009 & 5.9 & $2.6^{+0.5}_{-0.6}$ & dark/dark\\
IRDC 18151-1208 MM2 & 18 17 50.4 & -12 07 55 & 18.319 & 1.792 & 5.9 & $2.7^{+0.4}_{-0.5}$ & dark/-\tablefootmark{a}\\
G015.31-00.16 MM3 & 18 18 45.3 & -15 41 58 & 15.281 & -0.093 & 5.6 & $3.0^{+0.4}_{-0.5}$ & dark/dark\\
IRDC 18182-1433 MM2 & 18 21 14.9 & -14 33 06 & 16.577 & -0.081 & 5.1 & $3.5^{+0.4}_{-0.4}$ & point/point\\
IRDC 18223-1243 MM3 & 18 25 08.3 & -12 45 27 & 18.605 & -0.075 & 5.2 & $3.5^{+0.3}_{-0.4}$ & dark/point\\
ISOSS J18364-0221 SMM1 & 18 36 36.0 & -02 21 45 & 29.135 & 2.220 & 6.3 & $2.5^{+0.4}_{-0.4}$ & dark/point\\
\hline 
\end{tabular} 
\tablefoot{Columns (2)--(7) give the equatorial and galactic coordinates 
[$(\alpha, \,\delta)_{2000.0}$ and $(l, \,b)$, respectively], the source 
galactocentric distance ($R_{\rm GC}$), and the near kinematic distance from 
the Sun ($d$). The distances are derived by following Reid et al. (2009) as 
described in Sect.~4.1.1. Because IRDCs are seen as dark extinction features 
against the bright Galactic MIR background, it is expected that these clouds 
lie at the near kinematic distance. In the last column we give the comments 
concerning the source appearance in the \textit{Spitzer} 8 and 24-$\mu$m 
images.\\
\tablefoottext{a}{No 24-$\mu$m image available.}}
\end{minipage}
\end{table*}

\subsection{APEX molecular-line observations}

In this subsection we describe the molecular-line observations 
carried out with the Atacama Pathfinder Experiment (APEX) 12-m telescope at 
Llano de Chajnantor (Chile). The observed lines, their spectroscopic 
properties, and observational parameters are listed in Table~\ref{table:lines}.

\subsubsection{C$^{17}$O$(2-1)$}

The C$^{17}$O$(2-1)$ observations towards all the sources except 
J18364 SMM1 were carried out on 25 March 2008 with APEX during the 
observational campaign for the Science Verification (SV) of the Swedish
Heterodyne Facility Instrument (SHFI or SHeFI; \cite{belitsky2007}; 
\cite{vassilev2008a}). The source J18364 SMM1 
was observed on 26 April 2009. As front\-end, we 
used APEX-1 of the SHFI (\cite{vassilev2008b}). APEX-1 ope\-rates in 
single-sideband (SSB) mode using sideband separation mixers, and it has 
a sideband rejection ratio $>10$ dB. The centre frequency for the image 
band is given by $\nu_{\rm image}=\nu_{\rm cen}\pm12$ GHz, where $\nu_{\rm cen}$ 
is the passband centre frequency, and the $\pm$ refers to the image band 
corresponding to the upper or lower sideband (USB or LSB), respectively. 
The backend was the Fast Fourier Transfrom Spectrometer 
(FFTS; \cite{klein2006}) with a 1 GHz bandwidth divided into 16\,384 spectral 
channels du\-ring SV and 8\,192 channels during observations towards J18364 
SMM1.

The observations were performed in the wobbler-switching (WS) mode with a 
$150\arcsec$ azimuthal throw (symmetric offsets) and a chopping rate of 0.5 Hz.
The telescope pointing and focusing were checked by 
continuum scans on the planet Jupiter and 
the pointing was found to be accurate to $\sim3\arcsec$. Calibration was 
done by the chopper-wheel technique, and the intensity scale given by the 
system is $T_{\rm A}^*$, the antenna temperature corrected for atmospheric
attenuation. 

The spectra were reduced using the CLASS90 programme of the IRAM's GILDAS 
software package\footnote{{\tt http://www.iram.fr/IRAMFR/GILDAS}}. 
The individual spectra were averaged and the resulting spectra were 
Hanning-smoothed in order to increase the signal-to-noise (S/N) ratio of the 
data. A first- or third-order polynomial was applied to correct the baseline 
in the resulting spectra. In one case (G015.05 MM1), a polynomial baseline 
of order six had to be applied because there appeared to be some emission 
in the wobbler off-beams which resulted in a very poor baseline. 
The resulting $1\sigma$ rms noise le\-vels are 40--60 mK at the smoothed 
resolutions. The final velocity resolution of the smoothed spectra 
is 0.16 km~s$^{-1}$, except for the spectrum towards J18364 SMM1, where it is 
0.32 km~s$^{-1}$.

The $J=2-1$ transition of C$^{17}$O contains nine hyperfine (hf) 
components. We fitted the hf structure of the transition using 
``method hfs'' of CLASS90 to derive the LSR velocity (${\rm v}_{\rm LSR}$) of 
the emission, and FWHM linewidth ($\Delta {\rm v}$). The hf-line 
fitting can also be used to derive the line optical thickness, $\tau$. 
However, in all spectra the hf components are blended together and 
thus the optical thickness cannot be reliably determined. For the rest 
frequencies of the hf components, we used the values from Ladd et al. 
(1998; Table 6 therein). The adopted central frequency, 224\,714.199 MHz, is 
that of the $J_F = 2_{9/2} \rightarrow 1_{7/2}$ hf component which has 
a relative intensity of $R_i=\frac{1}{3}$.

\subsubsection{H$^{13}$CO$^+(3-2)$ and DCO$^+(3-2)$}

The H$^{13}$CO$^+(3-2)$ and DCO$^+(3-2)$ observations were carried out on 
9--11 April and 3 June 2010. As frontend, we used SHFI/APEX-1 and the backend 
was the FFTS with a 1 GHz bandwidth divided into 8\,192 channels. 
The observations were performed in the WS mode with similar 
settings as in the case of C$^{17}$O$(2-1)$ observations.
The telescope pointing and focusing were checked by observing the planets 
Saturn and Jupiter and the pointing was again found to be accurate to 
$\sim3\arcsec$. 

The individual spectra were averaged, smoothed, and a first- or third-order 
polynomial was applied to correct the baseline. The resulting $1\sigma$ rms 
noise values are 20--80 mK in the case of H$^{13}$CO$^+(3-2)$ and 30--40 mK 
in the case of DCO$^+(3-2)$. The final velocity resolutions of the 
smoothed H$^{13}$CO$^+(3-2)$ and DCO$^+(3-2)$ spectra are 0.28 and 0.34 
km~s$^{-1}$, respectively. As shown in the last column of 
Table~\ref{table:lines}, the on-source integration times were very different 
between different sources. This is particularly the case for 
H$^{13}$CO$^+(3-2)$ where $t_{\rm int}$ varies by a factor of $\sim11$. 
This explains the large variation (factor of 4) in the noise values of this 
transition.

The spectral-line transitions of H$^{13}$CO$^+$ and DCO$^+$ also have 
hf structure (e.g., \cite{schmid2004}; \cite{caselli2005}). 
The $J=3-2$ transition we have observed is split up into eight and five 
hf components in the case of H$^{13}$CO$^+$ and DCO$^+$, respectively.
To fit the hf structure, we used the rest frequencies of the 
H$^{13}$CO$^+(3-2)$ and DCO$^+(3-2)$ lines from the Cologne Database for 
Molecular Spectroscpy (CDMS; 
\cite{muller2005})\footnote{{\tt http://www.astro.uni-koeln.de/cdms/catalog}}.
The adopted central frequency of H$^{13}$CO$^+(3-2)$, 260\,255.35250 MHz, is 
that of the $J_{F_1, F} = 3_{7/2,4} \rightarrow 2_{5/2,3}$
hf component which has a relative intensity of $R_i\simeq0.322$. We 
note that this frequency is 13.5 kHz higher than the value determined by 
Schmid-Burgk et al. (2004; Table 3 therein). In the case of DCO$^+(3-2)$, 
the central frequency used, 216\,112.57790 MHz, is that of the 
$J_F = 3_4 \rightarrow 3_3$ hf component; its relative intensity is 
$R_i=\frac{3}{7}$. Caselli \& Dore (2005; their Table~5) determined a 2.1 
kHz higher frequency for DCO$^+(3-2)$. The frequency intervals of the 
hf components for H$^{13}$CO$^+(3-2)$ and DCO$^+(3-2)$ are very small. 
Therefore, the lines overlap significantly which causes the hf 
structure to be heavily blended. 

\subsubsection{N$_2$H$^+(3-2)$ and N$_2$D$^+(3-2)$}

The N$_2$H$^+(3-2)$ observations towards three of our target positions 
(I18102 MM1, G015.05 MM1, and G015.31 MM3) were carried 
out on 29 April 2009. The N$_2$D$^+(3-2)$ observations were performed during 
16--17 and 19--20 October 2008, 26 and 28 April 2009, and 7 June 2010. 
As frontend, we used SHFI/APEX-2 and APEX-1 for N$_2$H$^+(3-2)$ and 
N$_2$D$^+(3-2)$, respectively. The backend was the 1 GHz FFTS with 
8\,192 channels. The observations were performed in the WS mode with the 
similar settings as described above. The pointing and focusing were 
done by continuum scans on the planets Jupiter, Mars, and Neptune. 
The pointing was found to be accurate to $\sim3\arcsec-5\arcsec$. 

The data were reduced as described above, and a first- or third-order 
polynomial baseline correction resulted in $1\sigma$ rms 
noise values of 40--100 mK and 20--50 mK in the case of N$_2$H$^+$ and 
N$_2$D$^+$, respectively. The final velocity resolutions of the 
smoothed N$_2$H$^+$ and N$_2$D$^+$ spectra are 0.26 and 0.32 km~s$^{-1}$, 
respectively. Again, the on-source integration times were very 
different for different sources, particularly for N$_2$D$^+(3-2)$ 
[see Col.~(10) of Table~\ref{table:lines}].

The $J=3-2$ transitions of both N$_2$H$^+$ and N$_2$D$^+$ contain 38 hf
components. The hf lines were fitted using the rest frequencies 
from Pagani et al. (2009b; Tables 4 and 10 therein). The adopted central 
frequencies of N$_2$H$^+(3-2)$ and N$_2$D$^+(3-2)$, 279\,511.832 and 
231\,321.912 MHz, are those of the $J_{F_1 F} = 3_{45} \rightarrow 2_{34}$
hf component which has a relative intensity of $R_i=\frac{11}{63}$.
Also in these cases, the hf components are blended and thus the value 
of $\tau$ cannot be reliably determined through hf fitting.

\begin{table*}
\caption{Observed spectral-line transitions and observational parameters.}
\begin{minipage}{2\columnwidth}
\centering
\renewcommand{\footnoterule}{}
\label{table:lines}
\begin{tabular}{c c c c c c c c c c}
\hline\hline 
Transition & $\nu$ & $E_{\rm u}/k_{\rm B}$ & $n_{\rm crit}$ & HPBW & $\eta_{\rm MB}$ & $T_{\rm sys}$ & \multicolumn{2}{c}{Channel spacing\tablefootmark{a}} & $t_{\rm int}$\\
      & [MHz] & [K] & [$10^6$ cm$^{-3}$] & [\arcsec] & & [K] & [kHz] & [km~s$^{-1}$] & [min]\\
\hline        
DCO$^+(3-2)$ & 216\,112.57790\tablefootmark{b} & 20.7 & 1.8 & 28.9 & 0.75 & 150-184  & 122.07 & 0.17 & 5.5-13.5 \\
C$^{17}$O$(2-1)$ & 224\,714.199\tablefootmark{c} & 16.2 & 0.01 & 27.8 & 0.75 & 163-218  & 61.04\tablefootmark{d} & 0.08\tablefootmark{d} & 5-7.5\\
N$_2$D$^+(3-2)$ & 231\,321.912\tablefootmark{e} & 22.2 & 1.7 & 27.0 & 0.75 & 172-239 & 122.07 & 0.16 & 6.5-48\\ 
H$^{13}$CO$^+(3-2)$ & 260\,255.35250\tablefootmark{b} & 25.0 & 3.1 & 24.0 & 0.74 & 232-243 & 122.07 & 0.14 & 2.5-27\\
N$_2$H$^+(3-2)$ & 279\,511.832\tablefootmark{e} & 26.8 & 3.0 & 22.3 & 0.74 & 165-175 & 122.07 & 0.13 & 1.5-6.5\\
\hline 
\end{tabular} 
\tablefoot{Columns (2)--(4) give the rest frequencies of the observed 
transitions ($\nu$), their upper state energies ($E_{\rm u}/k_{\rm B}$, 
where $k_{\rm B}$ is the Boltzmann constant), and critical densities. Critical 
densities were calculated at $T\sim15$ K (typical value in IRDCs)  
using the collisional rate data available in the Leiden Atomic and Molecular 
Database (LAMDA; {\tt http://www.strw.leidenuniv.nl/$\sim$moldata/}) 
(\cite{schoier2005}). For N$_2$D$^+$, we used the Einstein $A-$coefficient 
from Pagani et al. (2009b) and the same collisional rate as for N$_2$H$^+$. 
Columns (5)--(10) give the APEX beamsize (HPBW) and the main beam efficiency 
($\eta_{\rm MB}$) at the observed frequencies, and the SSB 
system temperatures during the observations ($T_{\rm sys}$ in $T_{\rm A}^*$ 
scale, see text), channel widths (both in kHz and km~s$^{-1}$) of the 
original data, and the on-source integration times per position 
($t_{\rm int}$).\\
\tablefoottext{a}{The original channel spacings. The final spectra were 
Hanning smoothed which divides the number of channels by two.} 
\tablefoottext{b}{From the CDMS spectroscopic database (\cite{muller2005}).} \tablefoottext{c}{From Ladd et al. (1998).} \tablefoottext{d}{For the C$^{17}$O$(2-1)$ observations towards J18364 SMM1, channel width is 122.07 kHz or 0.16 km~s$^{-1}$.} \tablefoottext{e}{From Pagani et al. (2009b).}}
\end{minipage}
\end{table*}

\subsection{Archival dust continuum data}

Submillimetre and/or millimetre dust continuum maps of our sources are 
published in the papers by Beuther et al. (2002a; MAMBO 1.2 mm), Williams et 
al. (2004; SCUBA 450 \& 850 $\mu$m), Rathborne et al. (2006; MAMBO-II 1.2 mm), 
and Birkmann et al. (2006; SCUBA 450 \& 850 $\mu$m). For the sources I18102 
MM1, I18182 MM2, and J18364 SMM1 we retrieved 850 $\mu$m dust continuum data 
from the JCMT/SCUBA archive 
(\cite{difrancesco2008})\footnote{{\tt http://cadcwww.dao.nrc.ca/jcmt/}}. 
No archival SCUBA data was avai\-lable for the rest of our sources. Thus, for 
the sources G015.05 MM1, G015.31 MM3, and I18223 MM3 we extracted Bolocam 
1.1-mm maps from images produced by the Bolocam Galactic Plane Survey (BGPS; 
see, e.g., \cite{bally2010} and references 
therein)\footnote{{\tt http://irsa.ipac.caltech.edu/data/BOLOCAM$_{-}$GPS/}}. 
For I18151 MM2, no archival (sub)mm data was found and thus we used the 
MAMBO 1.2-mm map of this source published in the paper by Beuther et al. 
(2002a; kindly provided by H. Beuther).

In the present paper, the (sub)mm dust continuum data are used in particular 
to determine the H$_2$ column densities towards the line-observation 
positions. These are used to derive the fractional abundances of the observed 
molecules. 

\subsection{\textit {Spitzer} infrared data}

\begin{figure*}
\begin{center}
\includegraphics[width=6cm]{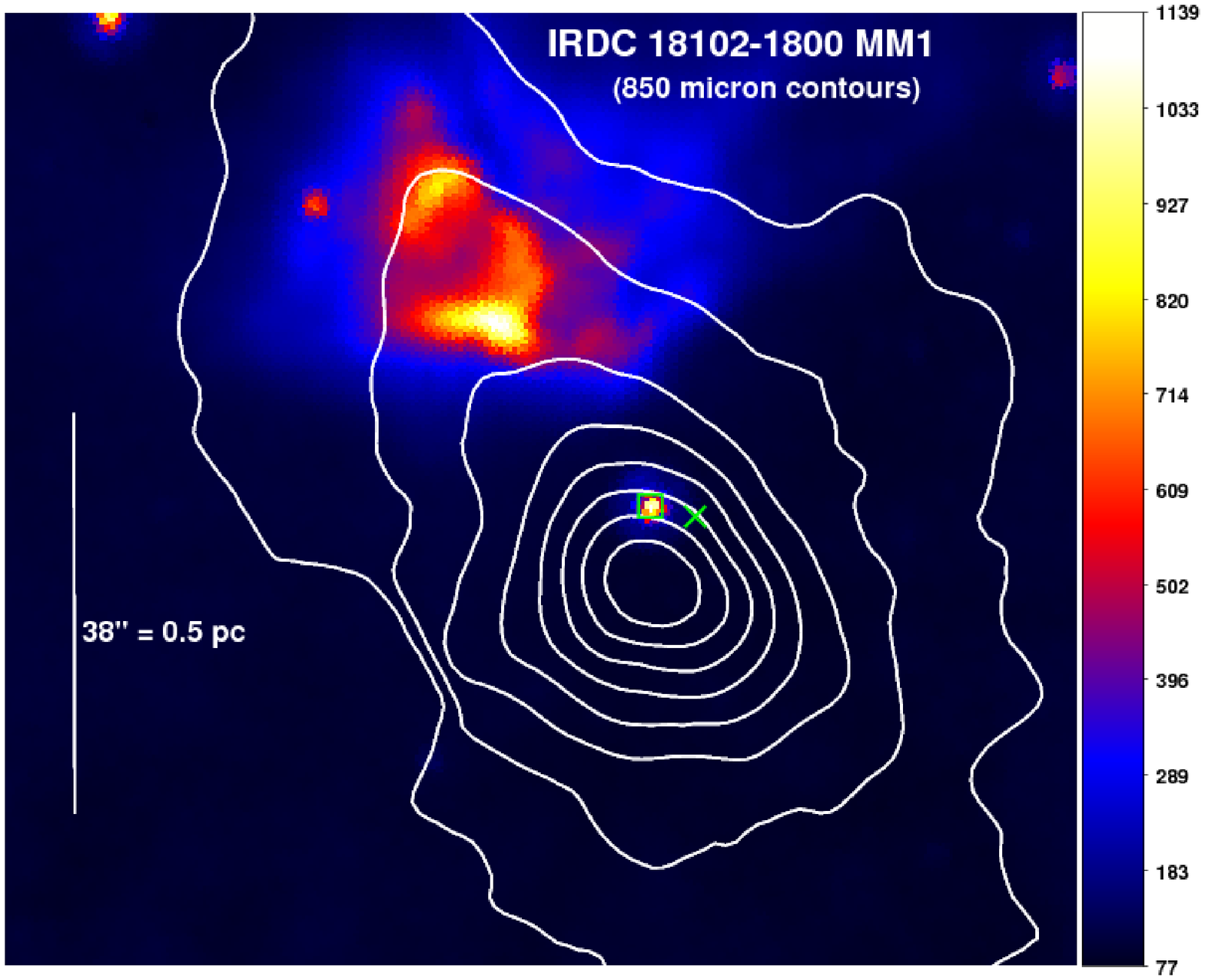}
\includegraphics[width=6cm]{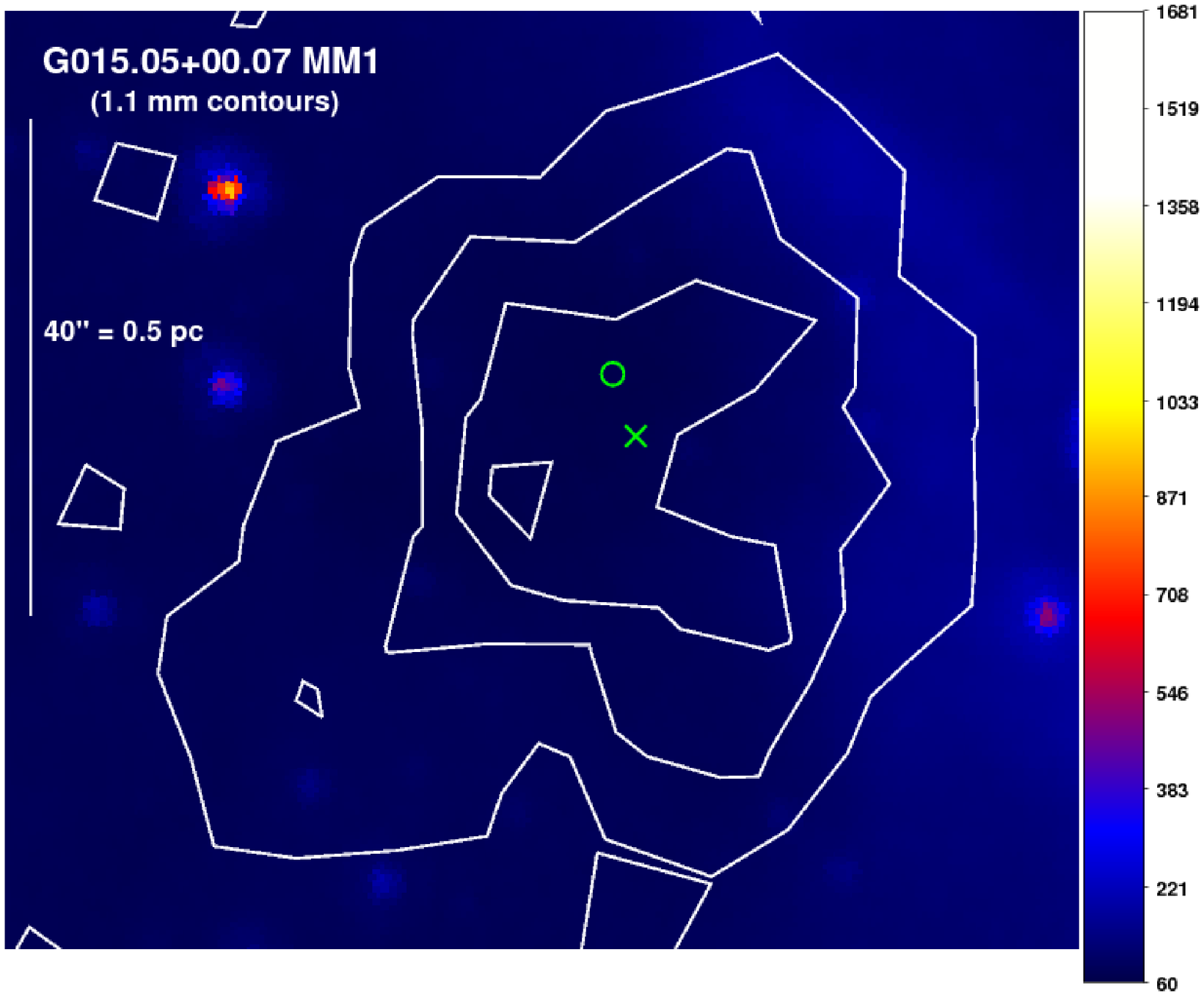}
\includegraphics[width=6cm]{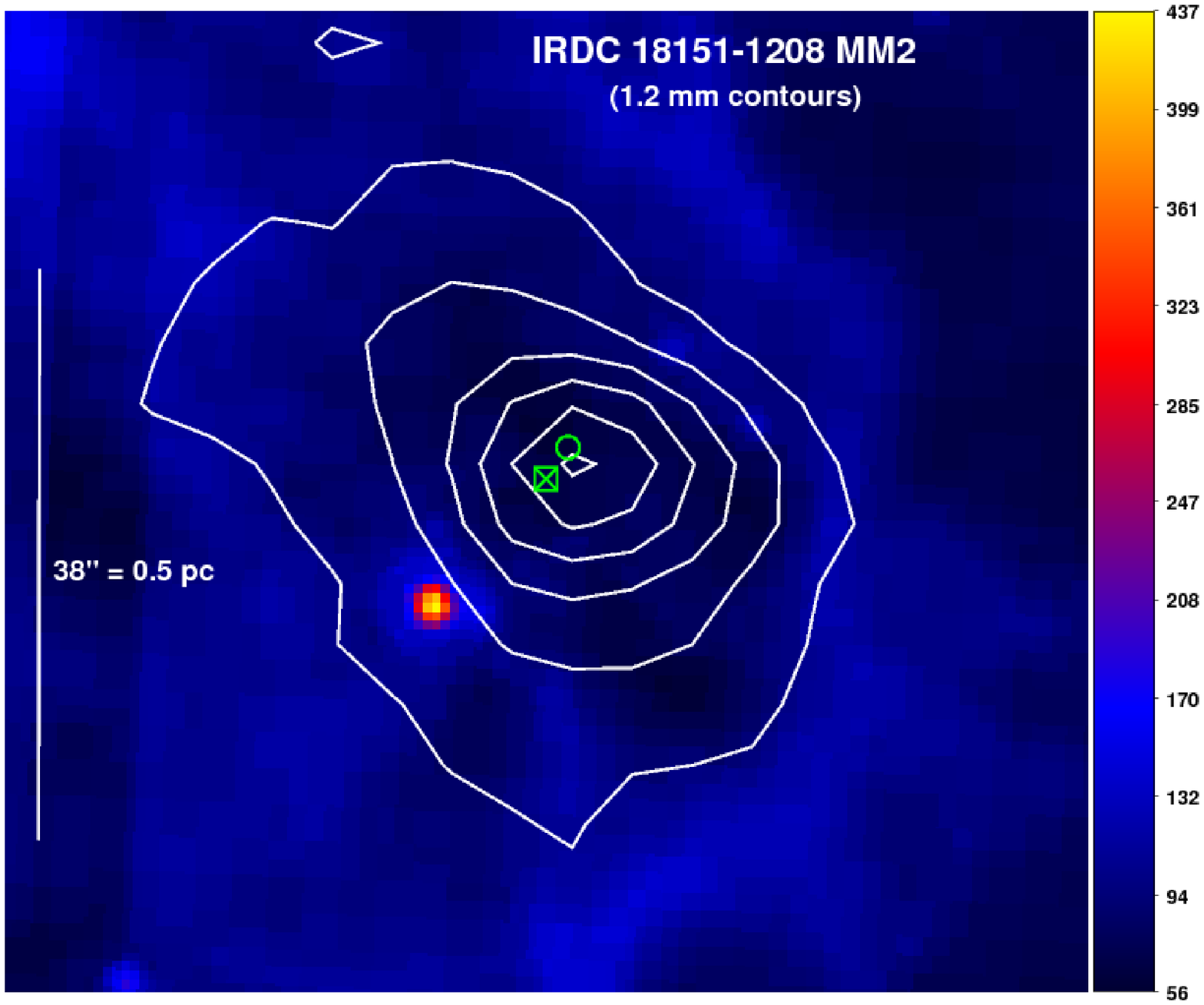}
\includegraphics[width=6cm]{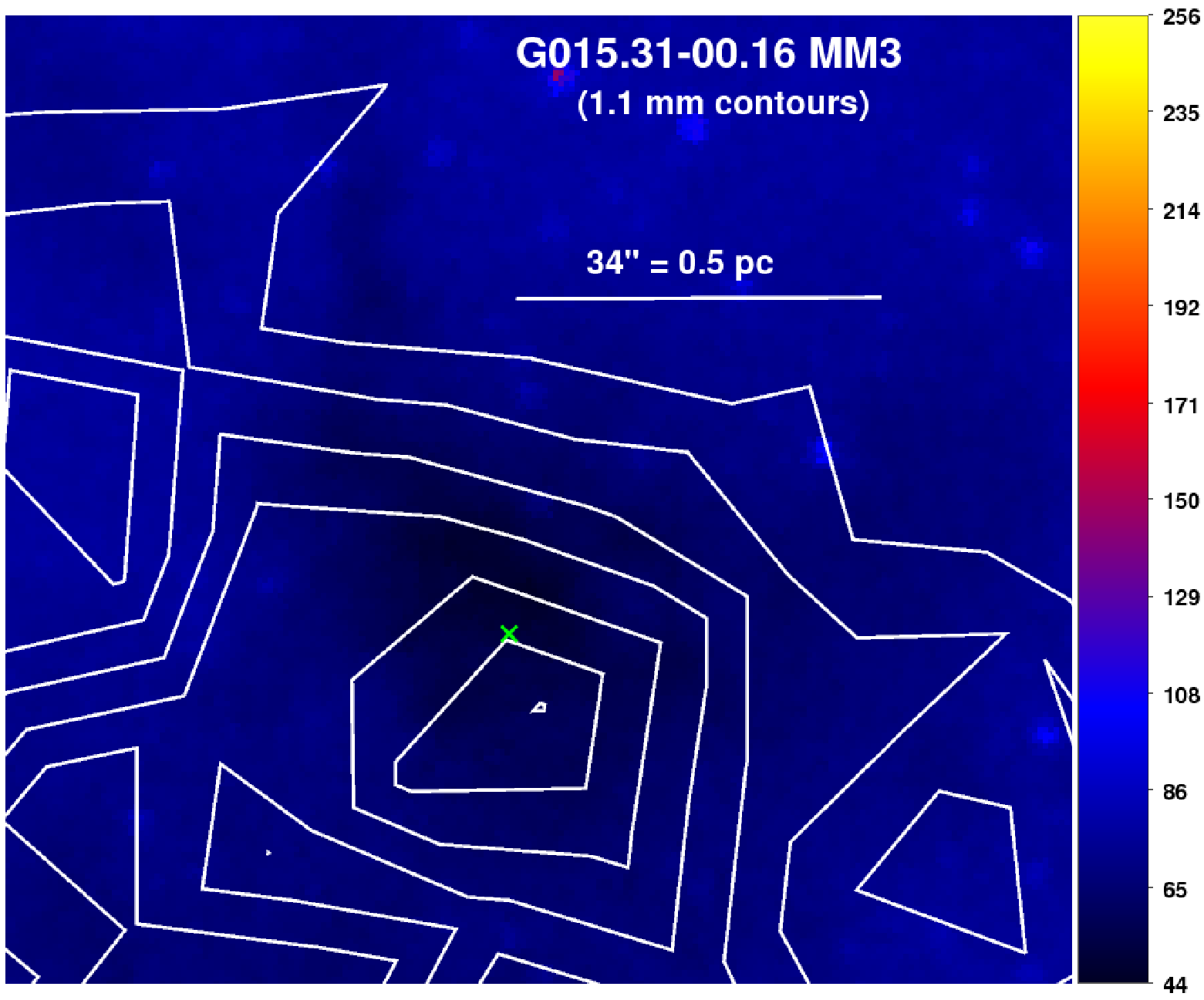}
\includegraphics[width=6cm]{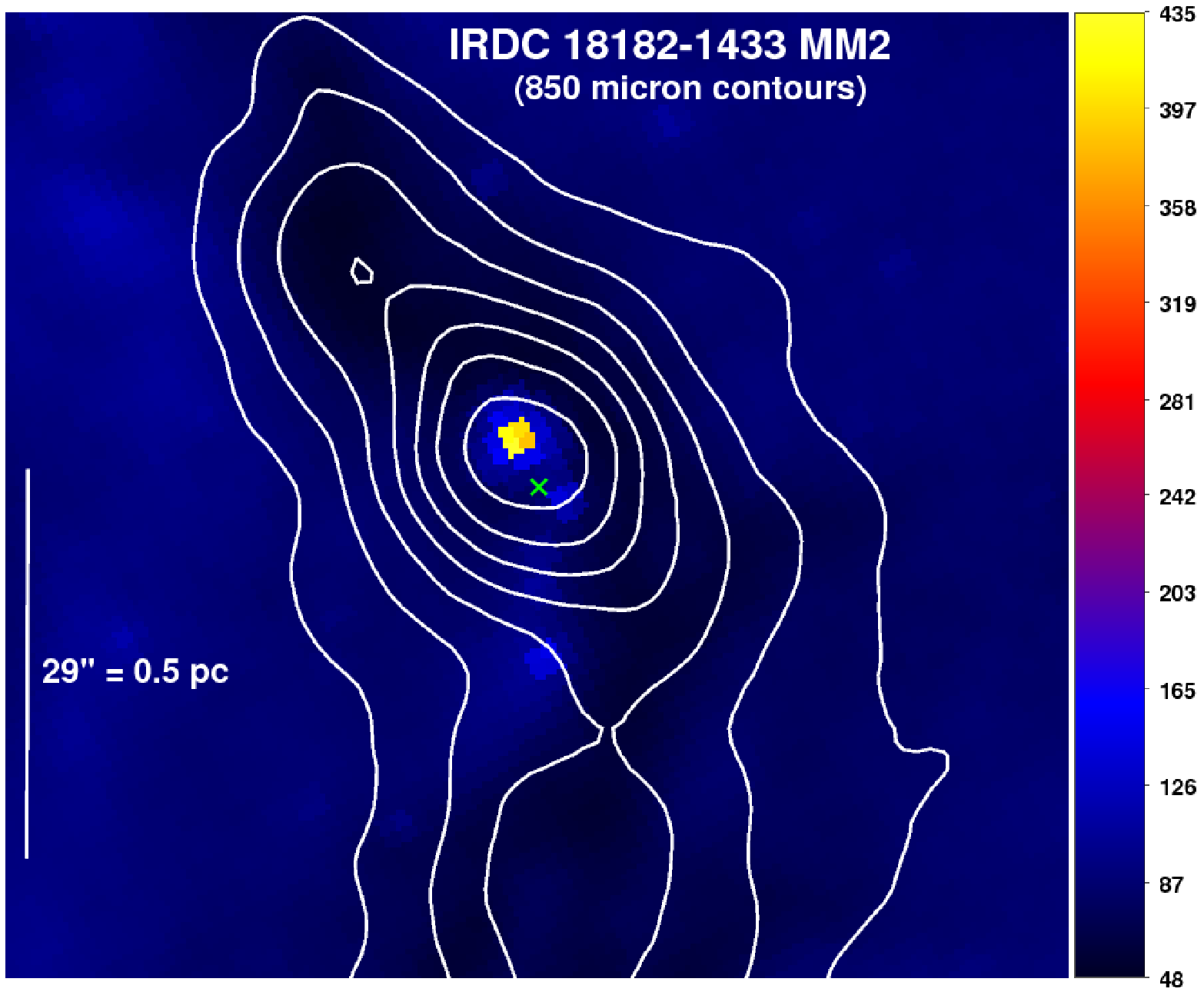}
\includegraphics[width=6cm]{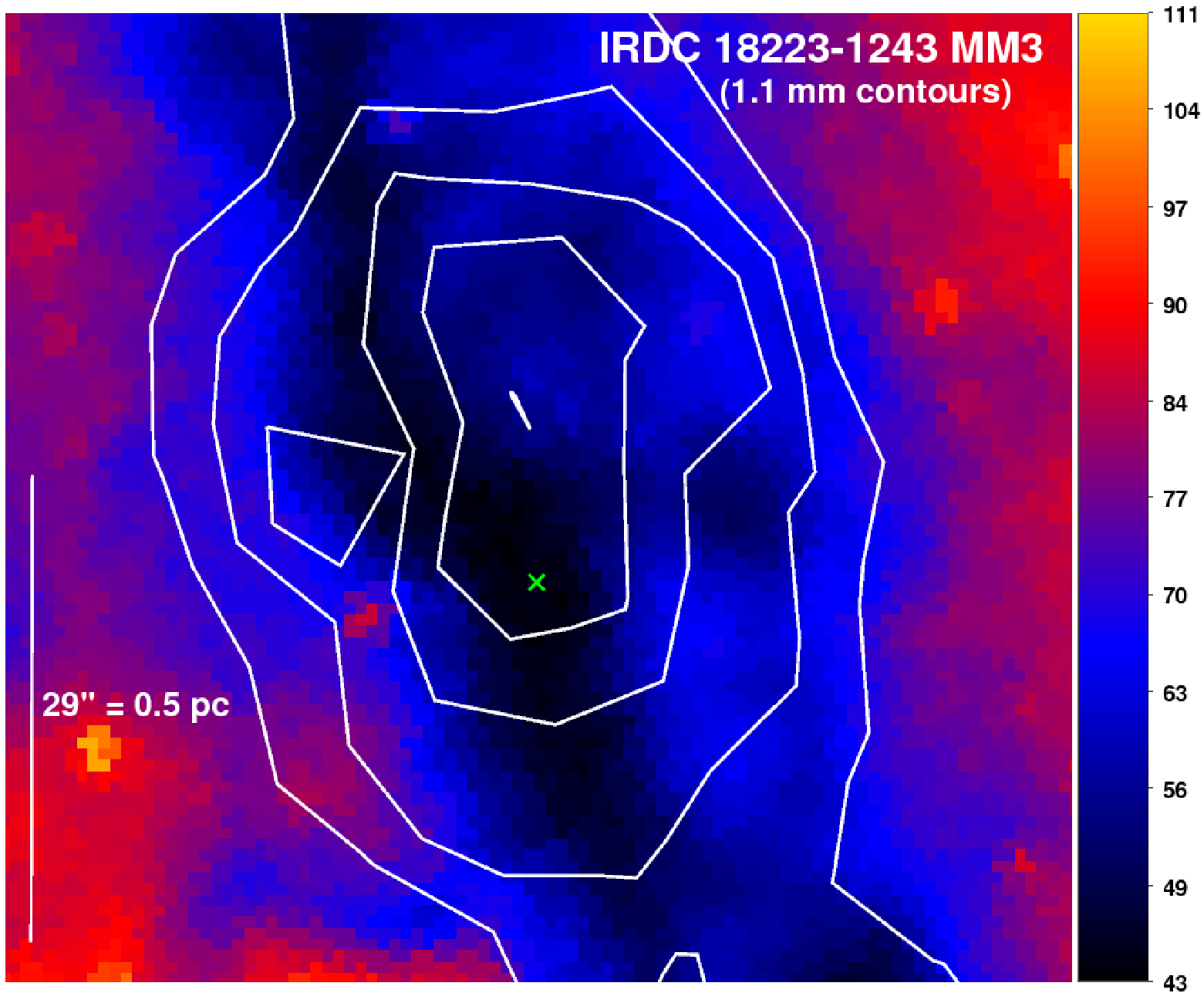}
\includegraphics[width=9cm]{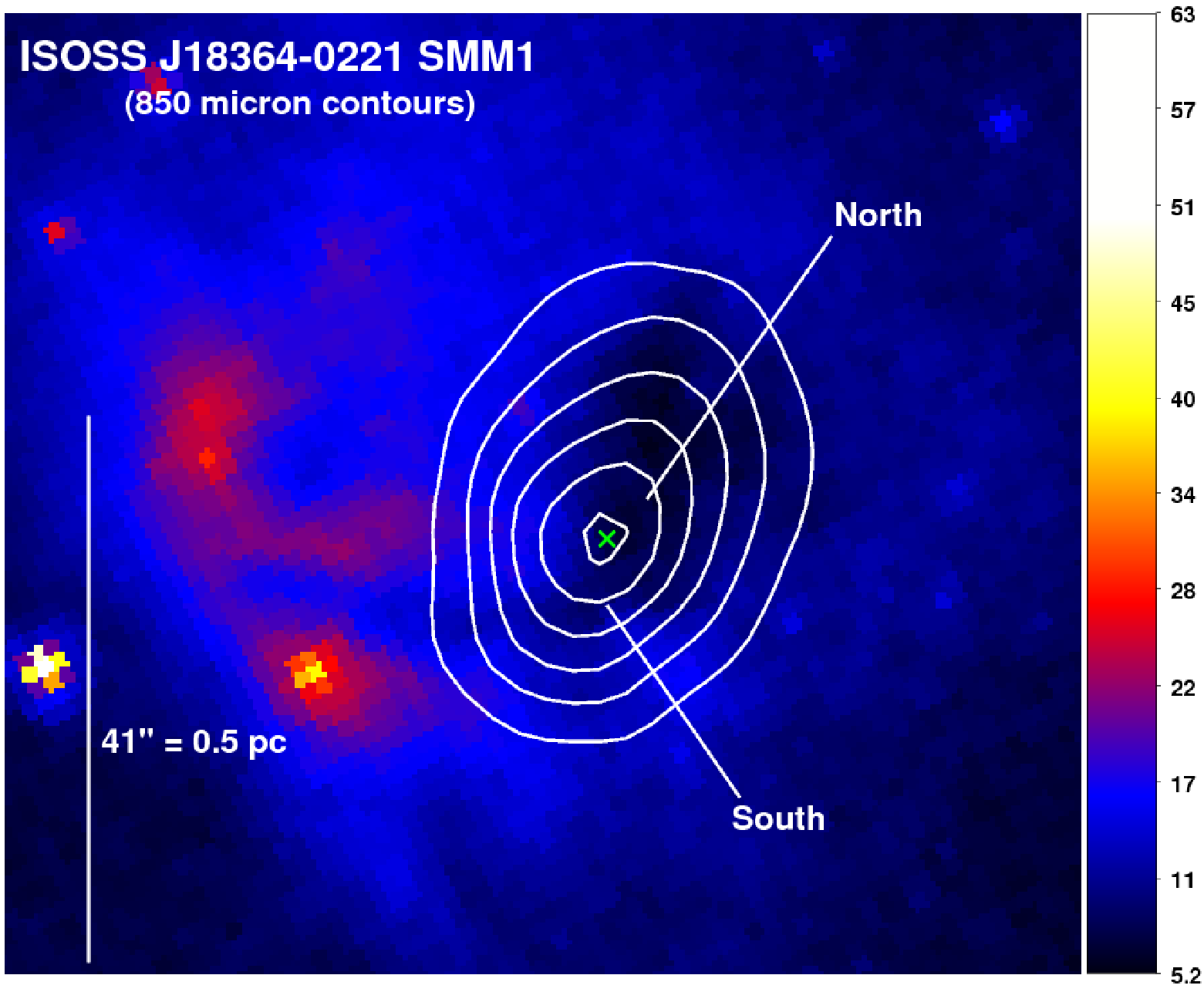}
\caption{\textit{Spitzer}/IRAC 8-$\mu$m images overlaid with contours of 
(sub)mm dust continuum emission (SCUBA 850 $\mu$m, Bolocam 1.1 mm, MAMBO 
1.2 mm). The colour bars indicate the 8-$\mu$m intensity scales in 
units of MJy~sr$^{-1}$. The contour step-size is 0.1 Jy~beam$^{-1}$ in all 
cases except G015.31 MM3 where it is 0.02 Jy~beam$^{-1}$. The contours go as 
follows: from 0.02 to 0.14 Jy~beam$^{-1}$ for G015.31 MM3; from 0.1 to 
0.7 Jy~beam$^{-1}$ for G015.05 MM1 and I18151 MM2; from 0.2 to 0.7 
Jy~beam$^{-1}$ for J18364 SMM1; from 0.2 to 0.9 Jy~beam$^{-1}$ for I18182 MM2; 
from 0.2 to 2.3 Jy~beam$^{-1}$ for I18102 MM1; and from 0.3 to 0.7 
Jy~beam$^{-1}$ for I18223 MM3. The crosses indicate the positions of our 
molecular-line observations. The box symbol towards I18102 MM1 
indicates the position of the 6.7 GHz Class {\footnotesize II} CH$_3$OH maser 
(\cite{beuther2002b}). For I18151 MM2, 
the box symbol shows the position of the 84.5 GHz Class {\footnotesize I} 
CH$_3$OH maser (\cite{marseille2010b}); it coincides with our target position. 
The circles indicate the positions of the 22.2 GHz H$_2$O masers 
associated with I18151 MM2 (\cite{beuther2002b}) and G015.05 MM1 
(\cite{wang2006}). The mm peak positions of the northern and southern cores 
within the clump J18364 SMM1 are indicated (\cite{hennemann2009}). 
The scale bar in each panel corresponds to 0.5 pc.}
\label{figure:maps}
\end{center}
\end{figure*}

In this study we also use \textit{Spitzer} IRAC 8-$\mu$m and MIPS 24-$\mu$m 
images of the sources to investigate the clump associations with 
protostellar activity. The 8-$\mu$m images are from the \textit{Spitzer} 
Galactic Legacy Infrared Mid-Plane Survey Extraordinaire (GLIMPSE; 
\cite{benjamin2003})\footnote{{\tt http://irsa.ipac.caltech.edu/data/SPITZER/GLIMPSE/}}. The 24-$\mu$m images (not shown in the paper) were taken from the 
\textit{Spitzer} MIPS GALactic plane survey (MIPSGAL; \cite{carey2009})\footnote{{\tt http://irsa.ipac.caltech.edu/data/SPITZER/MIPSGAL/}}.
In Fig.~\ref{figure:maps}, we show the \textit{Spitzer}/IRAC 8-$\mu$m images 
of the clumps overlaid with contours of (sub)mm dust continuum emission. 
We note that better quality \textit{Spitzer} data of the sources are 
available, and already published (see references in Appendix B). 
However, in the present work we do not use \textit{Spitzer} data to derive 
the phy\-sical properties of the sources; the data are used to illustrate the 
source appearance at MIR wavelengths.

\section{Observational results}

\subsection{Spectra}

The Hanning-smoothed spectra are shown in Fig.~\ref{figure:spectra}. Note 
that all data are presented in units of $T_{\rm A}^*$. The C$^{17}$O$(2-1)$ 
line is clearly detected towards all sources but the hf structure of 
the line is completely blended in all cases except G015.31 MM3 where 
it is partially resolved. The non-resolution of the hf structure is 
caused by the linewidths to be larger than the separation in velocity of 
individual hf components. 

The H$^{13}$CO$^+(3-2)$ line is also clearly detected towards all 
target positions but towards G015.31 MM3 the line is very weak (only 
$\sim3\sigma$). The ``double peak''-like profiles in some of the 
smoothed H$^{13}$CO$^+(3-2)$ spectra can be attributed to the noise in most 
cases. In the case of J18364 SMM1, the self-absorption dip is a distinctive 
feature also in the unsmoothed spectrum, and thus likely to be real.
DCO$^+(3-2)$ emission is seen towards all sources, but 
the line is very weak in I18102 MM1 and G015.31 MM3 (detected only at 
$\sim3\sigma$). 

N$_2$H$^+(3-2)$ line-observations were performed only towards three of our 
sources, and the line is quite strong in all cases. The N$_2$H$^+(3-2)$ 
line of I18102 MM1 exhibits red asymmetry with a central dip near the 
velocity of the strongest hf component; this feature is 
also clearly visible in the unsmoothed spectrum. The N$_2$D$^+(3-2)$ line is 
detected towards all sources except G015.31 MM3, I18182 MM2, and 
J18364 SMM1 in which cases the spectra have a low S/N ratio. 

\begin{figure*}
\begin{center}
\includegraphics[width=2.54cm, angle=-90]{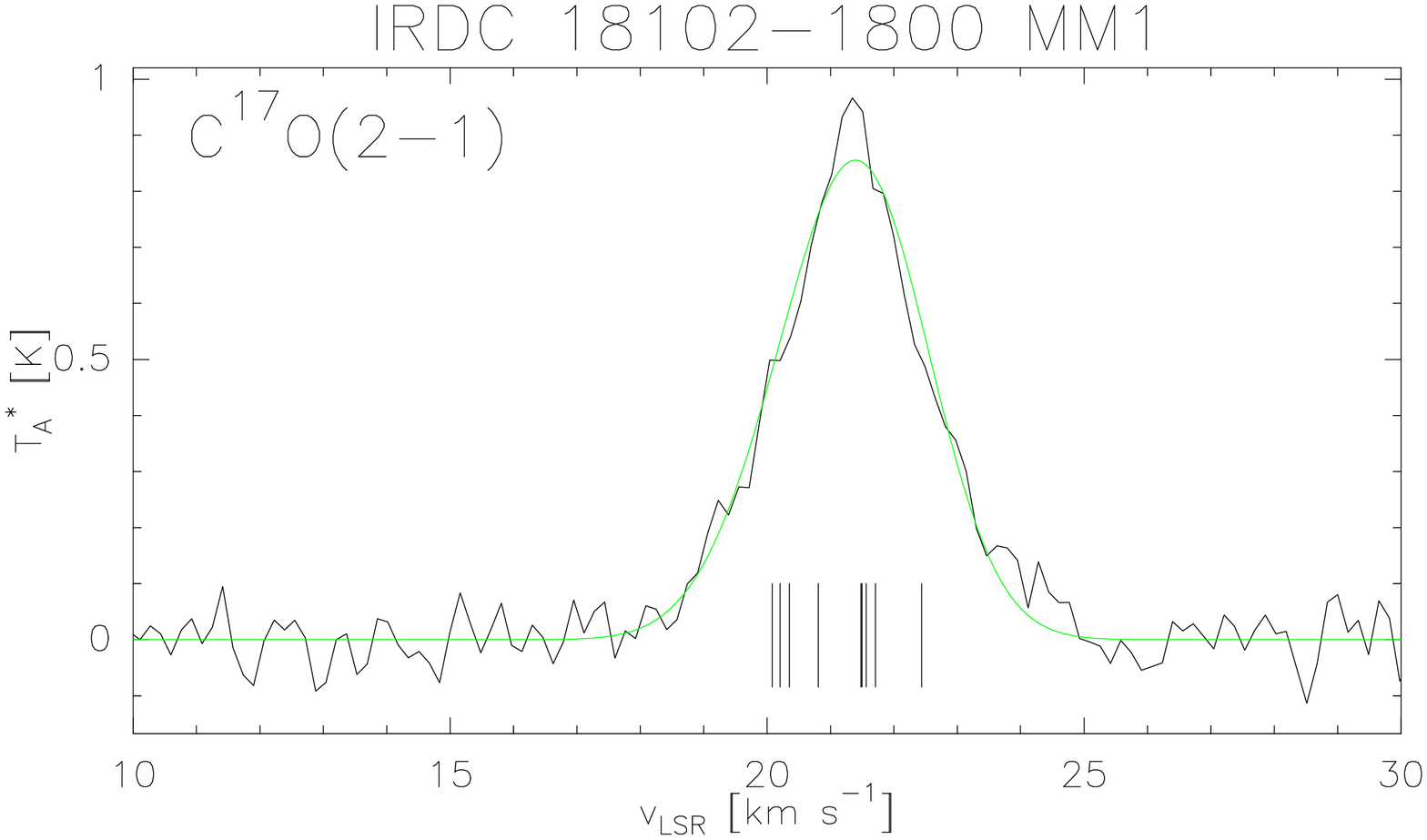}
\includegraphics[width=2.54cm, angle=-90]{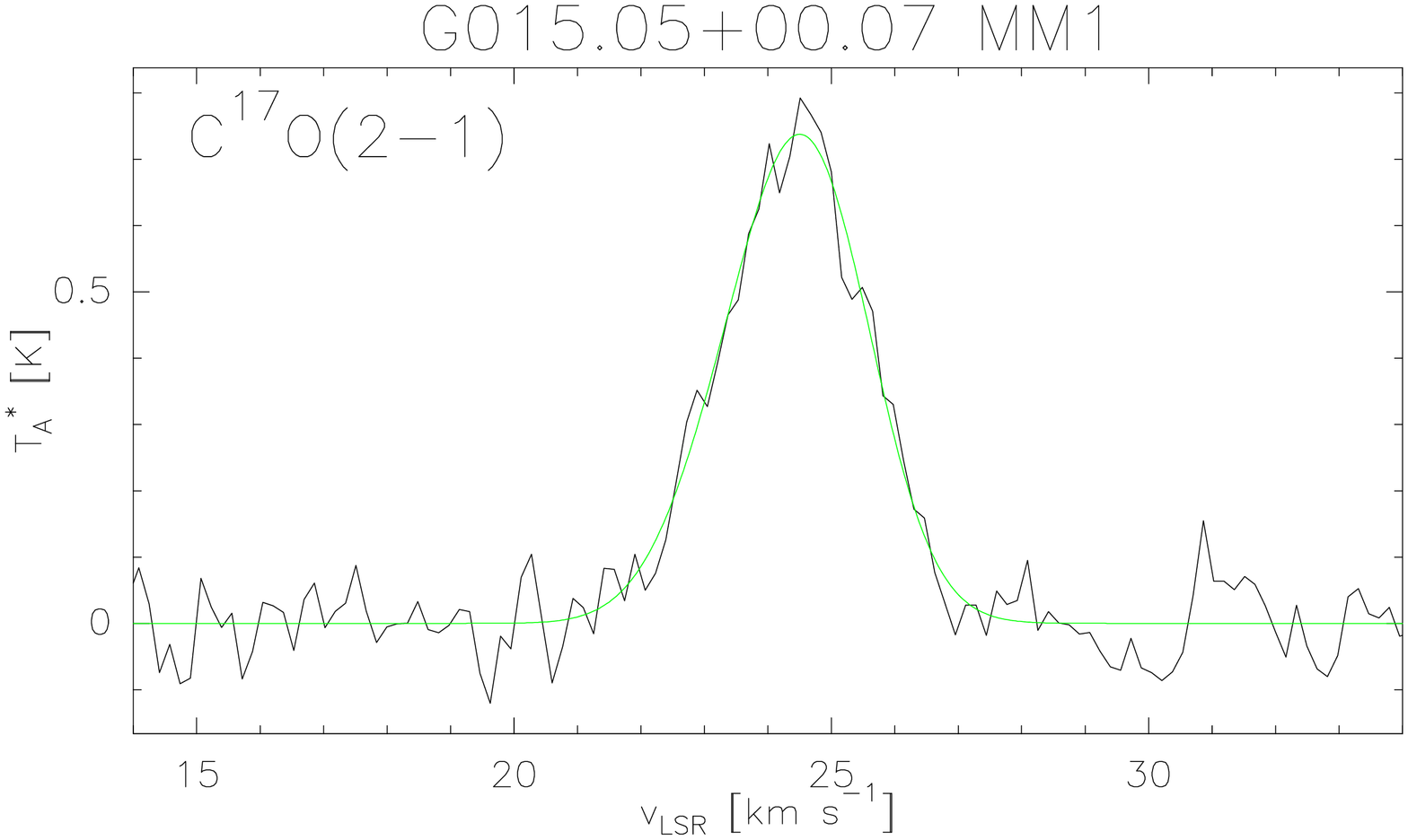}
\includegraphics[width=2.54cm, angle=-90]{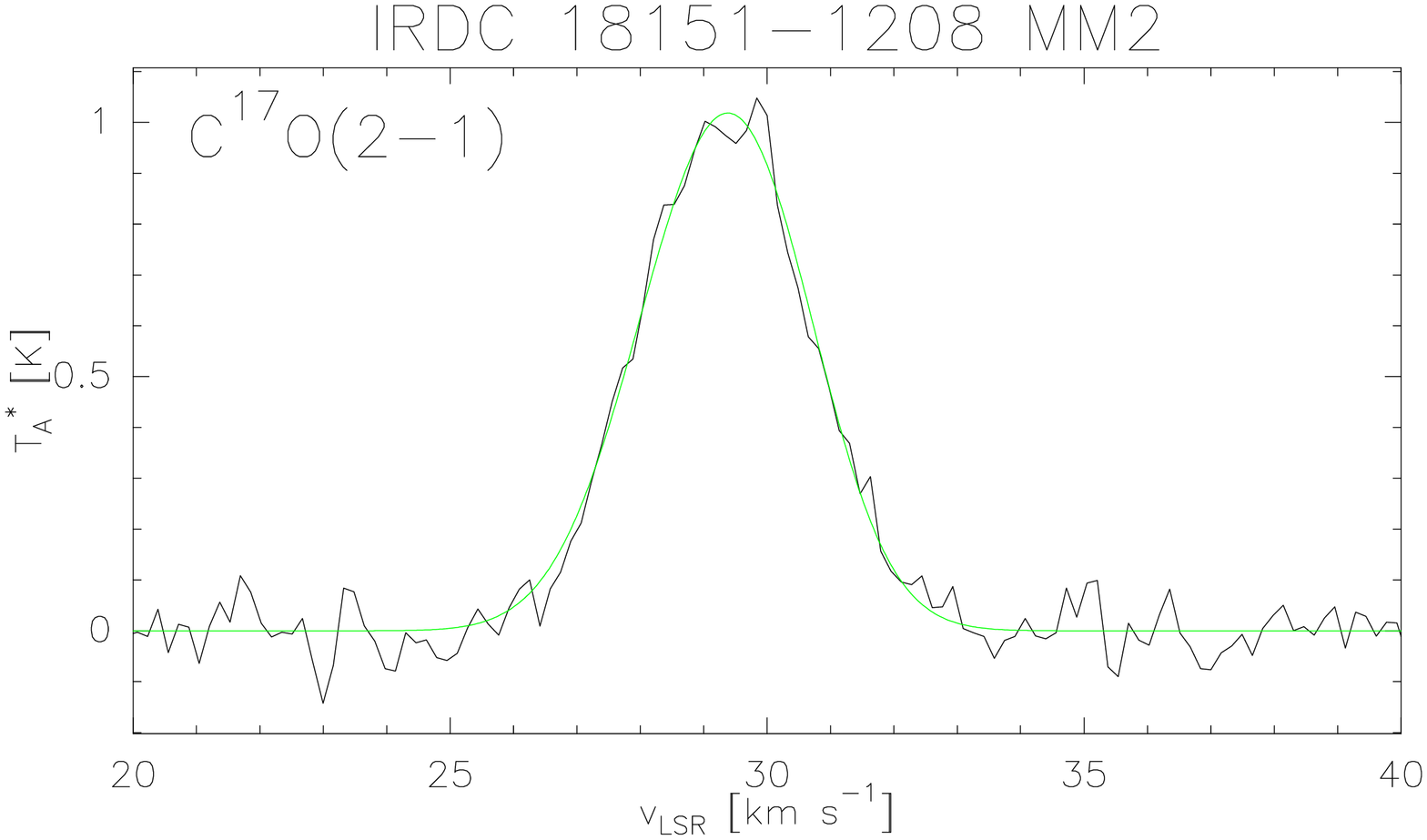}
\includegraphics[width=2.54cm, angle=-90]{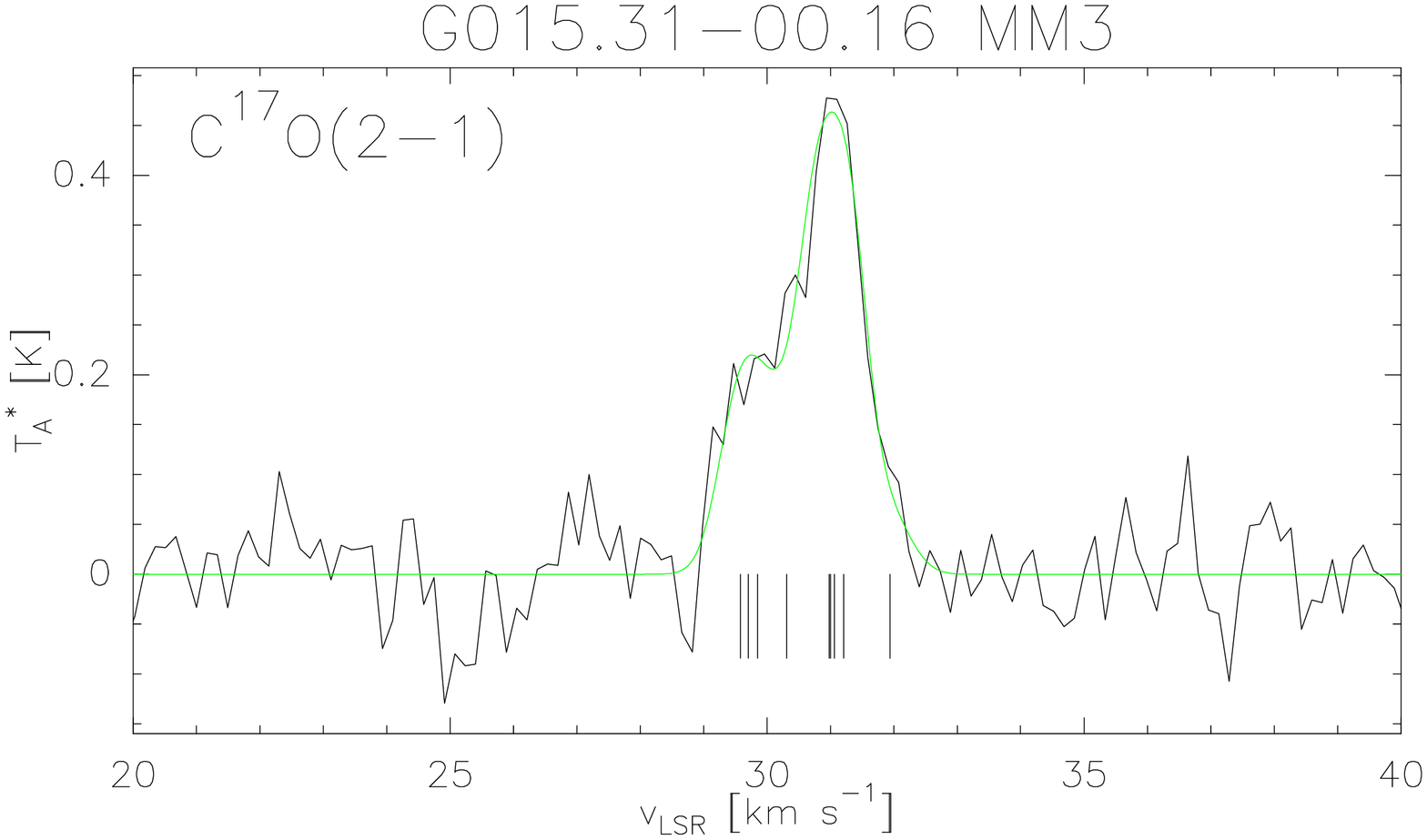}
\includegraphics[width=2.4cm, angle=-90]{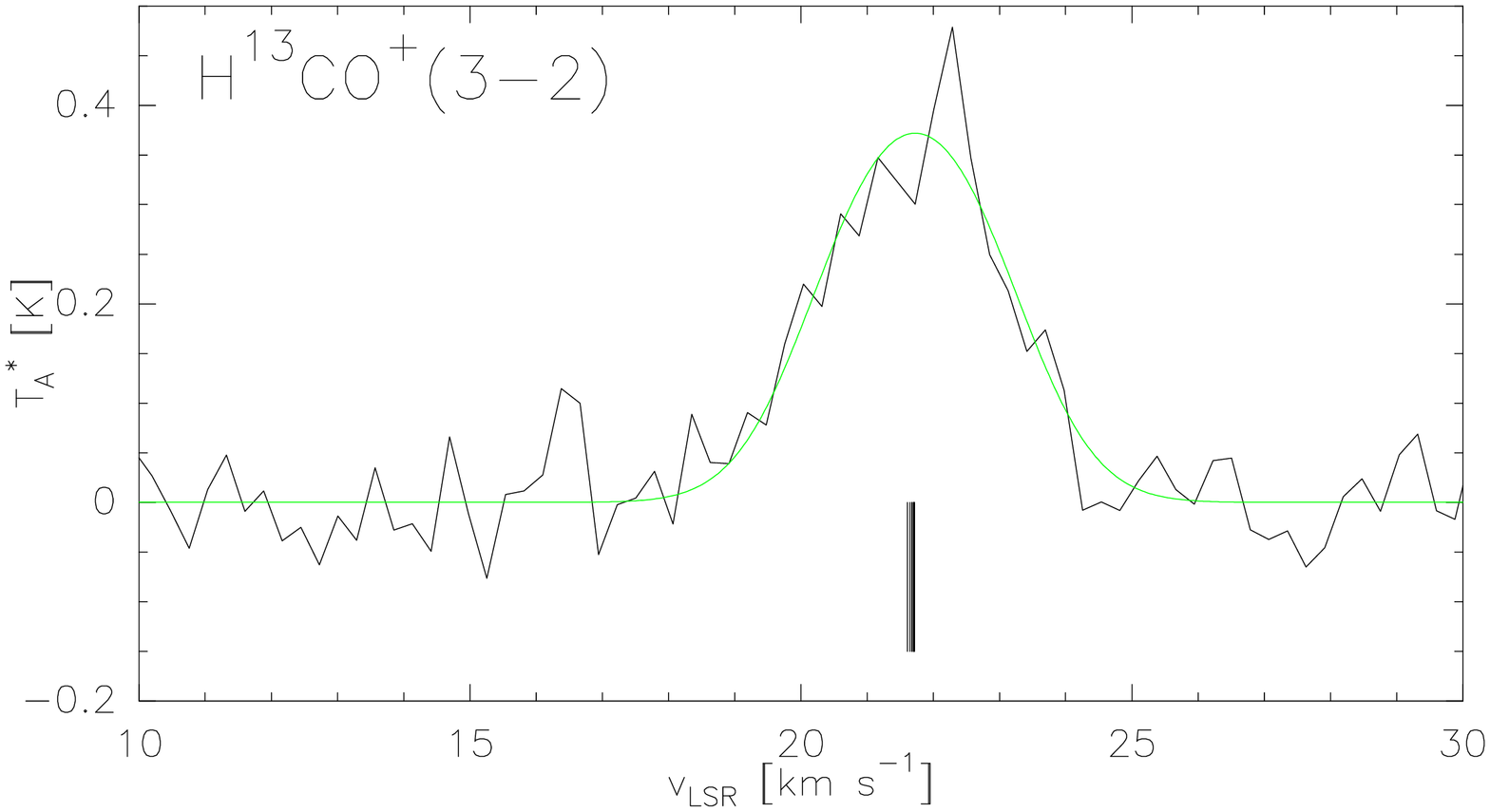}
\includegraphics[width=2.4cm, angle=-90]{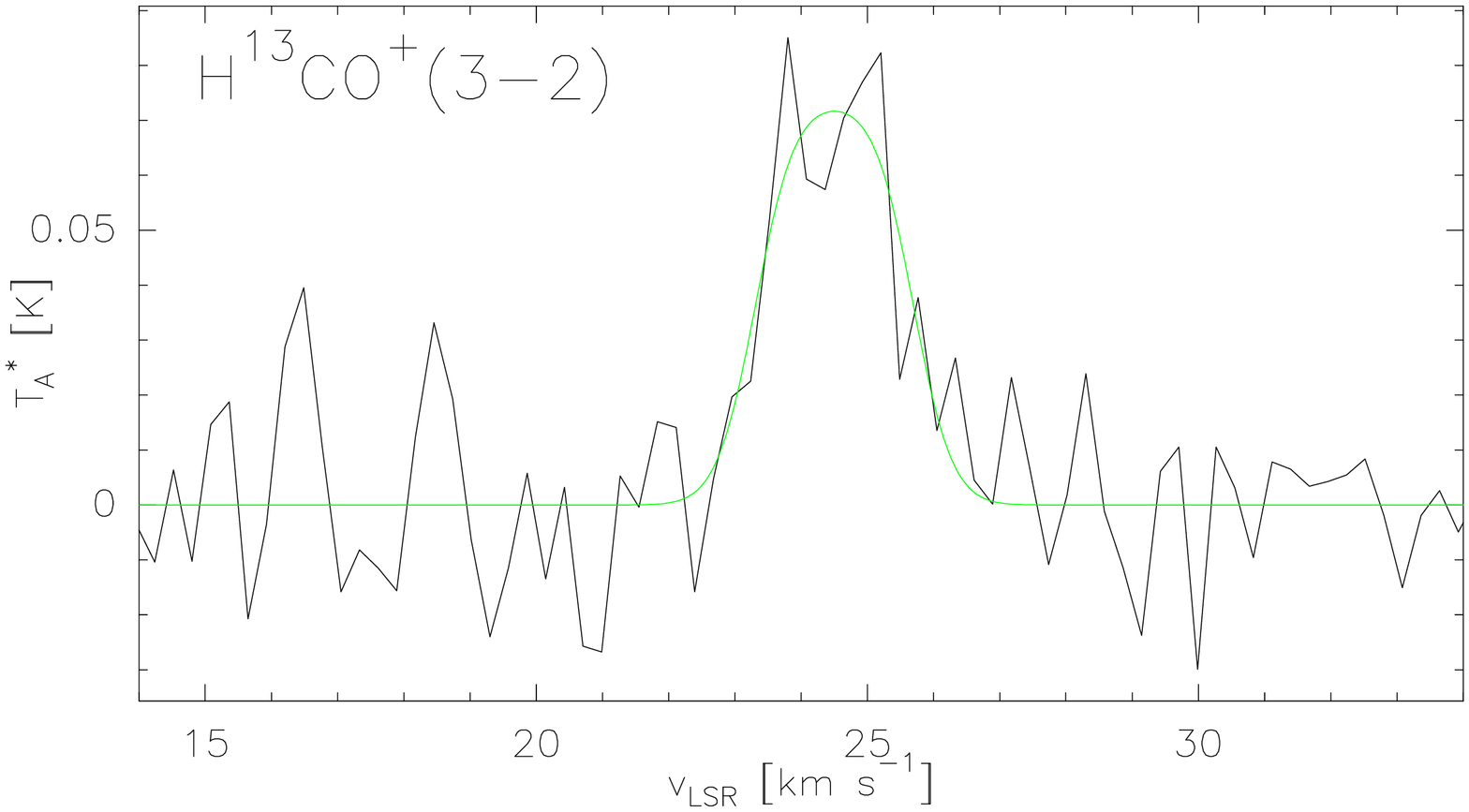}
\includegraphics[width=2.4cm, angle=-90]{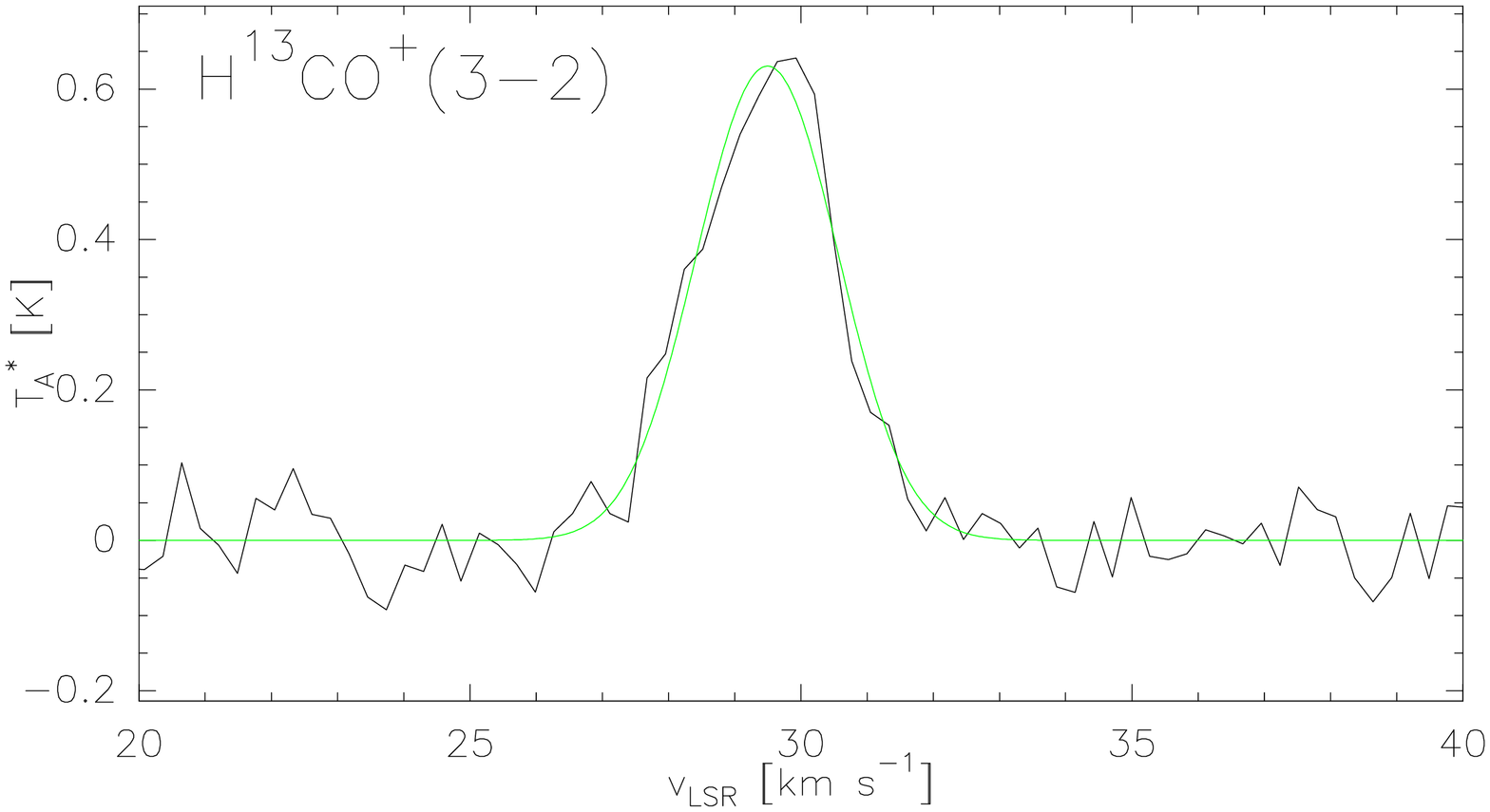}
\includegraphics[width=2.4cm, angle=-90]{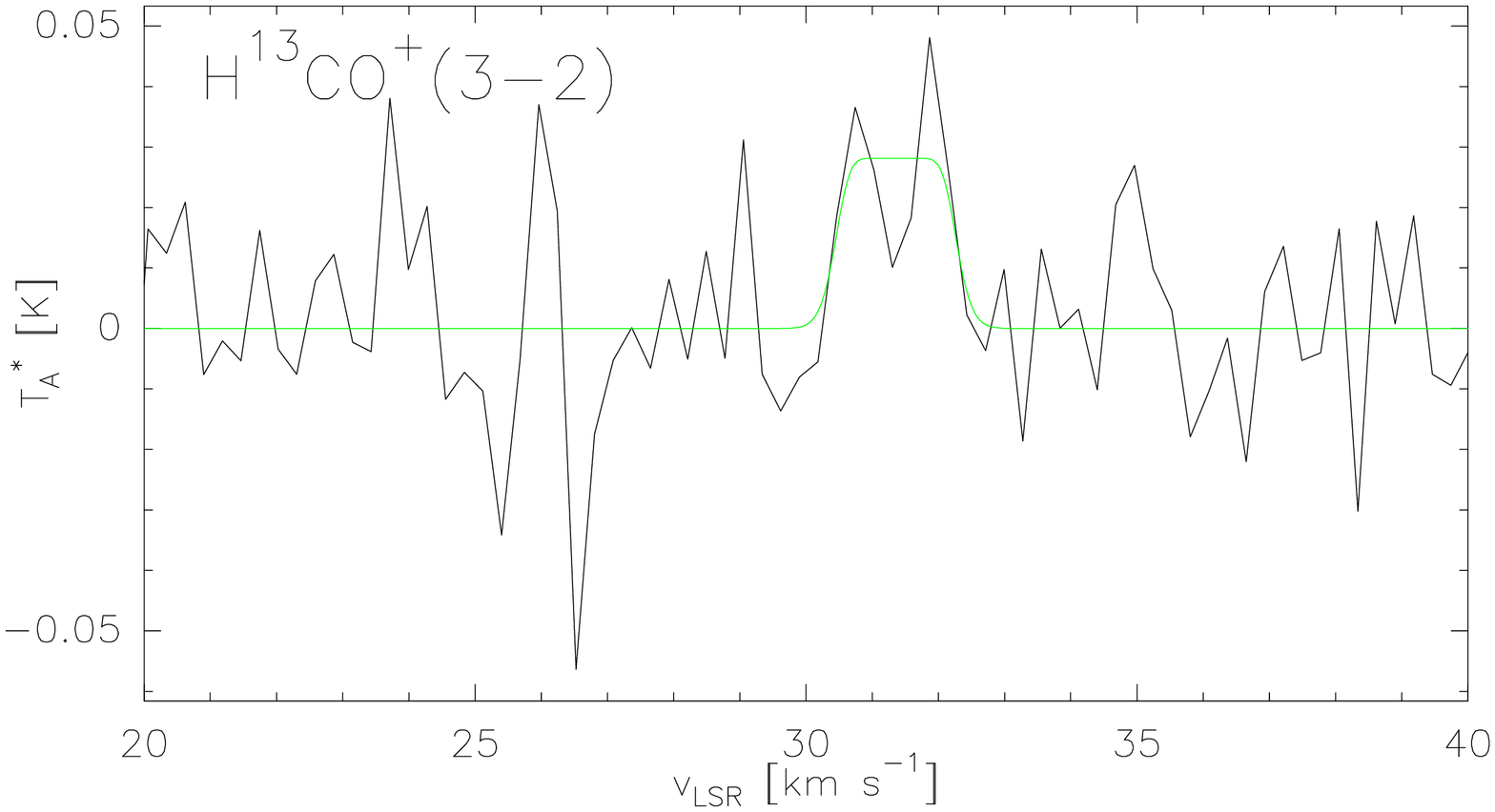}
\includegraphics[width=2.4cm, angle=-90]{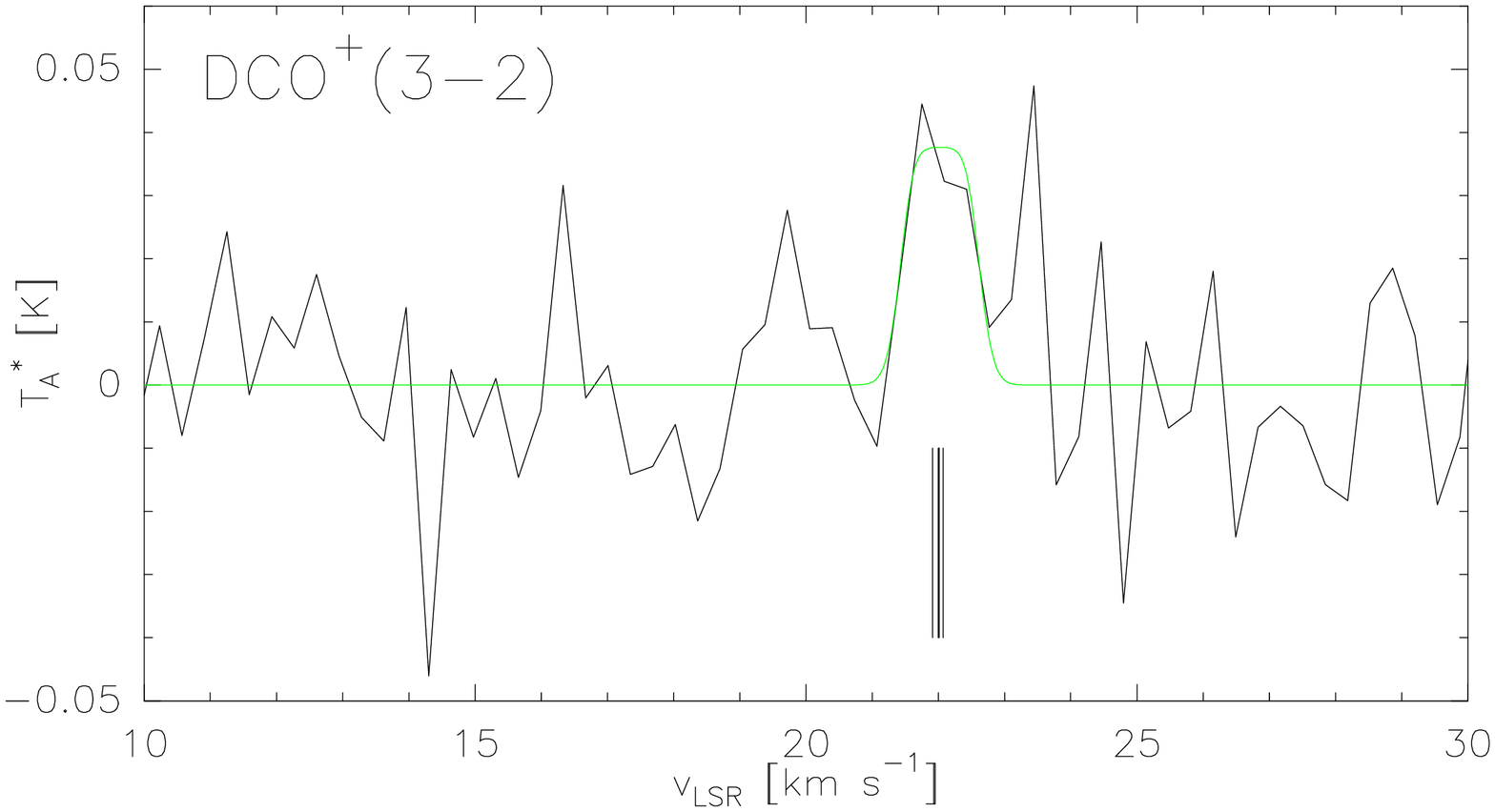}
\includegraphics[width=2.4cm, angle=-90]{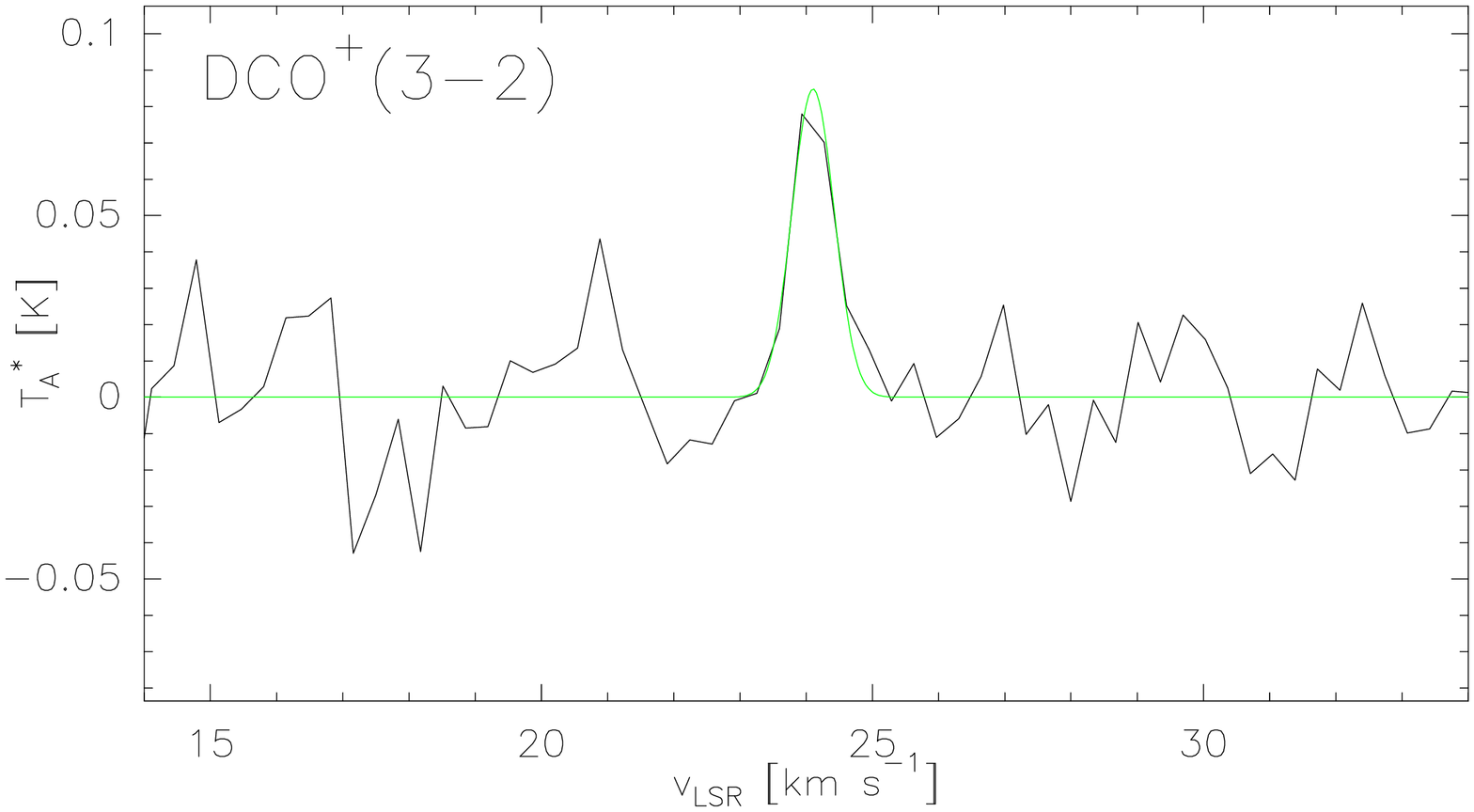}
\includegraphics[width=2.4cm, angle=-90]{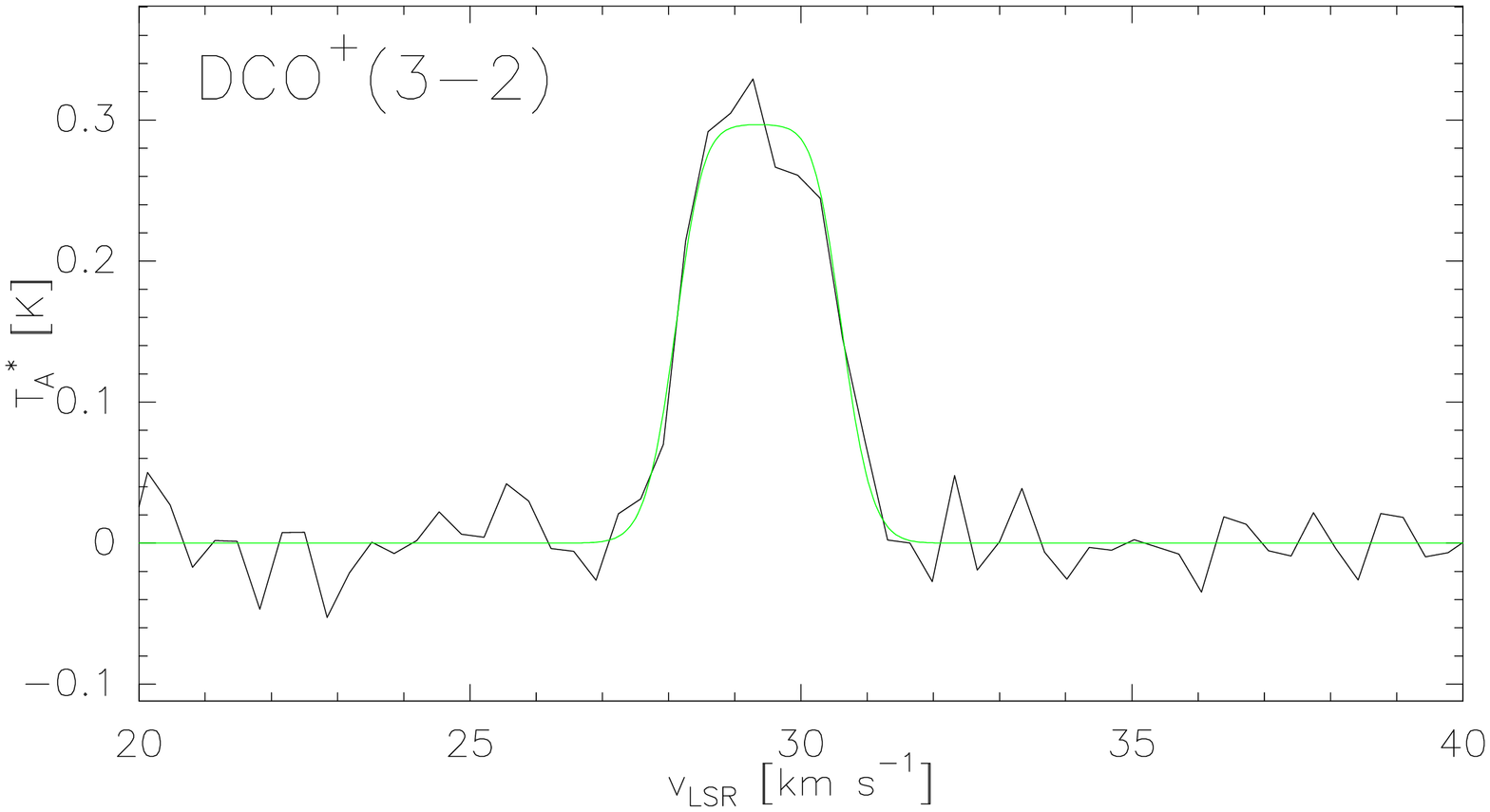}
\includegraphics[width=2.4cm, angle=-90]{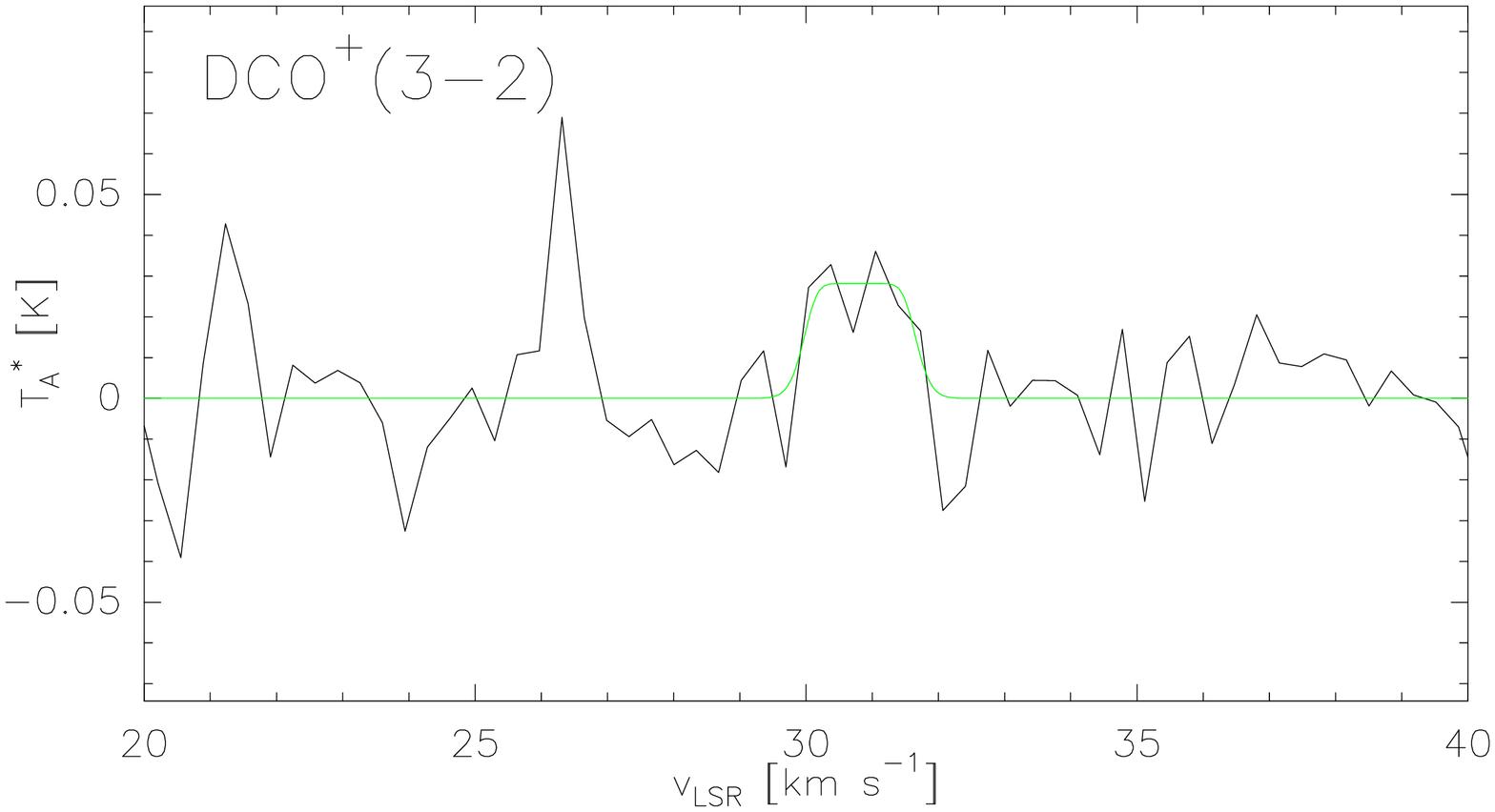}
\includegraphics[width=2.4cm, angle=-90]{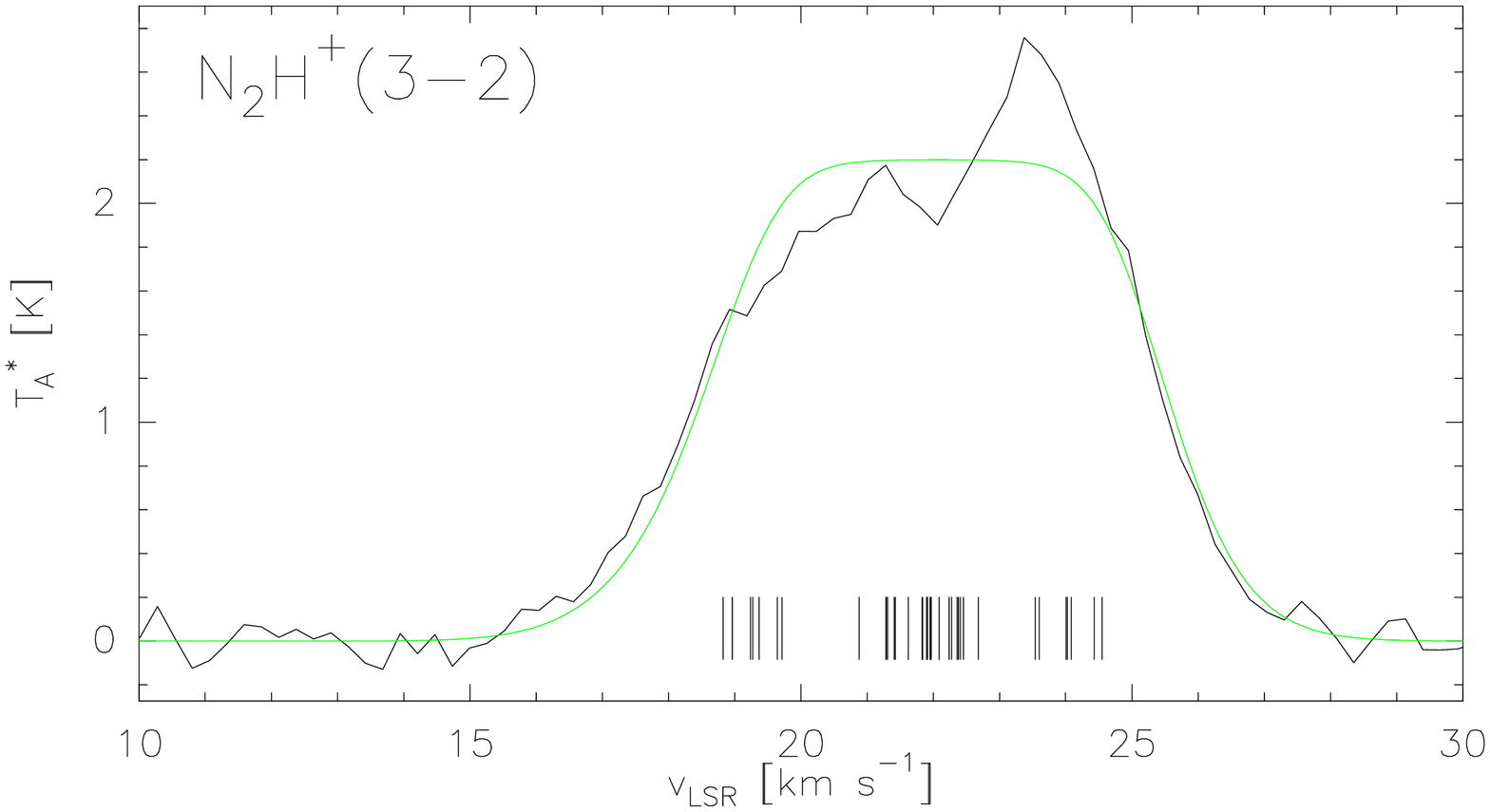}
\includegraphics[width=2.4cm, angle=-90]{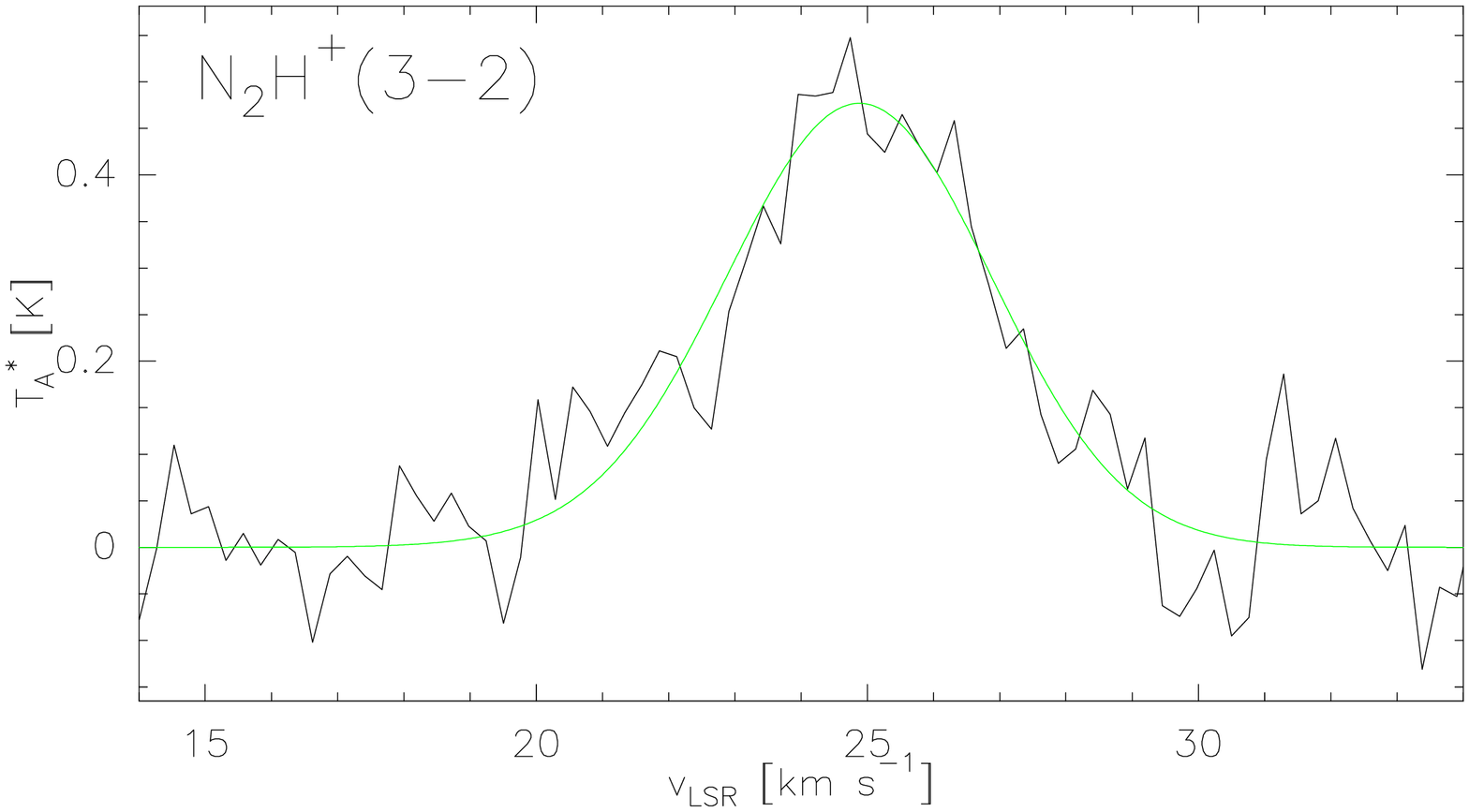}
\includegraphics[width=2.46cm, angle=-90]{blank.eps}
\includegraphics[width=2.4cm, angle=-90]{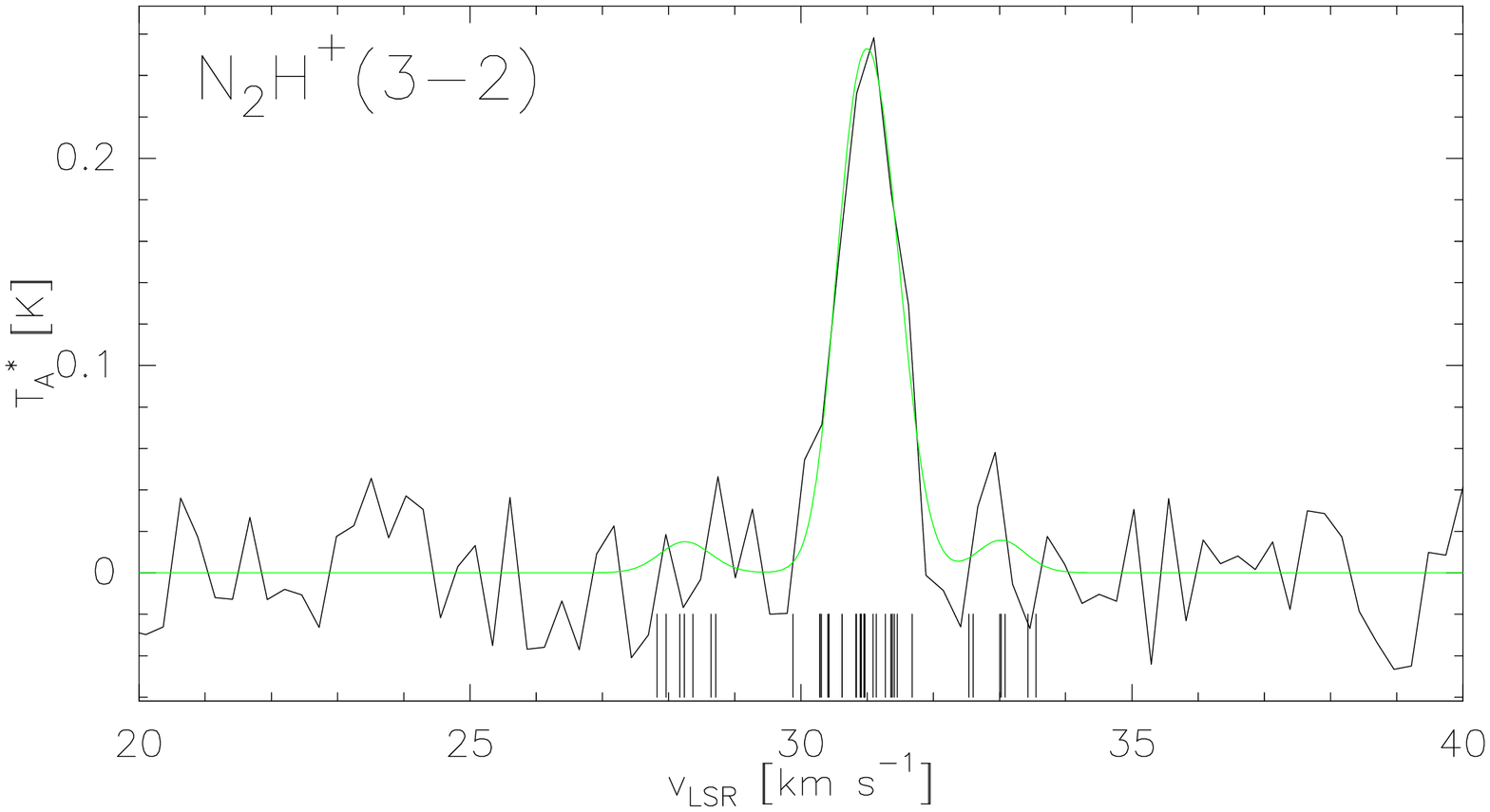}
\includegraphics[width=2.4cm, angle=-90]{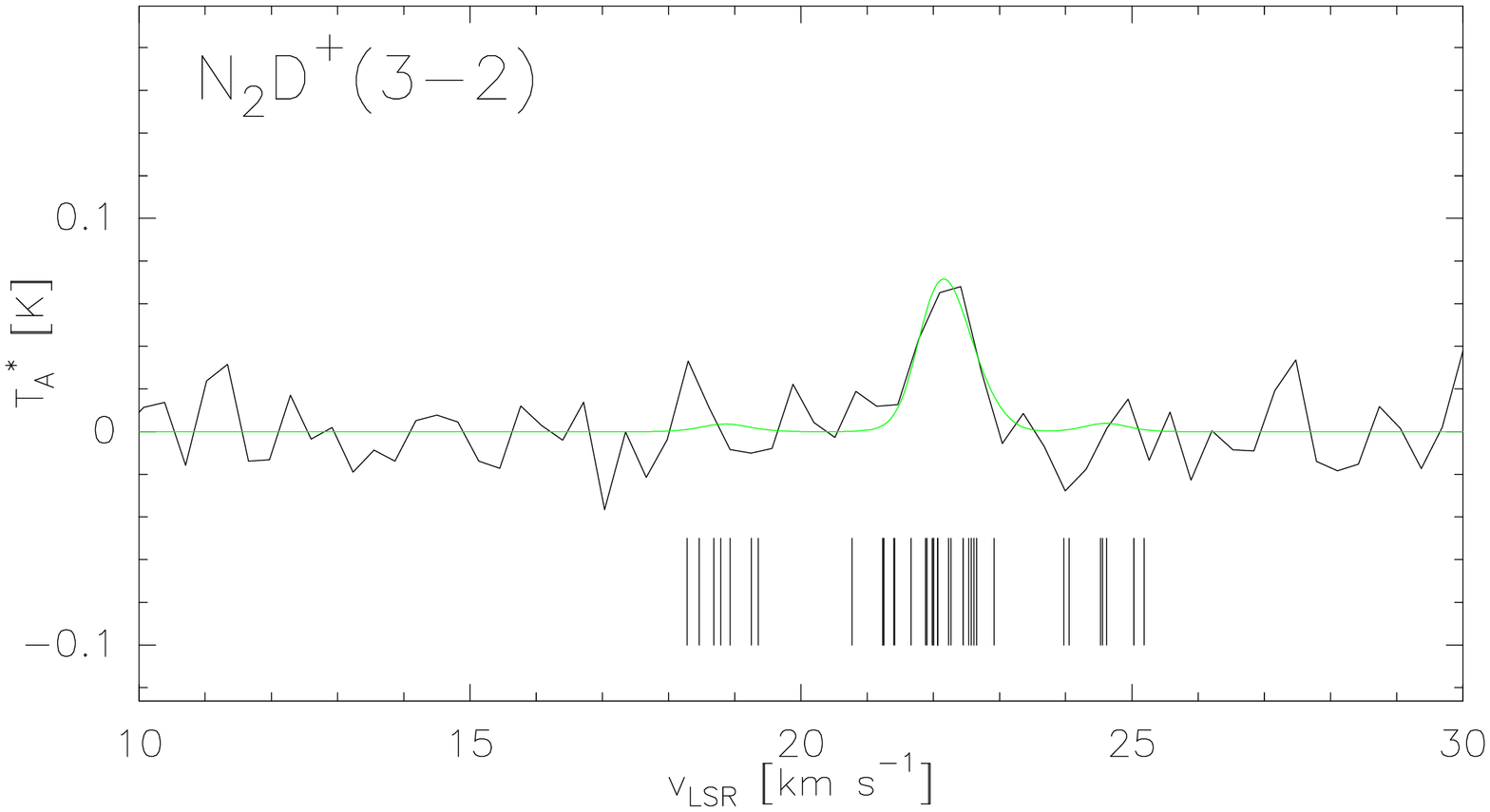}
\includegraphics[width=2.4cm, angle=-90]{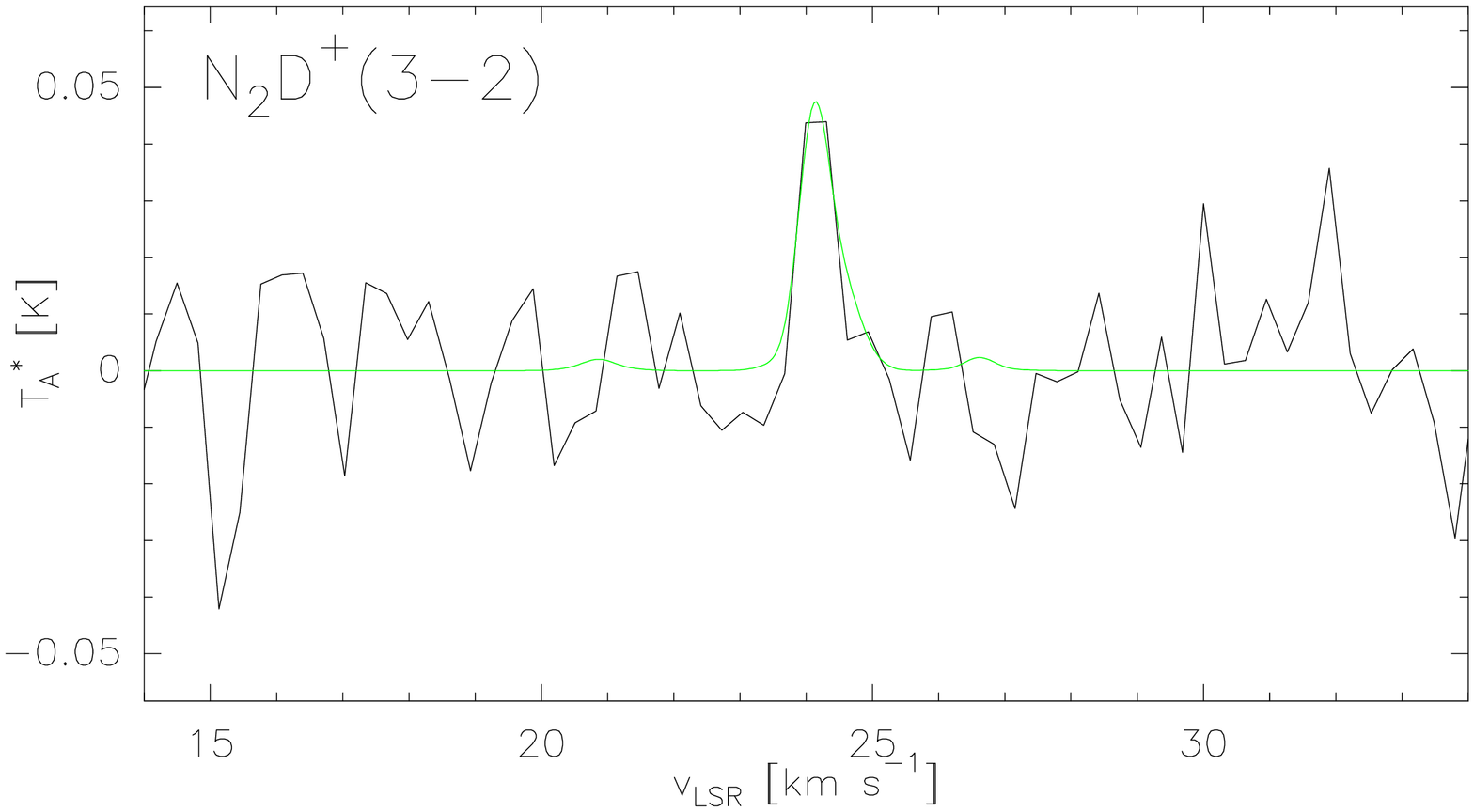}
\includegraphics[width=2.4cm, angle=-90]{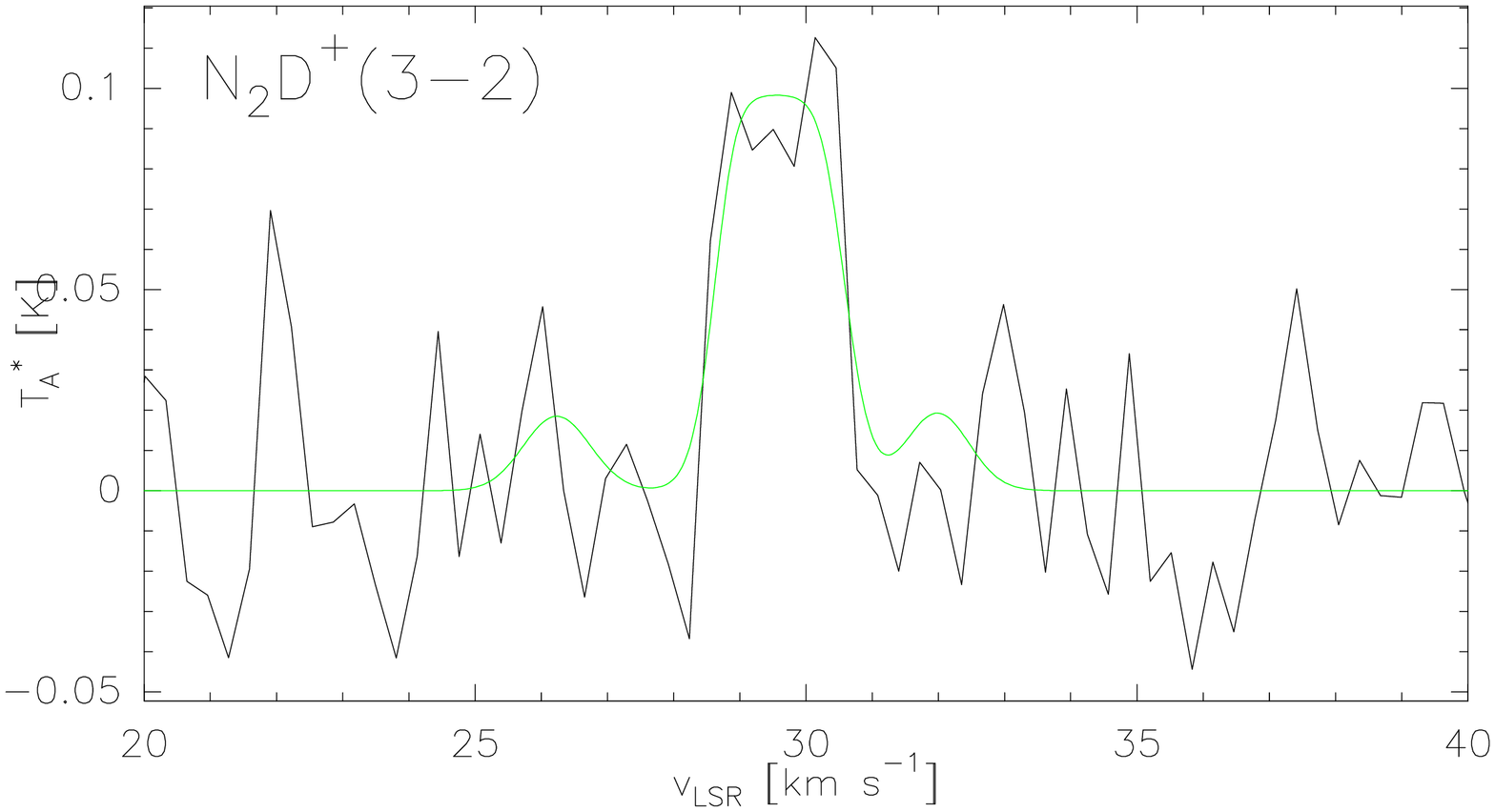}
\includegraphics[width=2.4cm, angle=-90]{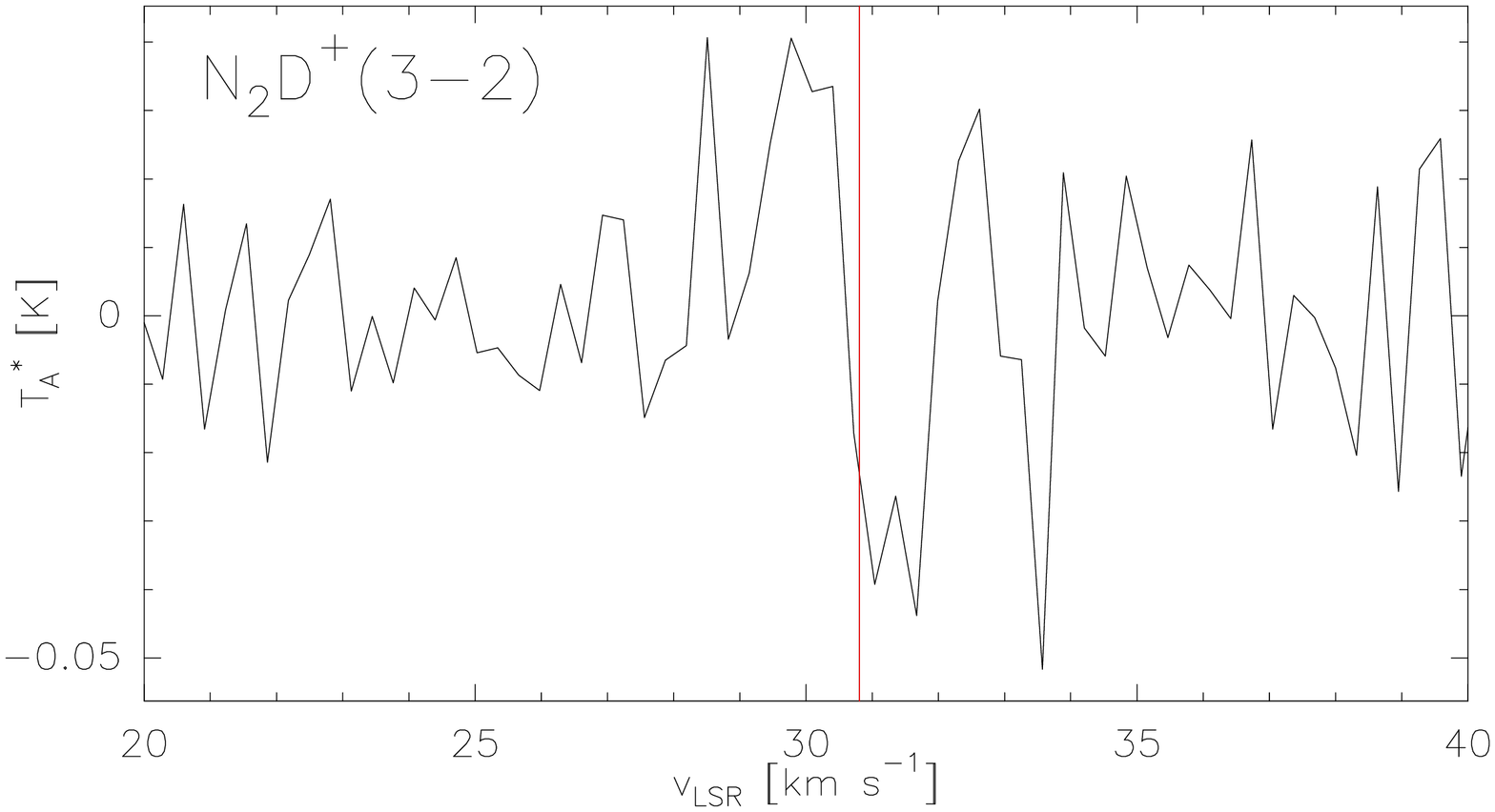}
\caption{Smoothed C$^{17}$O$(2-1)$, H$^{13}$CO$^+(3-2)$, DCO$^+(3-2)$, 
N$_2$H$^+(3-2)$, and N$_2$D$^+(3-2)$ spectra (\textit{top to bottom}) 
towards the clumps (\textit{left to right}). Overlaid on the spectra 
are the hf-structure fits. The relative velocities of each individual 
hf components in each observed transition are labelled with a short 
bar on the spectra towards I18102 MM1; these are also shown on the 
C$^{17}$O and N$_2$H$^+$ spectra towards G015.31 MM3. The red 
vertical line on the N$_2$D$^+$ spectrum towards G015.31 MM3 shows the 
${\rm v_{\rm LSR}}=30.81$ km~s$^{-1}$ of the clump as measured from 
N$_2$H$^+(1-0)$ by SSK08. Note that N$_2$H$^+(3-2)$ observations were carried 
out only towards three sources and that N$_2$D$^+(3-2)$ was not detected 
towards G015.31 MM3, I18182 MM2, and J18364 SMM1.}
\label{figure:spectra}
\end{center}
\end{figure*}

\begin{figure*}
\begin{center}
\includegraphics[width=2.7cm, angle=-90]{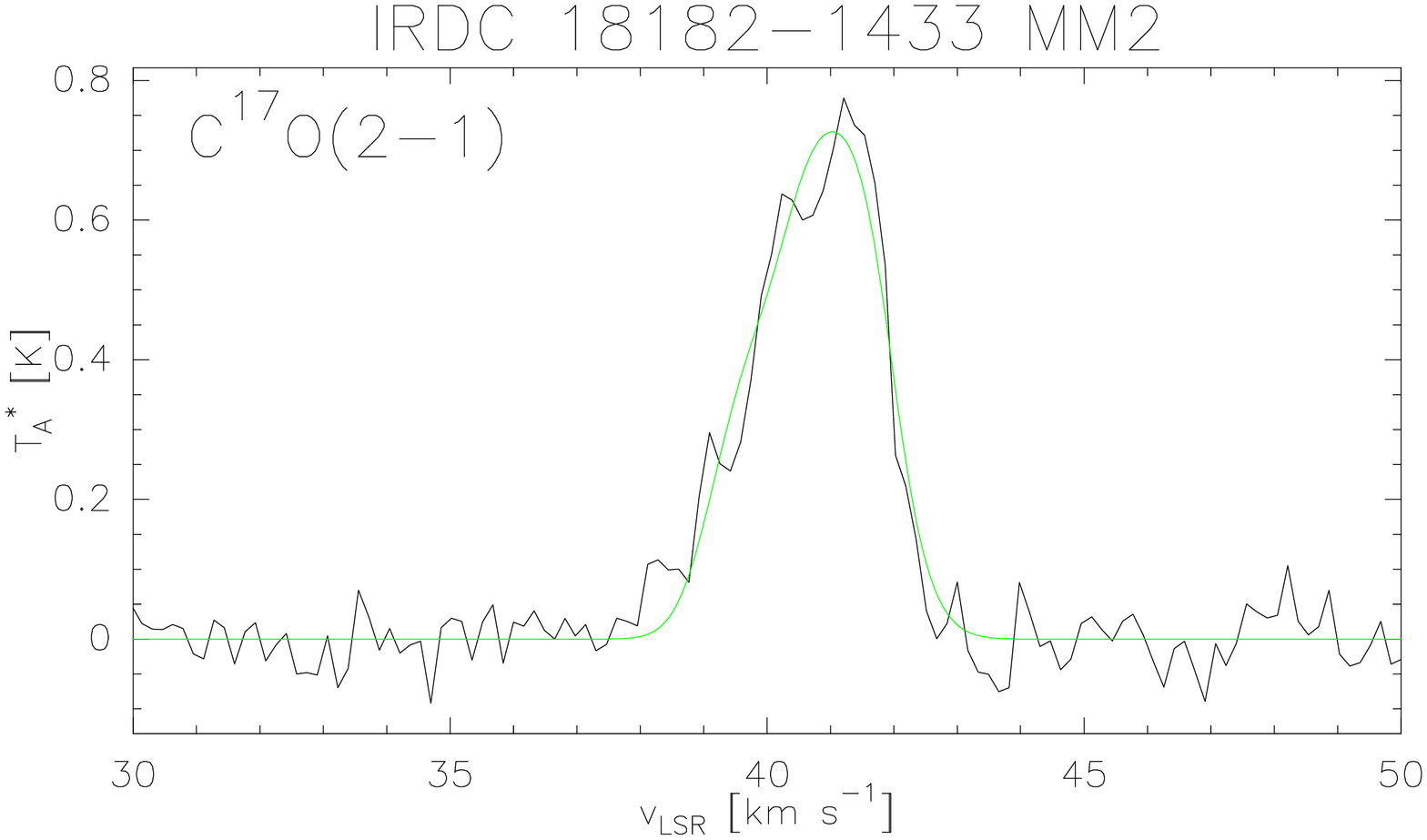}
\includegraphics[width=2.7cm, angle=-90]{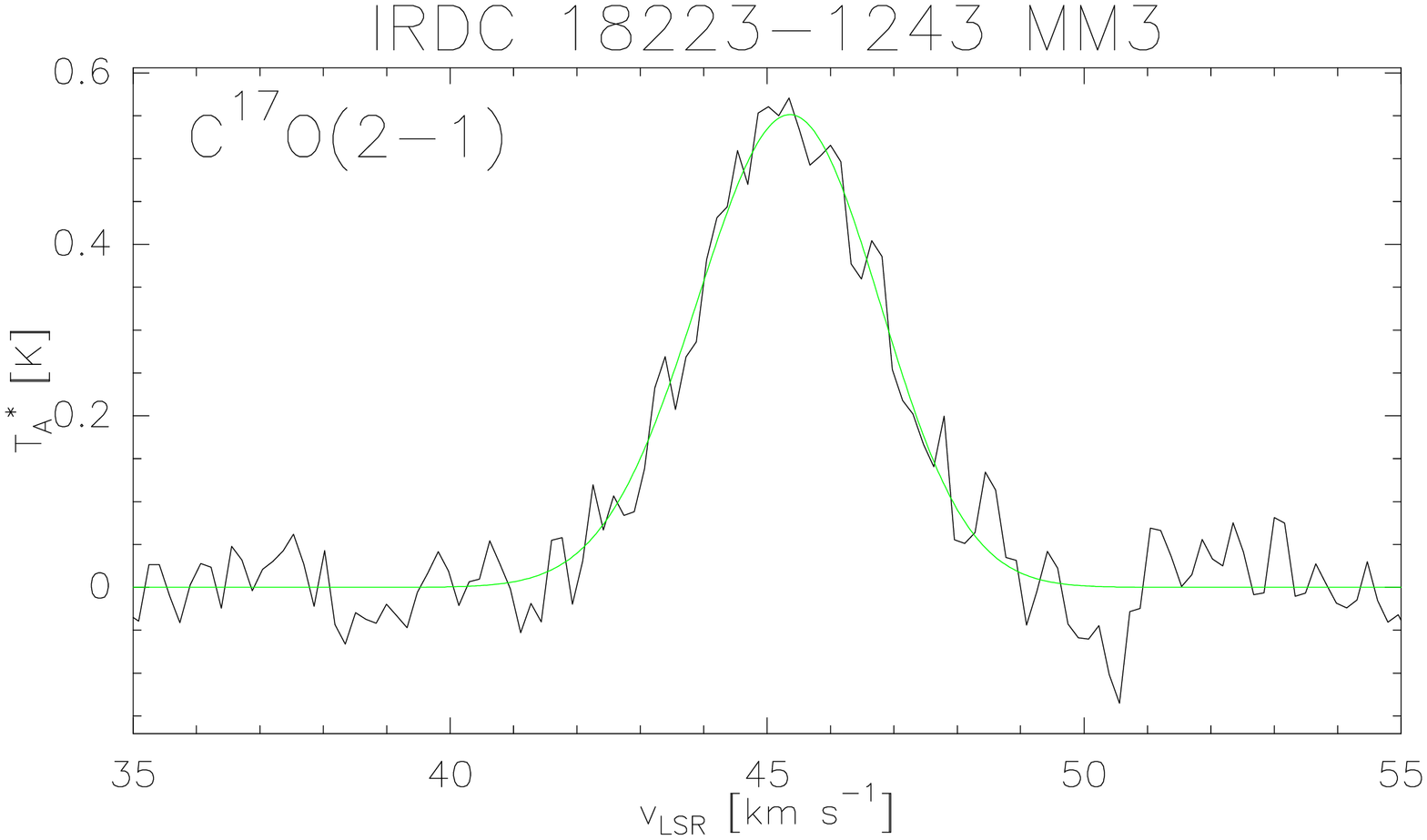}
\includegraphics[width=2.7cm, angle=-90]{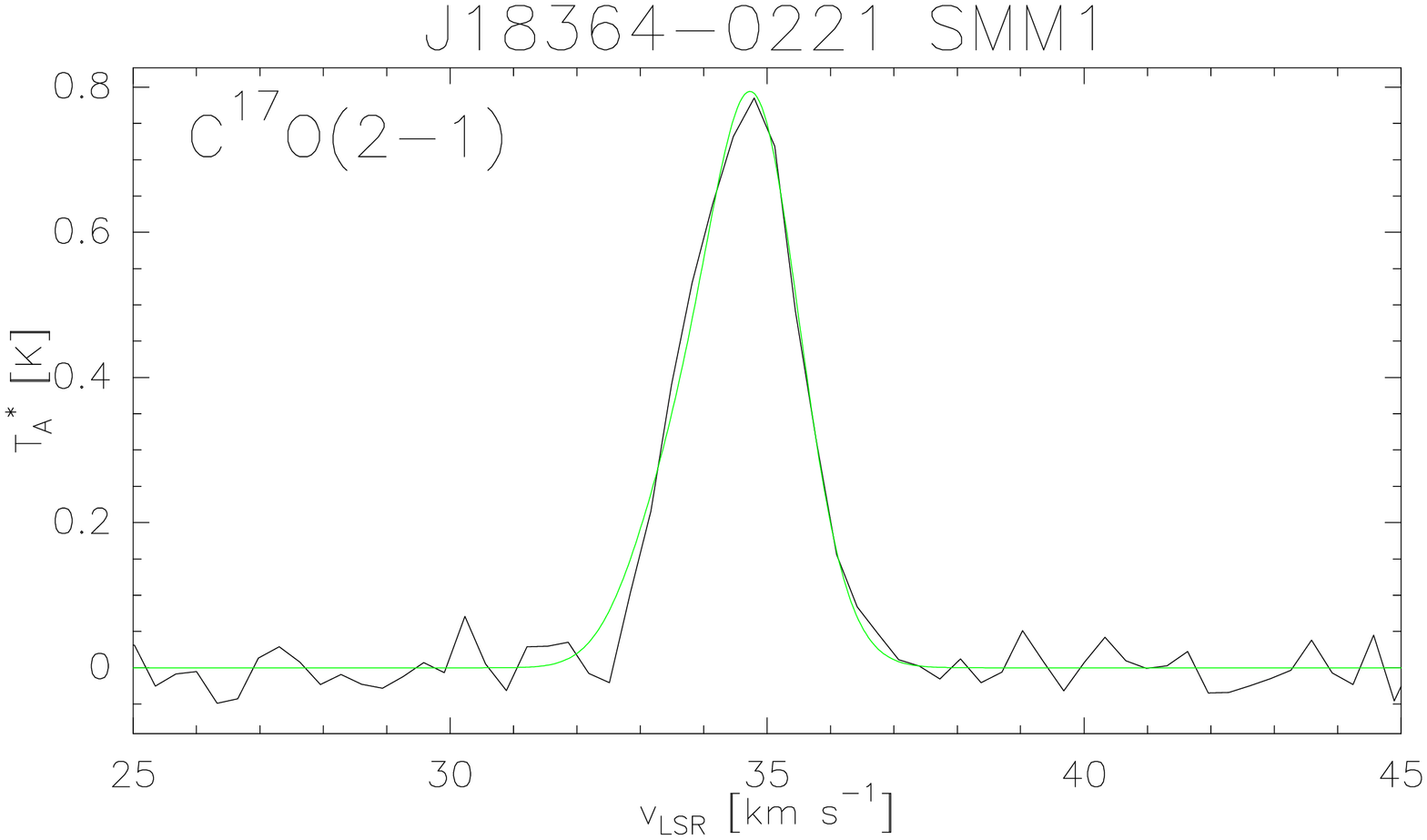}
\includegraphics[width=2.5cm, angle=-90]{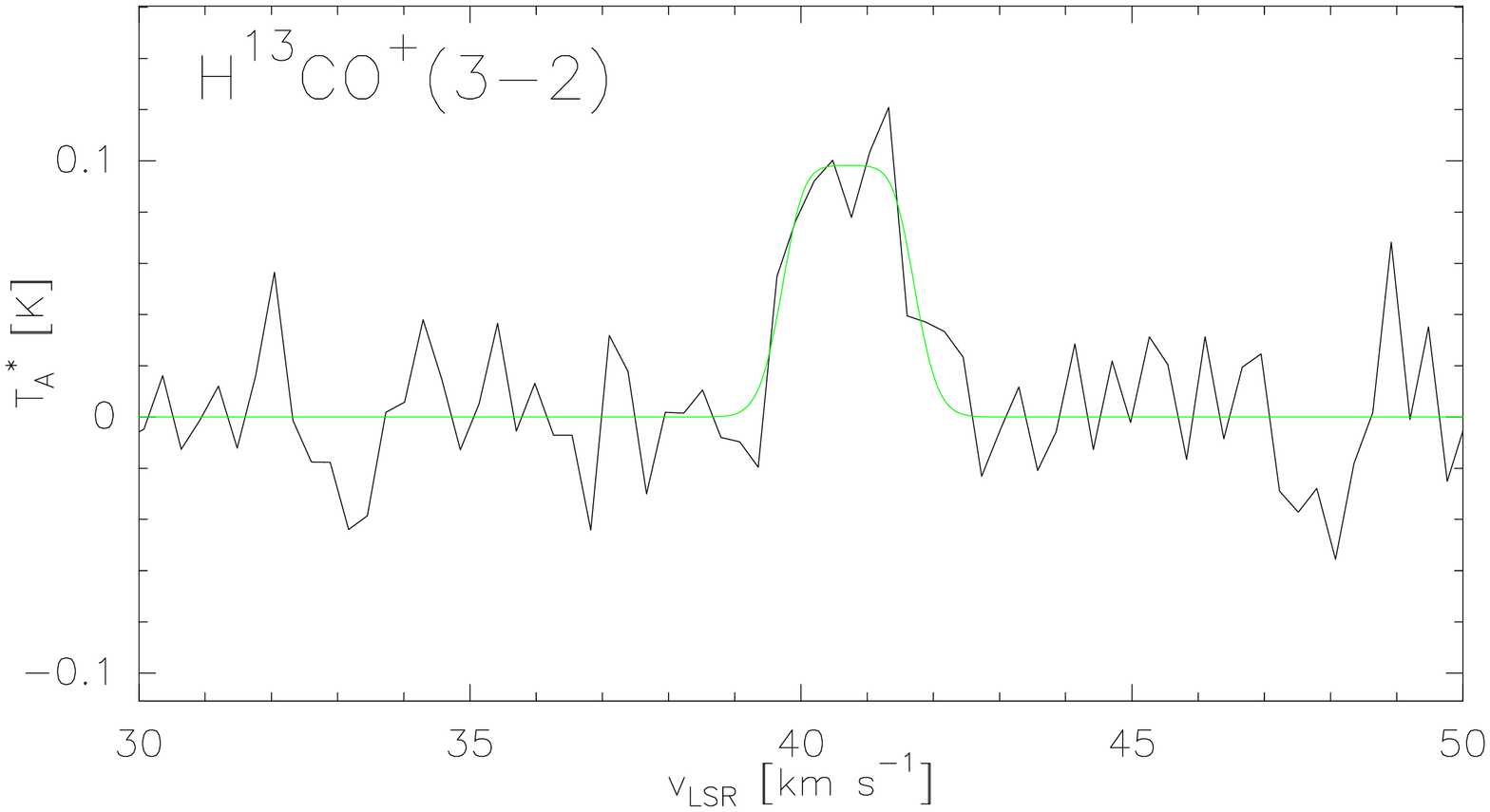}
\includegraphics[width=2.5cm, angle=-90]{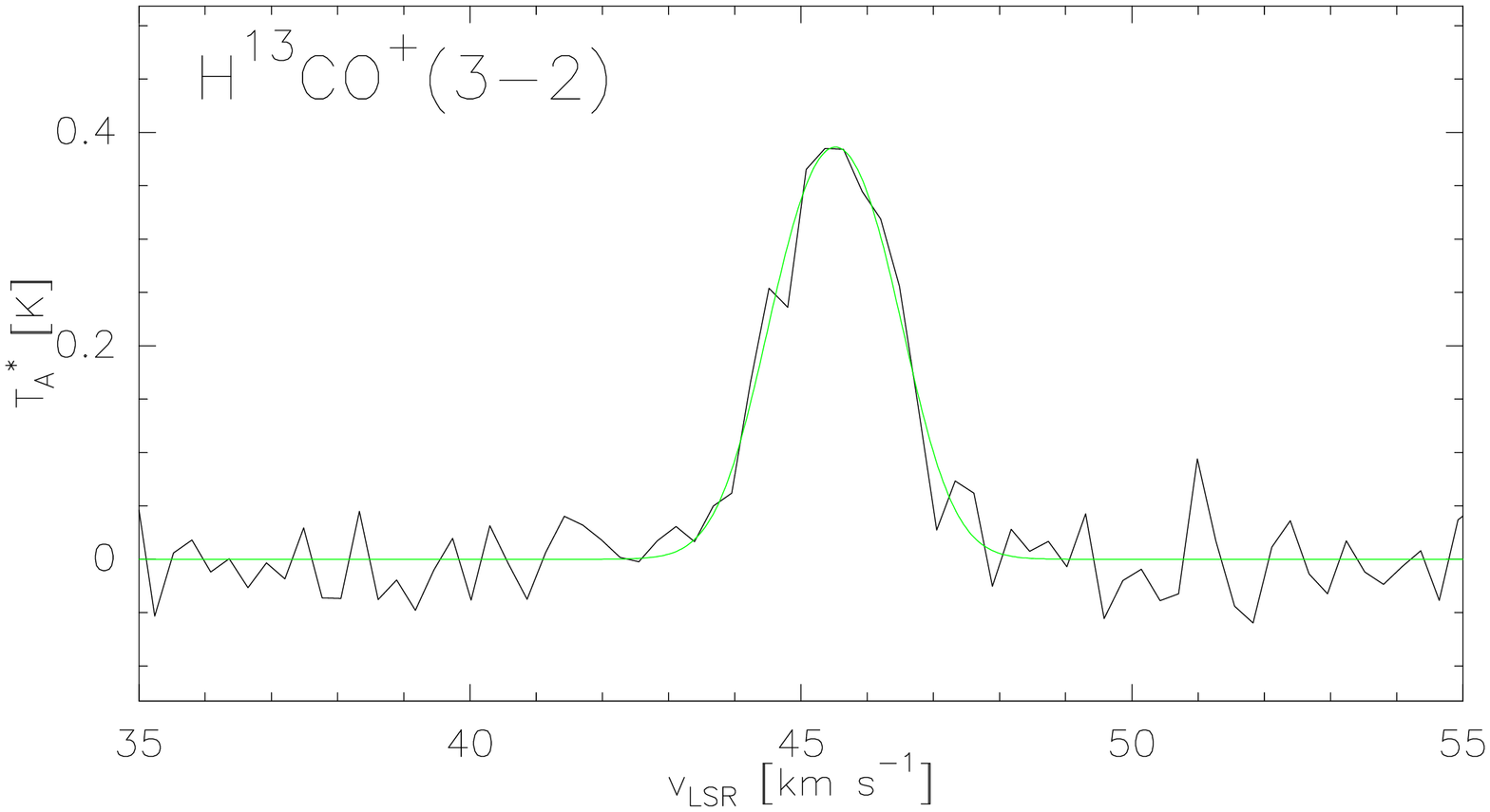}
\includegraphics[width=2.5cm, angle=-90]{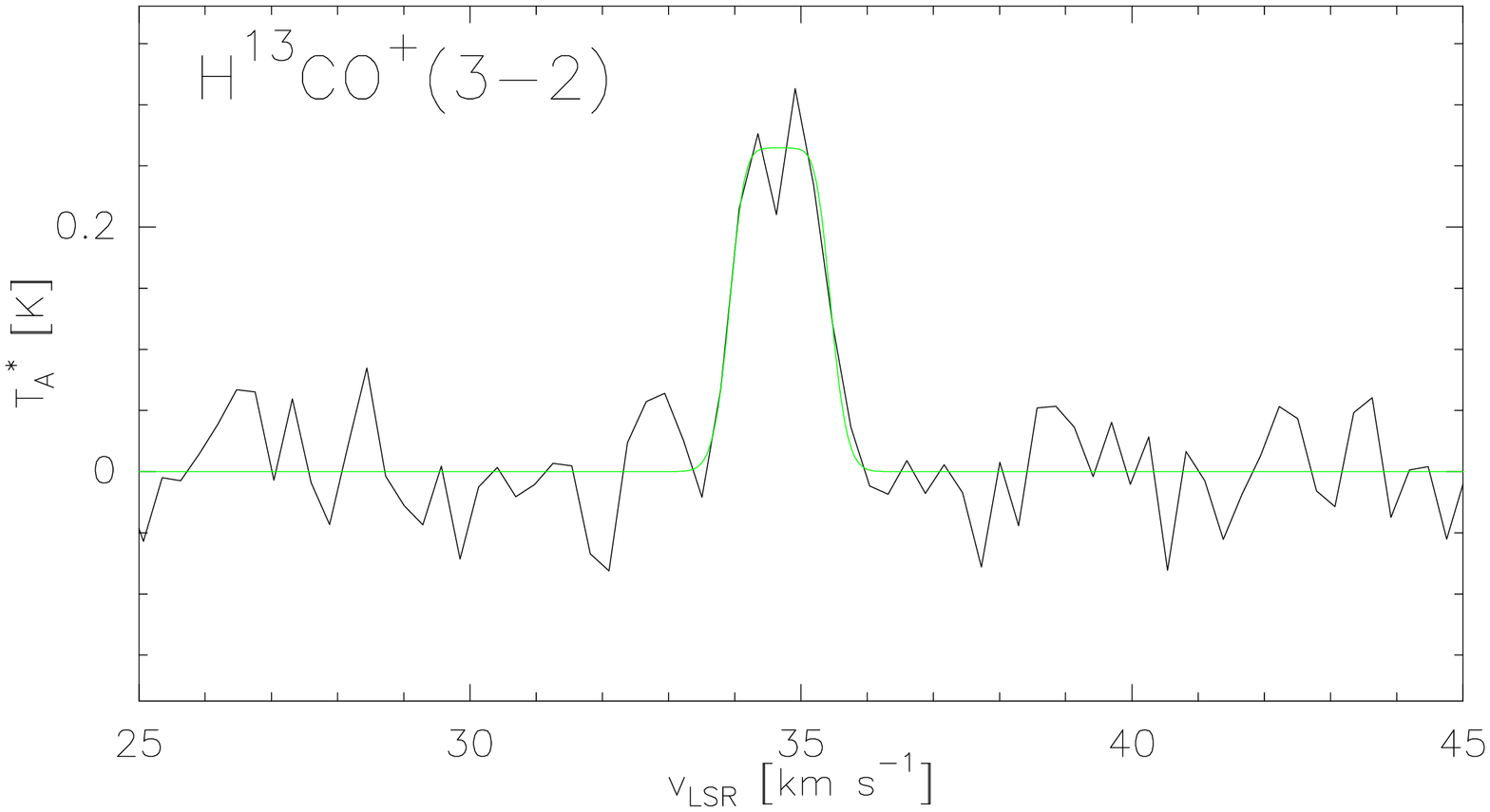}
\includegraphics[width=2.5cm, angle=-90]{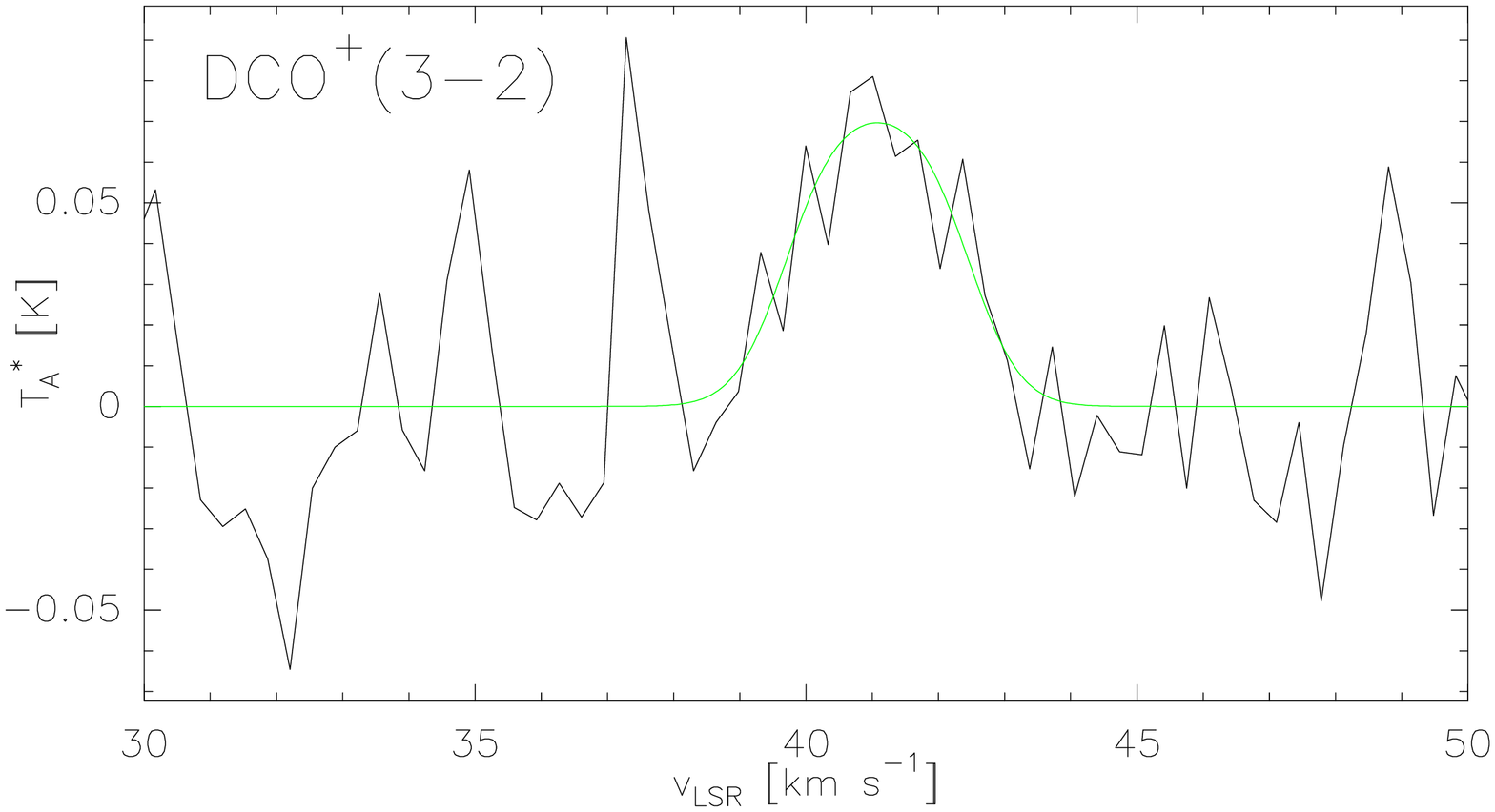}
\includegraphics[width=2.5cm, angle=-90]{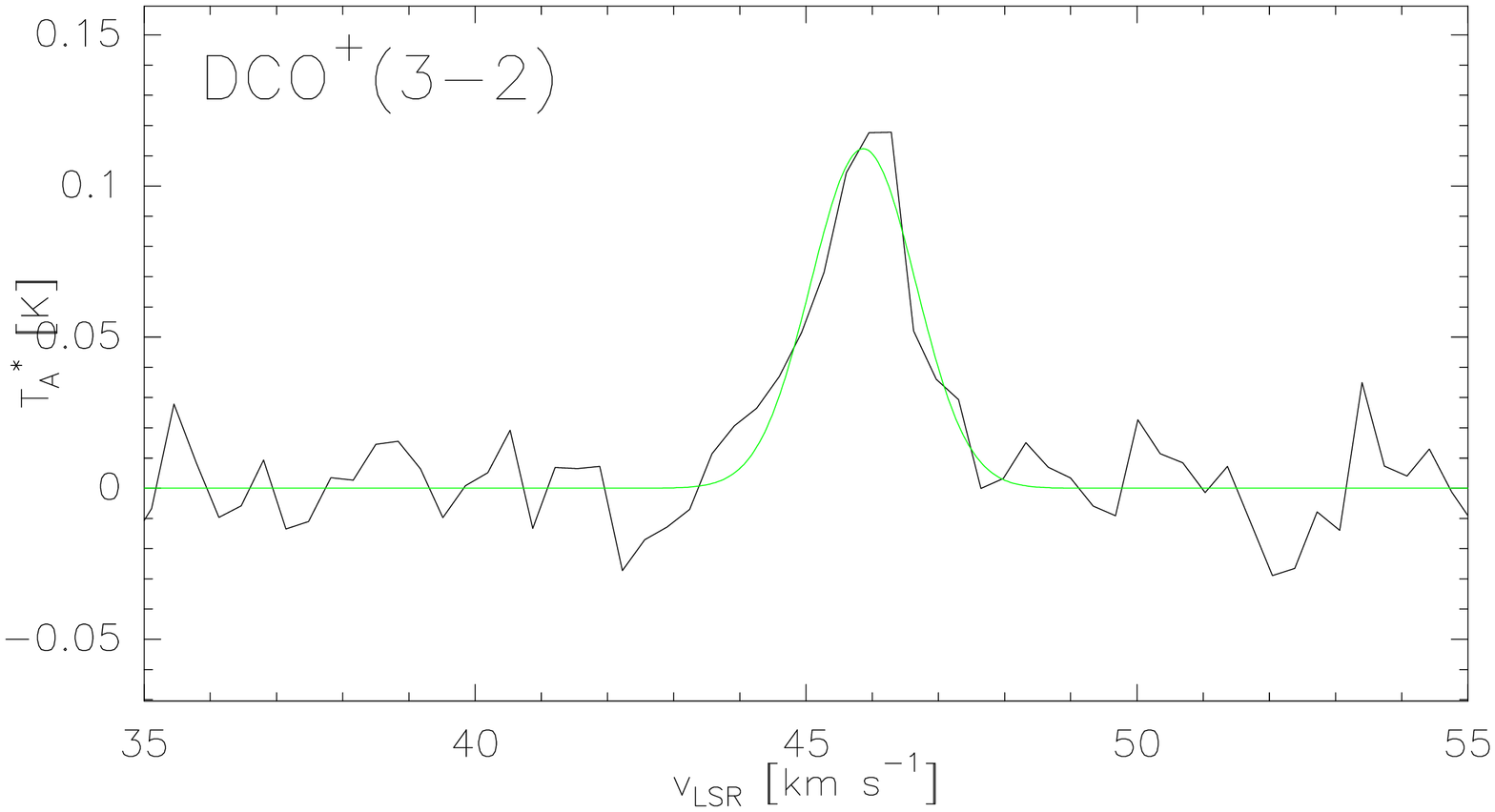}
\includegraphics[width=2.5cm, angle=-90]{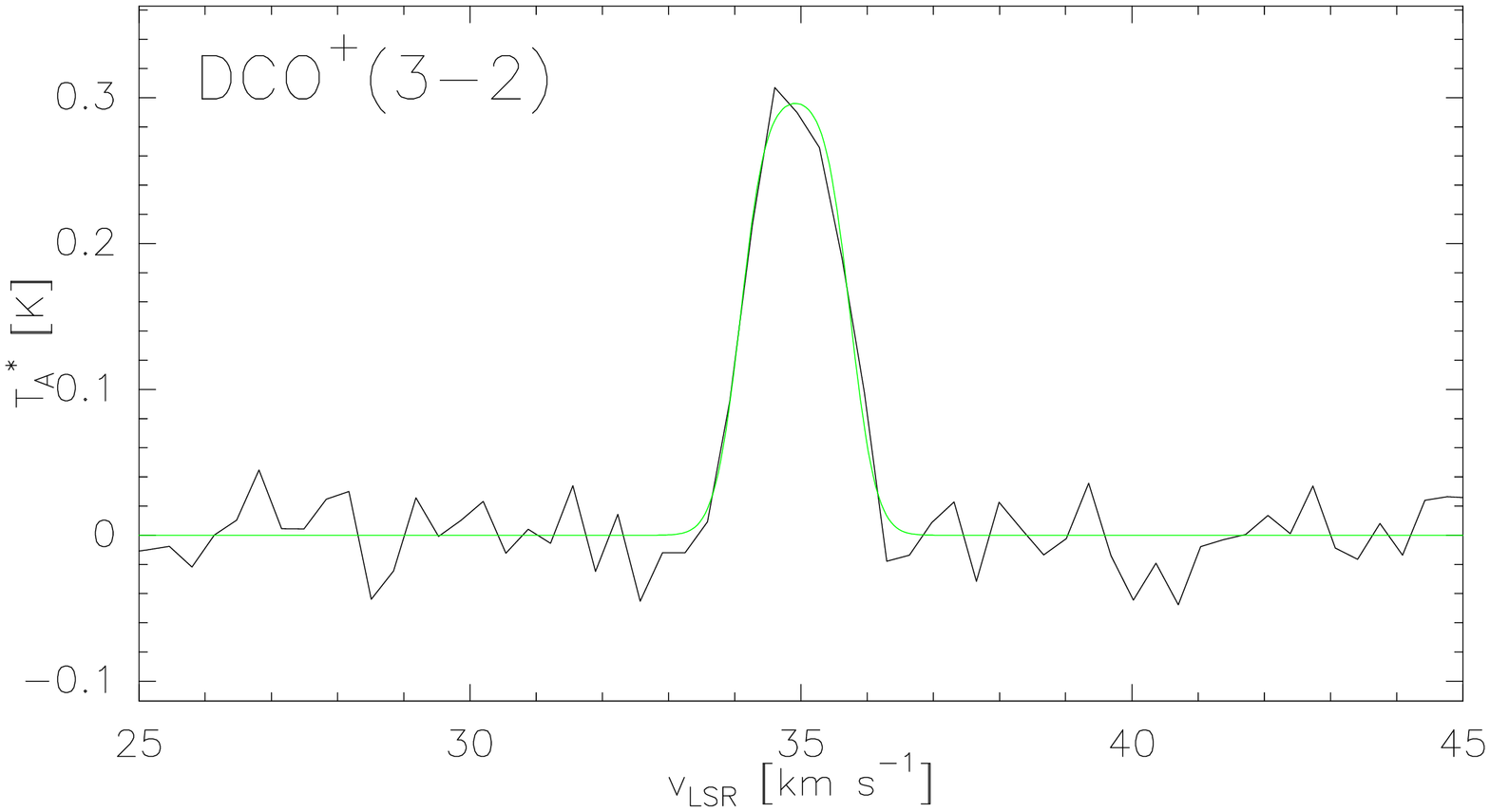}
\includegraphics[width=2.5cm, angle=-90]{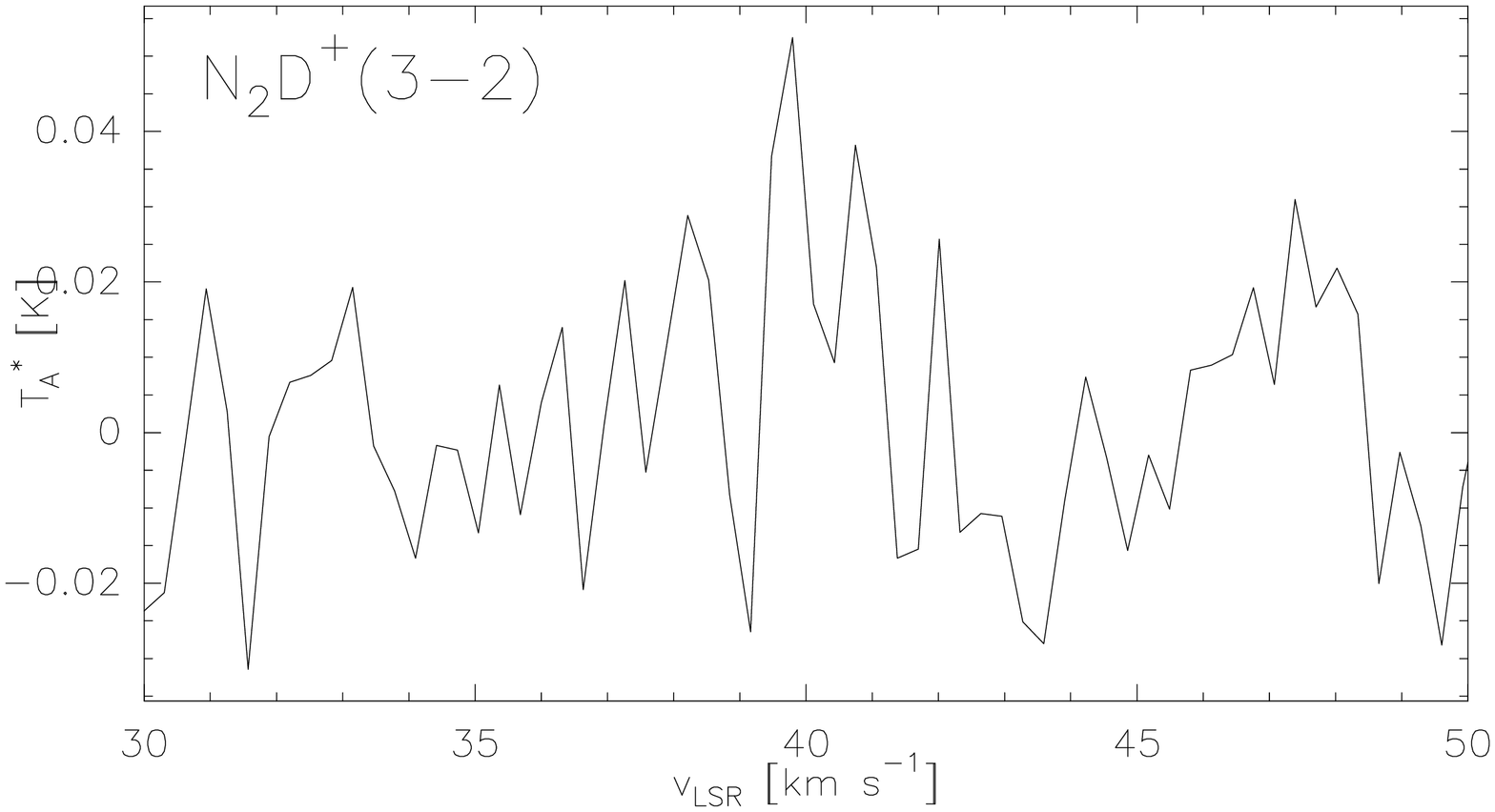}
\includegraphics[width=2.5cm, angle=-90]{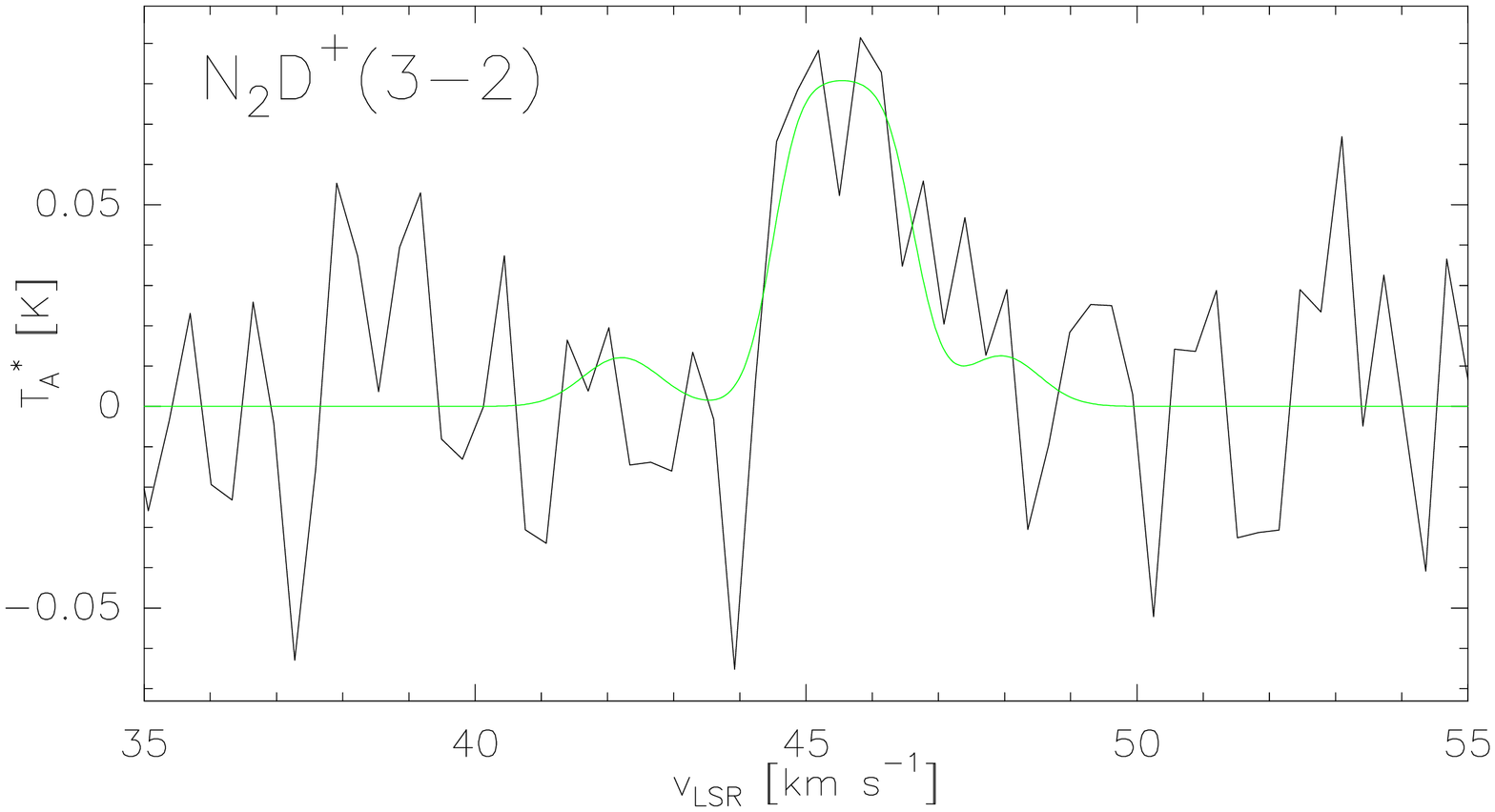}
\includegraphics[width=2.5cm, angle=-90]{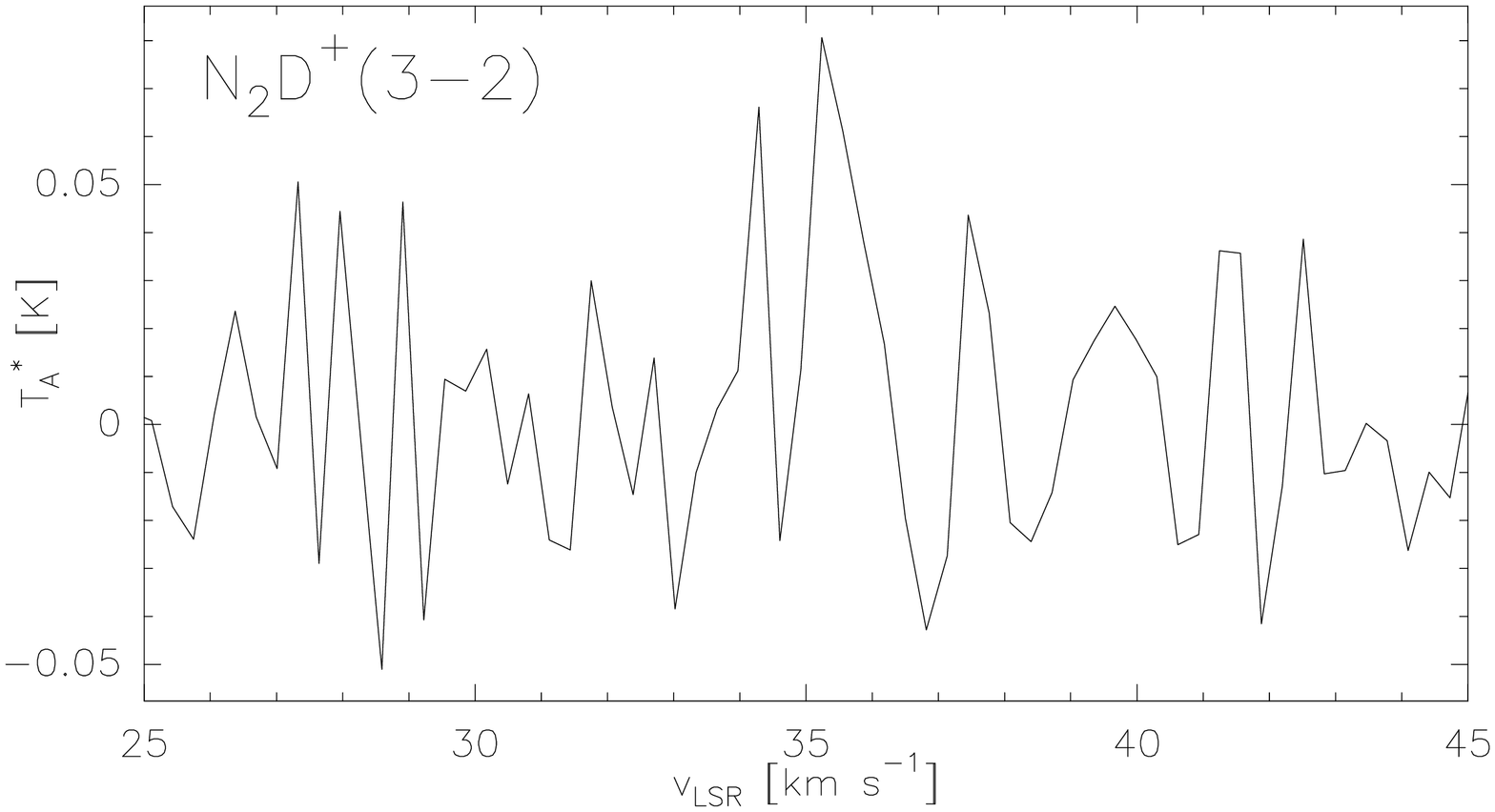}
\addtocounter{figure}{-1}
\caption{continued.}
\label{figure:spectra}
\end{center}
\end{figure*}

\subsubsection{Other line detections}

All the observed sources show additional spectral lines in the observed 
frequency bands, mostly in the spectra covering the DCO$^+(3-2)$ and 
N$_2$D$^+(3-2)$ lines. The line identification was done by using Weeds, 
which is an extension of CLASS (\cite{maret2011}), and applying the Jet 
Propulsion Laboratory\footnote{{\tt http://spec.jpl.nasa.gov/}} 
(JPL; \cite{pickett1998}) and CDMS spectroscopic databases. We used the LTE 
modelling application of Weeds to check if all predicted lines of a 
candidate molecule are present in the observed spectrum (see Sect.~4.2). 
In some cases, we were able to reject some line candidates on the basis of 
non-detection of other transitions expected at nearby frequencies.

\begin{figure}[!h]
\resizebox{0.8\hsize}{!}{\includegraphics[angle=-90]{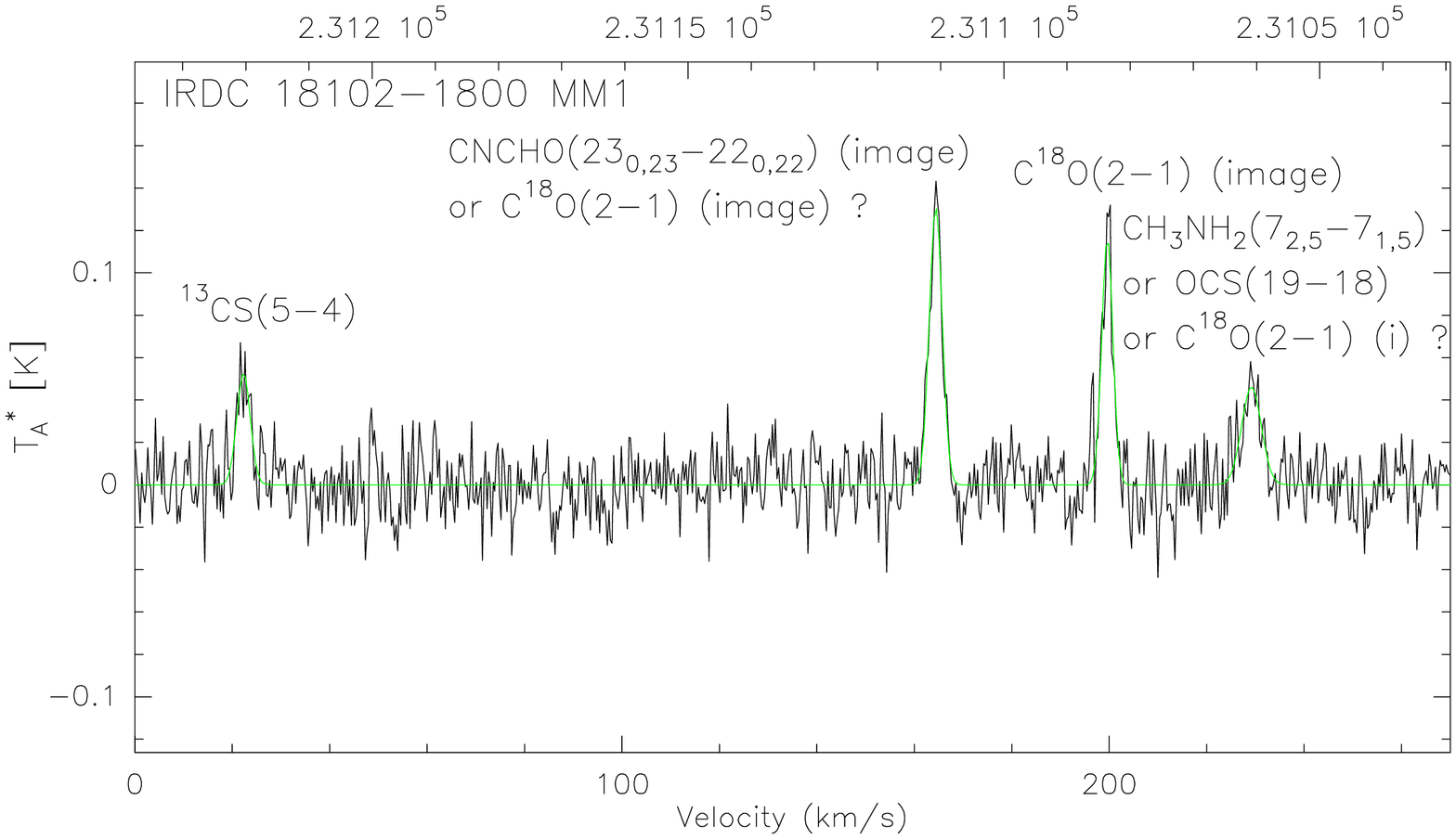}}
\resizebox{0.82\hsize}{!}{\includegraphics[angle=-90]{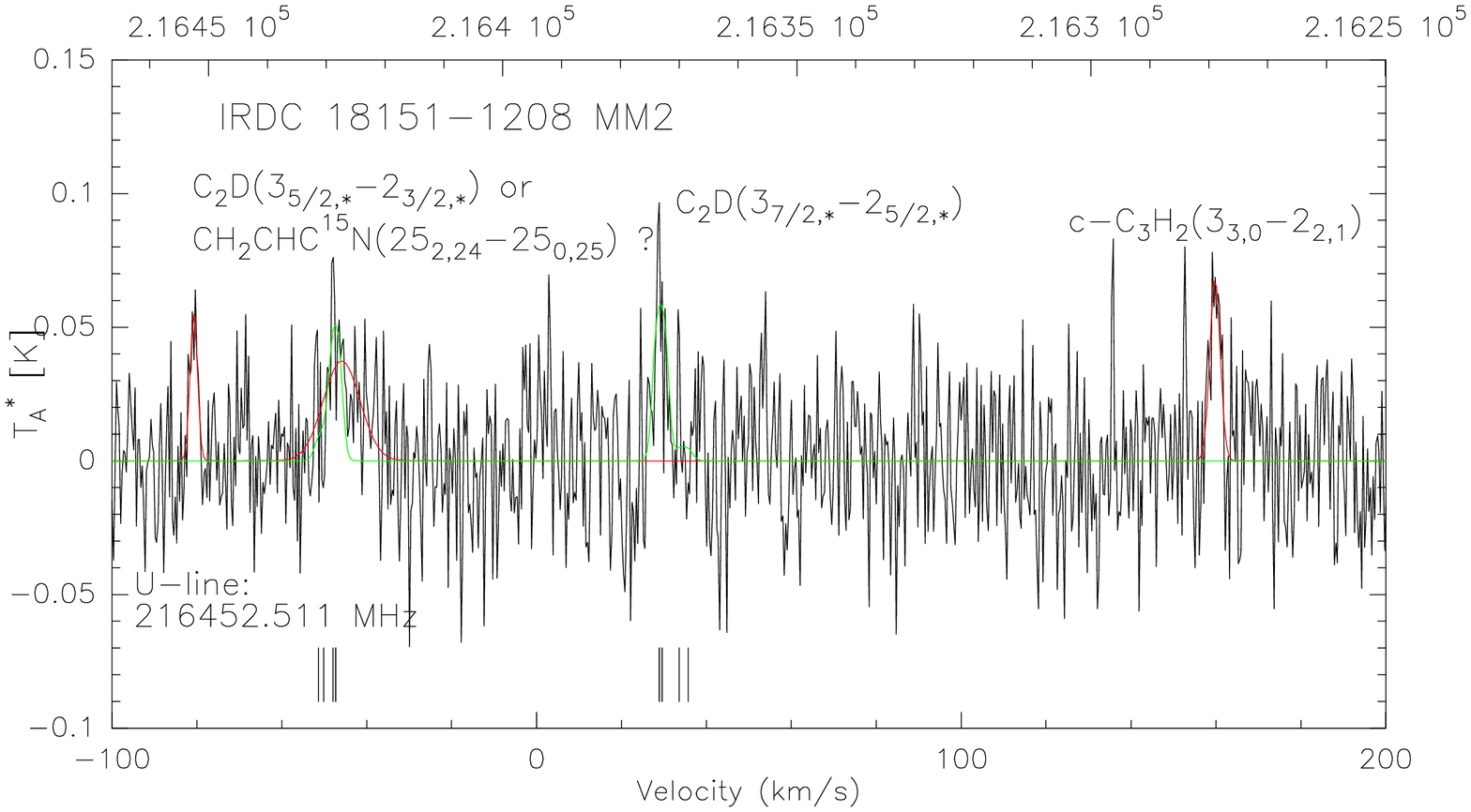}}
\resizebox{0.8\hsize}{!}{\includegraphics[angle=-90]{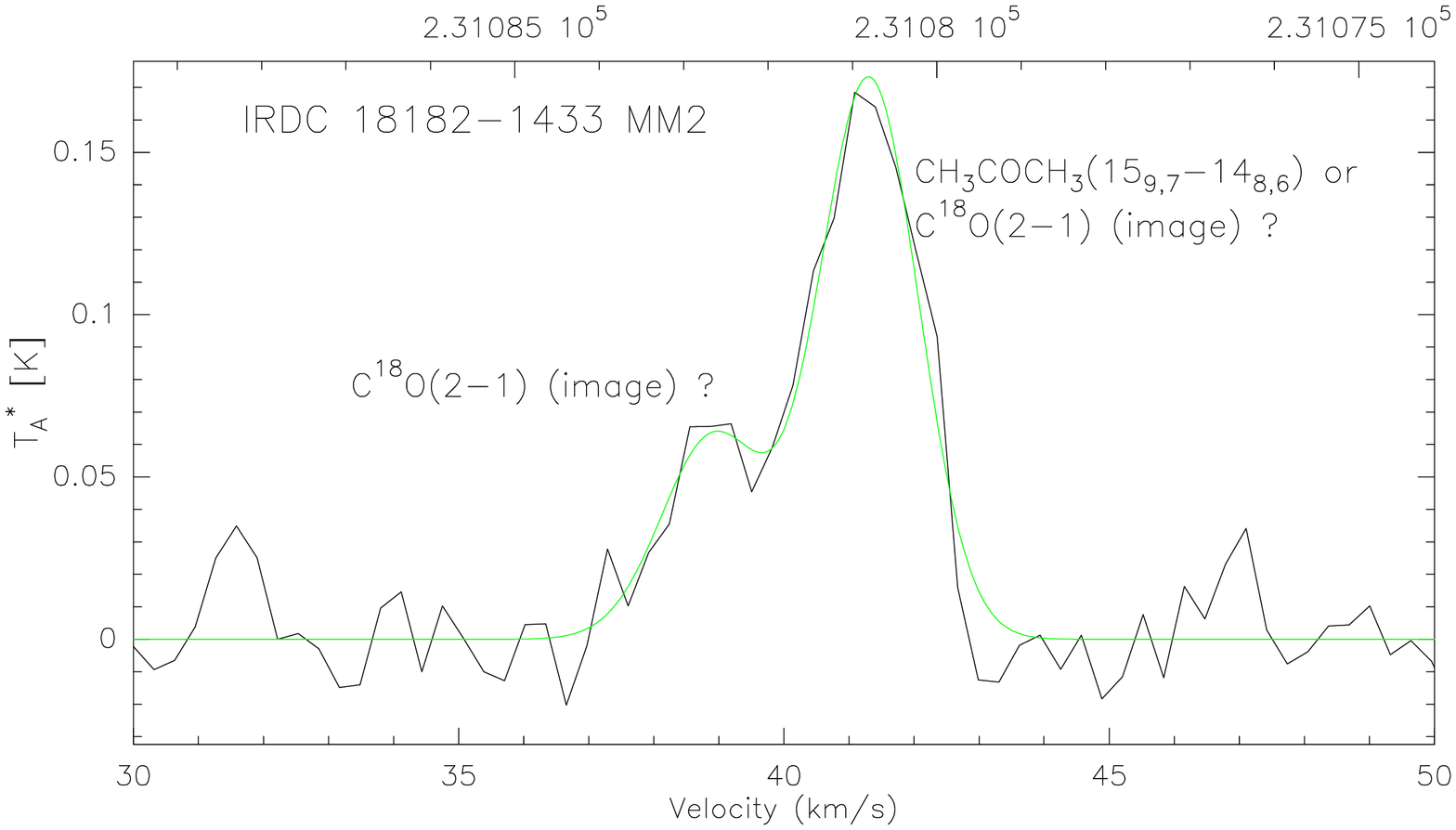}}
\caption{Multi-line spectra of I18102 MM1, I18151 MM2, and I18182 MM2. 
On the spectrum towards I18151 MM2, the green line shows the 
hf-structure fit to C$_2$D$(3-2)$, and the red line indicates single Gaussian 
fits to the other lines. The relative velocities of the C$_2$D$(3-2)$ hf 
components are labelled with a short bar. On the other spectra, the green line 
represents single Gaussian fits.}
\label{figure:multiline}
\end{figure}

I18102 MM1 and I18151 MM2 show relatively line-rich spectra 
at $\sim231$ GHz and $\sim216$ GHz, in the USB and LSB, respectively. 
Two additional lines at $\sim231$ GHz (USB) are
also visible in the spectrum of I18182 MM2 (see Fig.~\ref{figure:multiline}).
The only unambiguous line identification in the spectrum of I18102 MM1 is 
$^{13}$CS$(5-4)$. The second strongest line ($T_{\rm A}^*=0.12$ K) can 
be assigned to C$^{18}$O$(2-1)$ at $\sim219.56$ GHz seen in the image side 
band (marked with ``image''). The strongest line in the spectrum 
($T_{\rm A}^*=0.13$ K) could be due to CNCHO$(J_{K_a,K_c}=23_{0,23}-22_{0,22})$ at 
$\sim219.53$ GHz arising from the image band, or it could represent an 
additional C$^{18}$O$(2-1)$ velocity component along the line of sight. The 
latter is likely to be true due to the relative strong line intensity. 
The line at $\sim231.06$ GHz can be assigned to either 
OCS$(19-18)$ or CH$_3$NH$_2$-E$(J_{K_a,K_c}=7_{2,5}-7_{1,5})$ because these two 
transitions are blended. However, it could also be 
caused by yet another velocity component of C$^{18}$O$(2-1)$ seen towards 
I18102 MM1. 
From the spectrum towards I18151 MM2 we identified the C$_2$D$(N=3-2)$ line 
sepa\-rated into two hf groups. This transition is split up into 13 
hf components of which the detected two groups contain 10 components 
together. The hf group at $\sim216.43$ GHz is blended with 
CH$_2$CHC$^{15}$N$(J_{K_a,K_c}=25_{2,24}-25_{0,25})$. 
The $J_{K_a,K_c}=3_{3,0}-2_{2,1}$ transition of \textit{ortho}-c-C$_3$H$_2$ 
is seen towards I18151 MM2, but the line at $\sim216.45$ GHz is unidentified 
(marked with ``U''). 
The CH$_3$COCH$_3$-EA$(J_{K_a,K_c}=15_{9,7}-14_{8,6})$ line in 
the USB is possibly detected towards I18182 MM2. However, it is blended with 
C$^{18}$O$(2-1)$ entering via the image band. The relatively 
strong intensity of $T_{\rm A}^*=0.17$ K suggests the line is due to 
C$^{18}$O$(2-1)$. The weaker nearby line could be related to an 
additional C$^{18}$O$(2-1)$ velocity component.

\begin{figure*}
\begin{center}
\includegraphics[width=3cm,angle=-90]{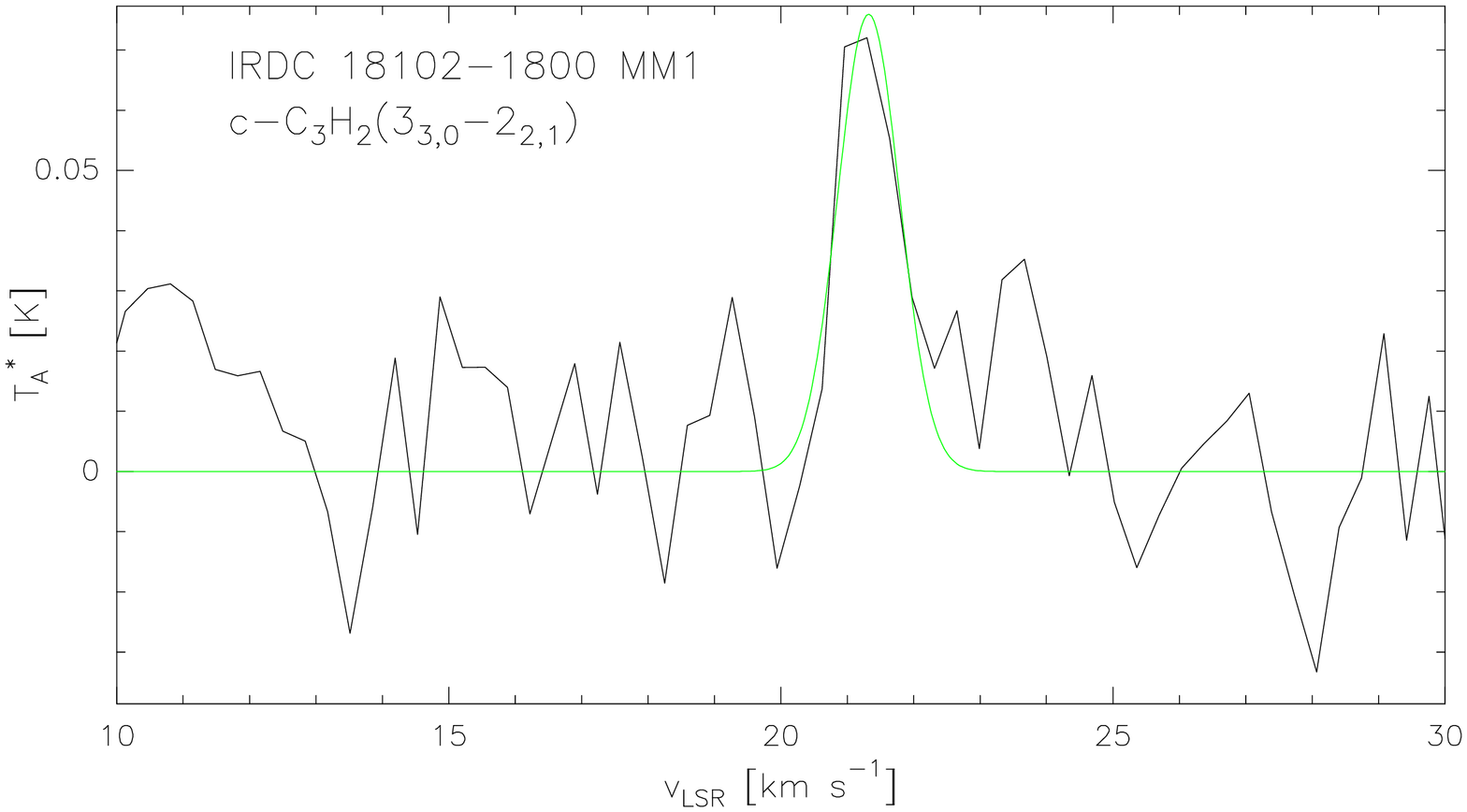}
\includegraphics[width=3cm,angle=-90]{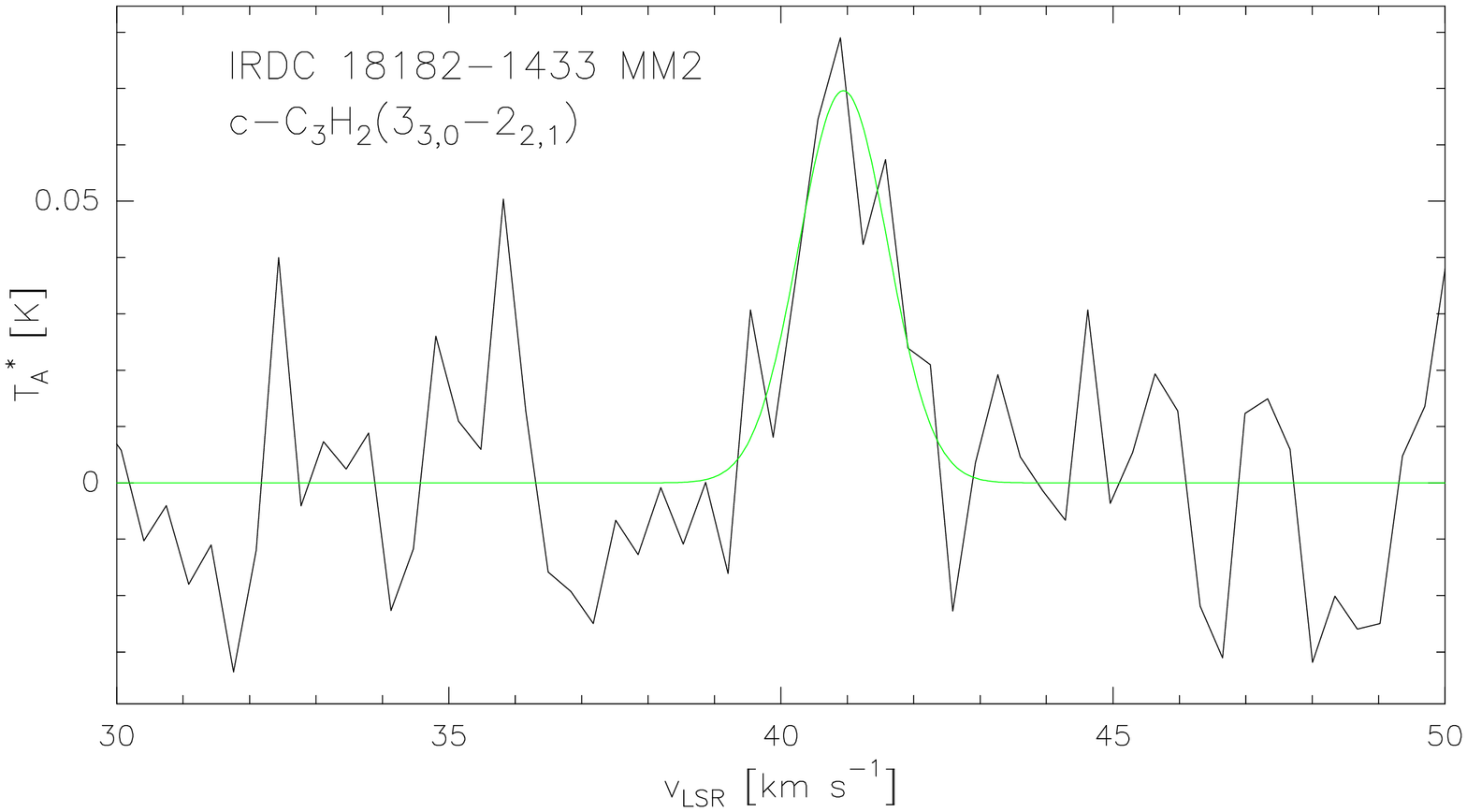}
\includegraphics[width=3cm,angle=-90]{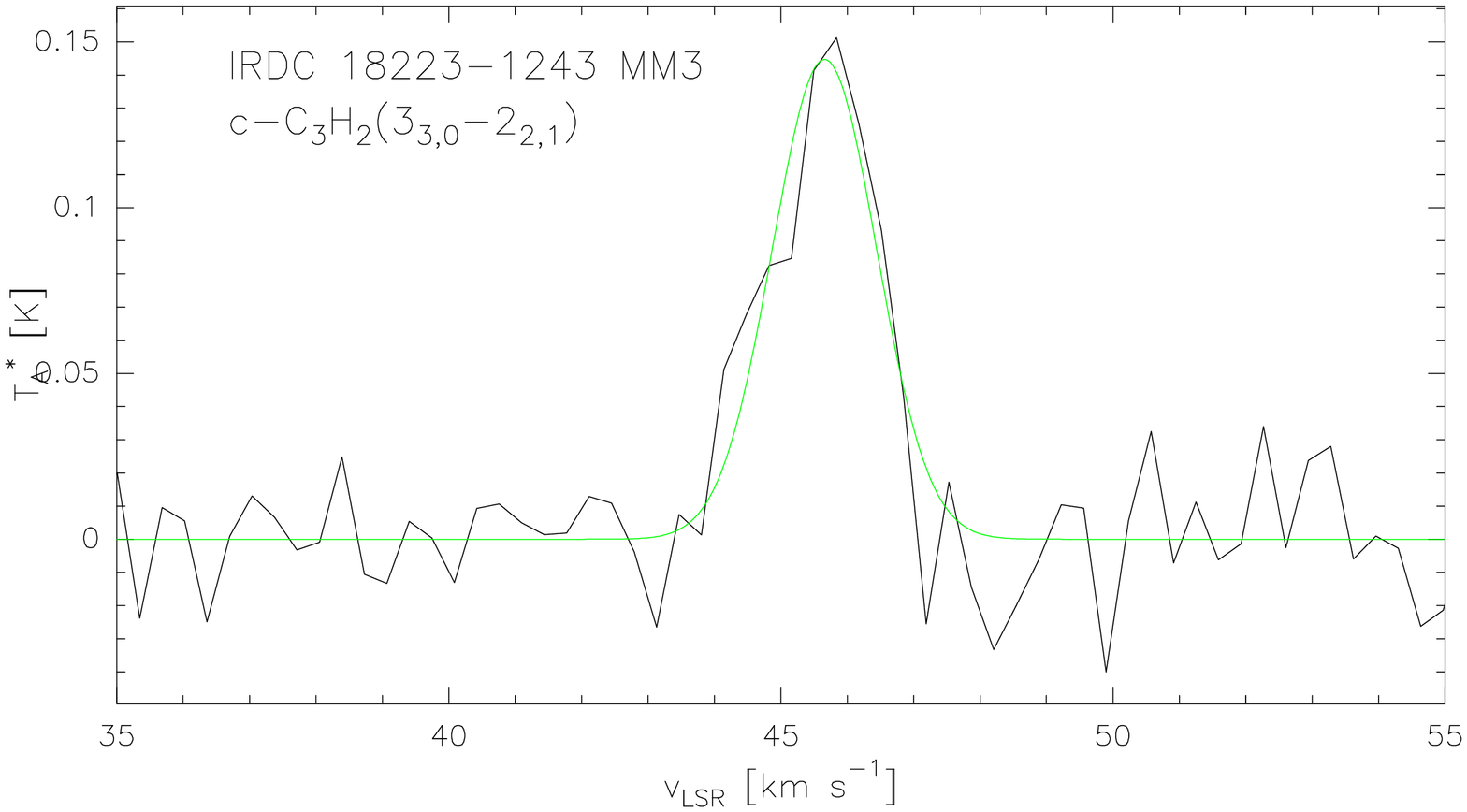}
\caption{\textit{ortho}-c-C$_3$H$_2(3_{3,0}-2_{2,1})$ spectra overlaid with 
Gaussian fits.}
\label{figure:C3H2}
\end{center}
\end{figure*}

\begin{figure*}
\begin{center}
\includegraphics[width=3cm,angle=-90]{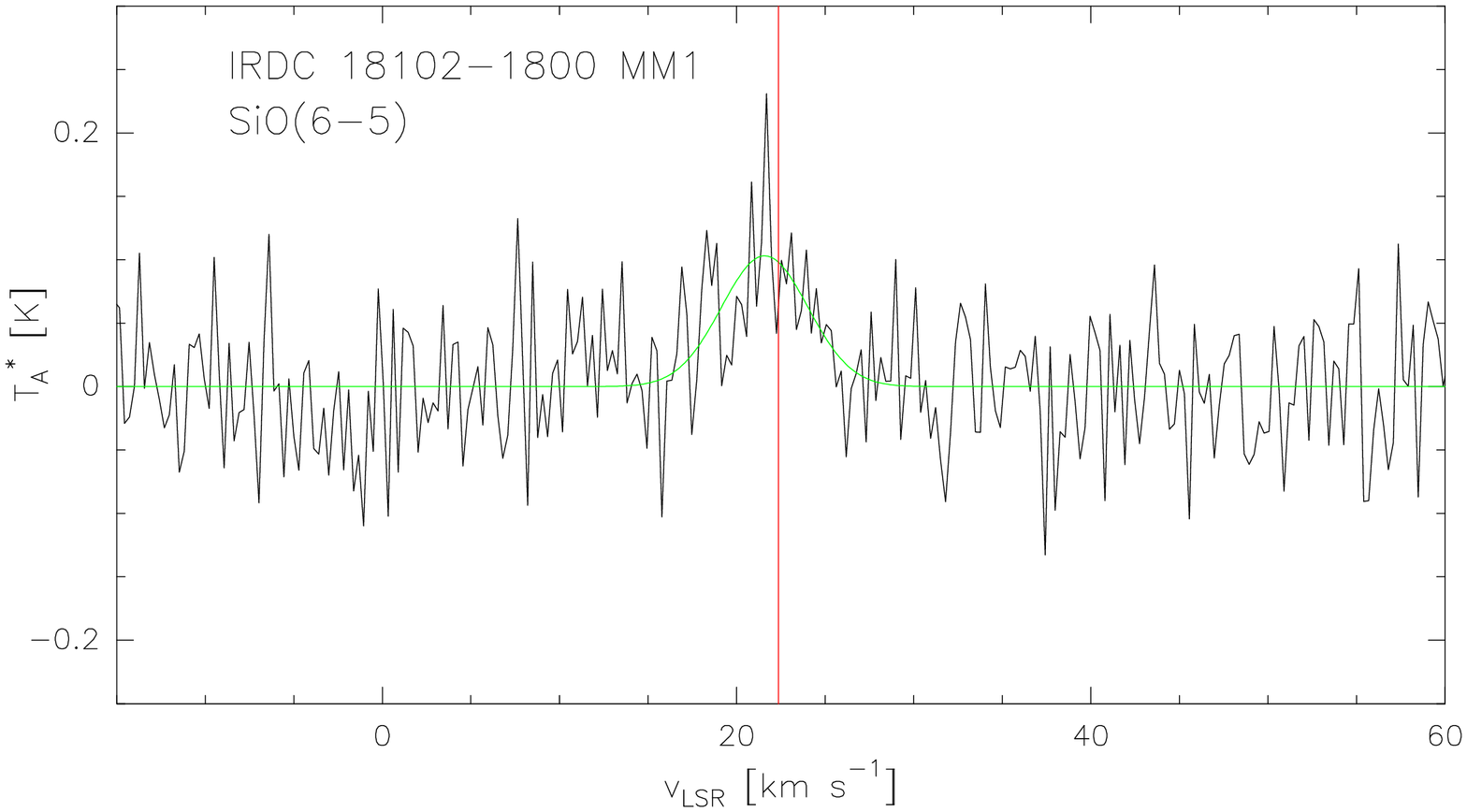}
\includegraphics[width=3cm,angle=-90]{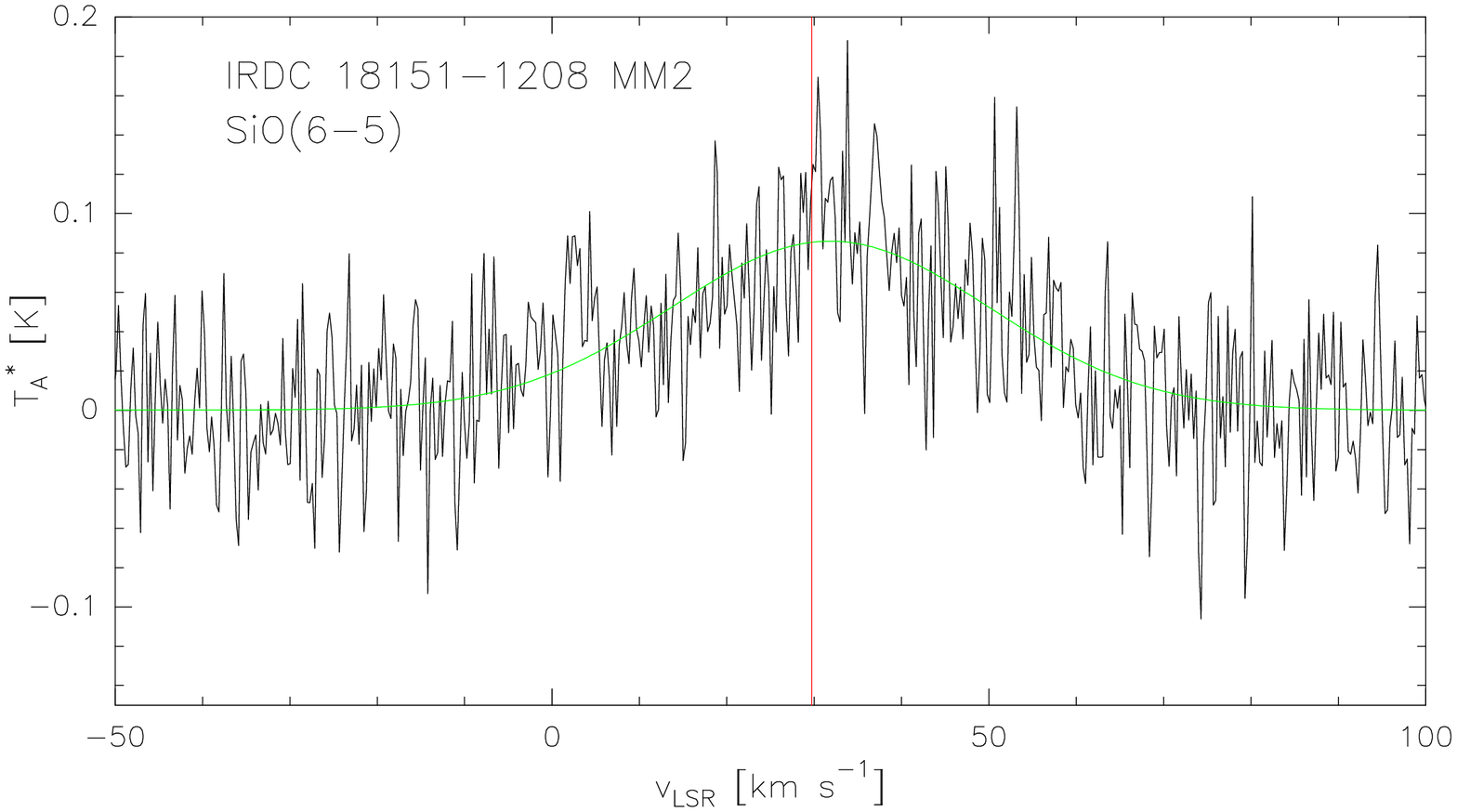}
\includegraphics[width=3cm,angle=-90]{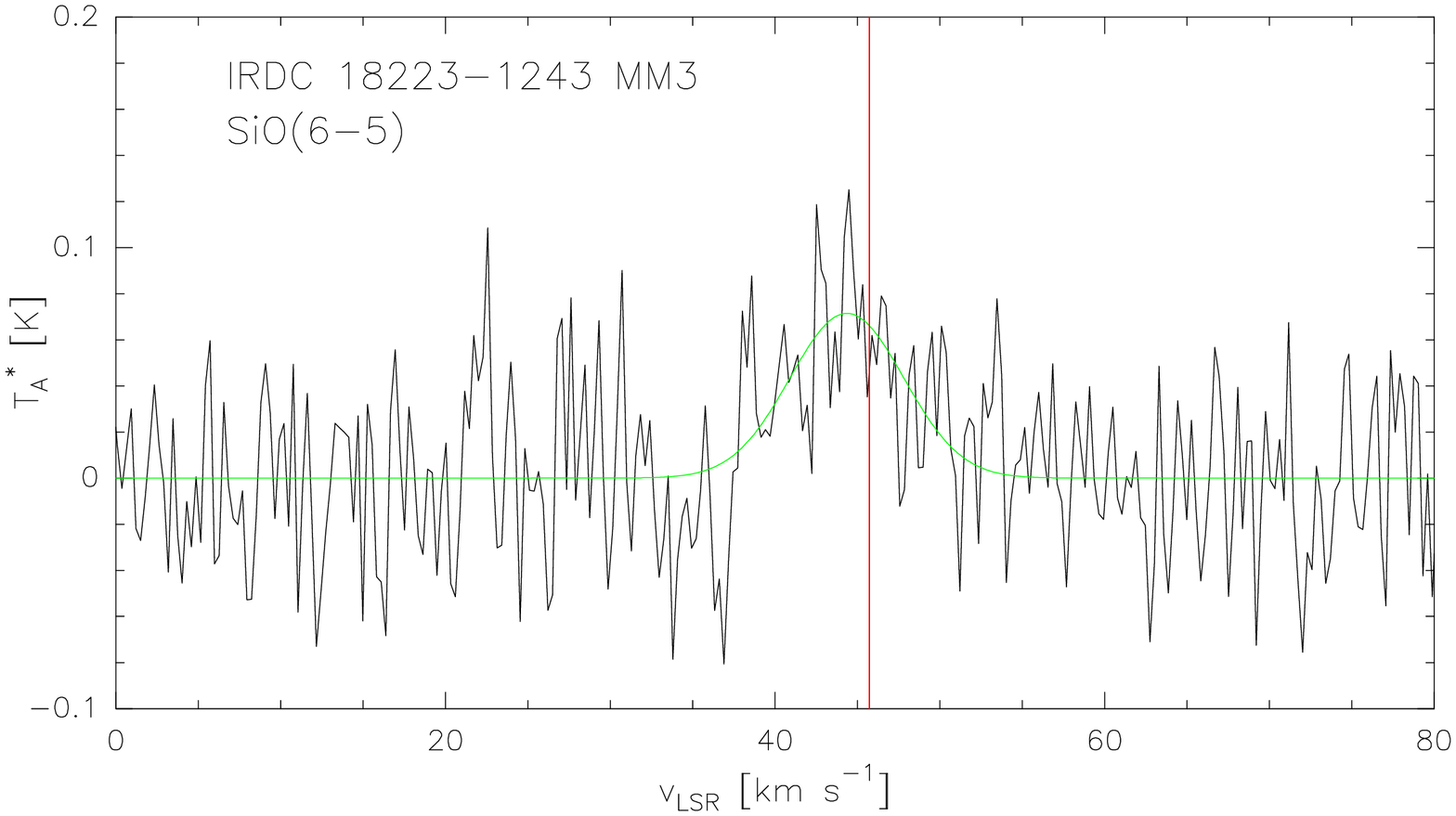}
\caption{SiO$(6-5)$ spectra overlaid with Gaussian fits. The vertical 
line in each panel indicates the ${\rm v_{\rm LSR}}$ of the clump as measured 
from N$_2$H$^+(1-0)$ by SSK08, i.e., 22.36, 29.74, and 45.73 
km~s$^{-1}$ for I18102 MM1, I18151 MM2, and I18223 MM3, respectively. 
The velocity range in each panel is wider relative to that of 
Fig.~\ref{figure:spectra} for a better illustration.}
\label{figure:sio}
\end{center}
\end{figure*}

\begin{figure*}
\begin{center}
\includegraphics[width=2.12cm,angle=-90]{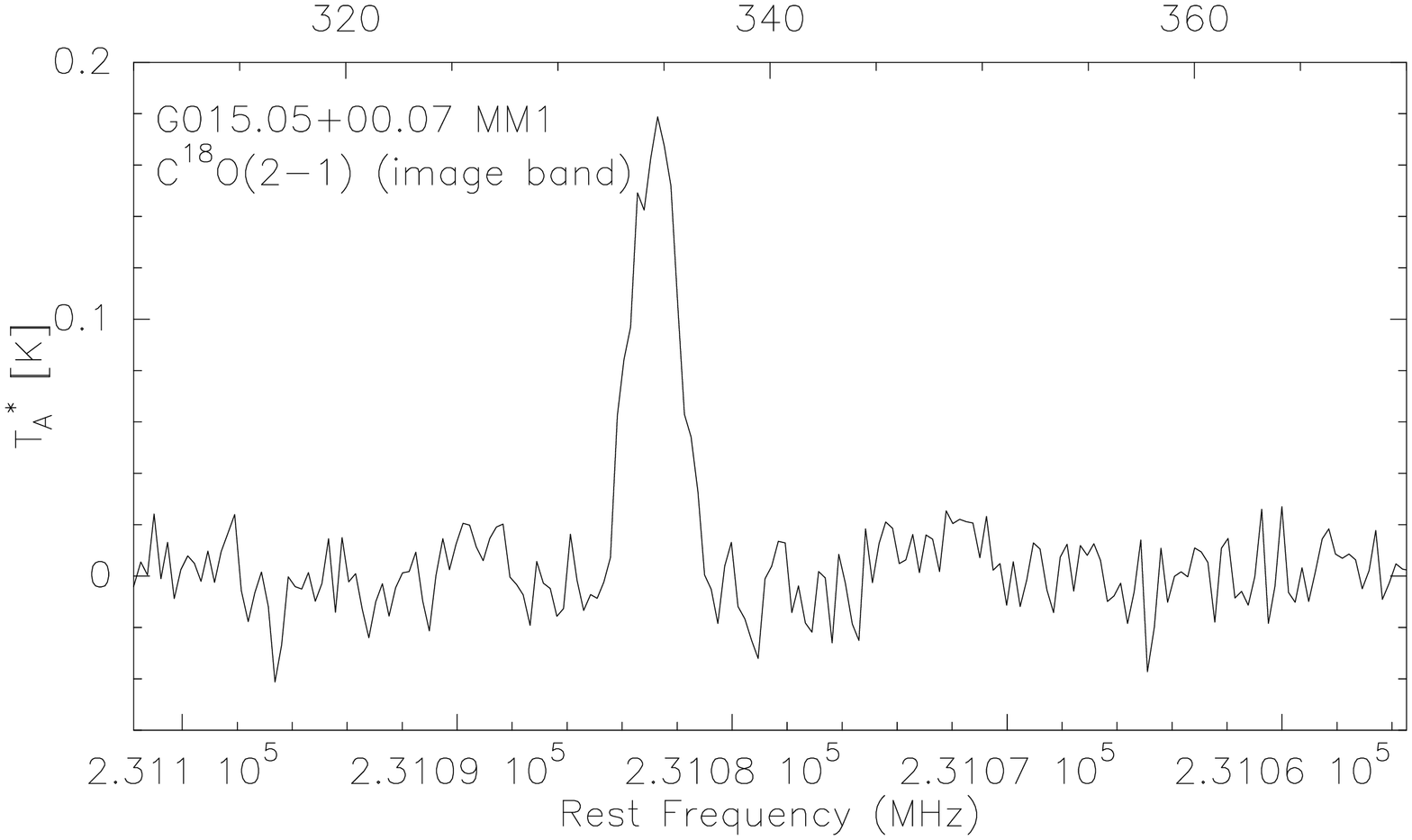}
\includegraphics[width=2.12cm,angle=-90]{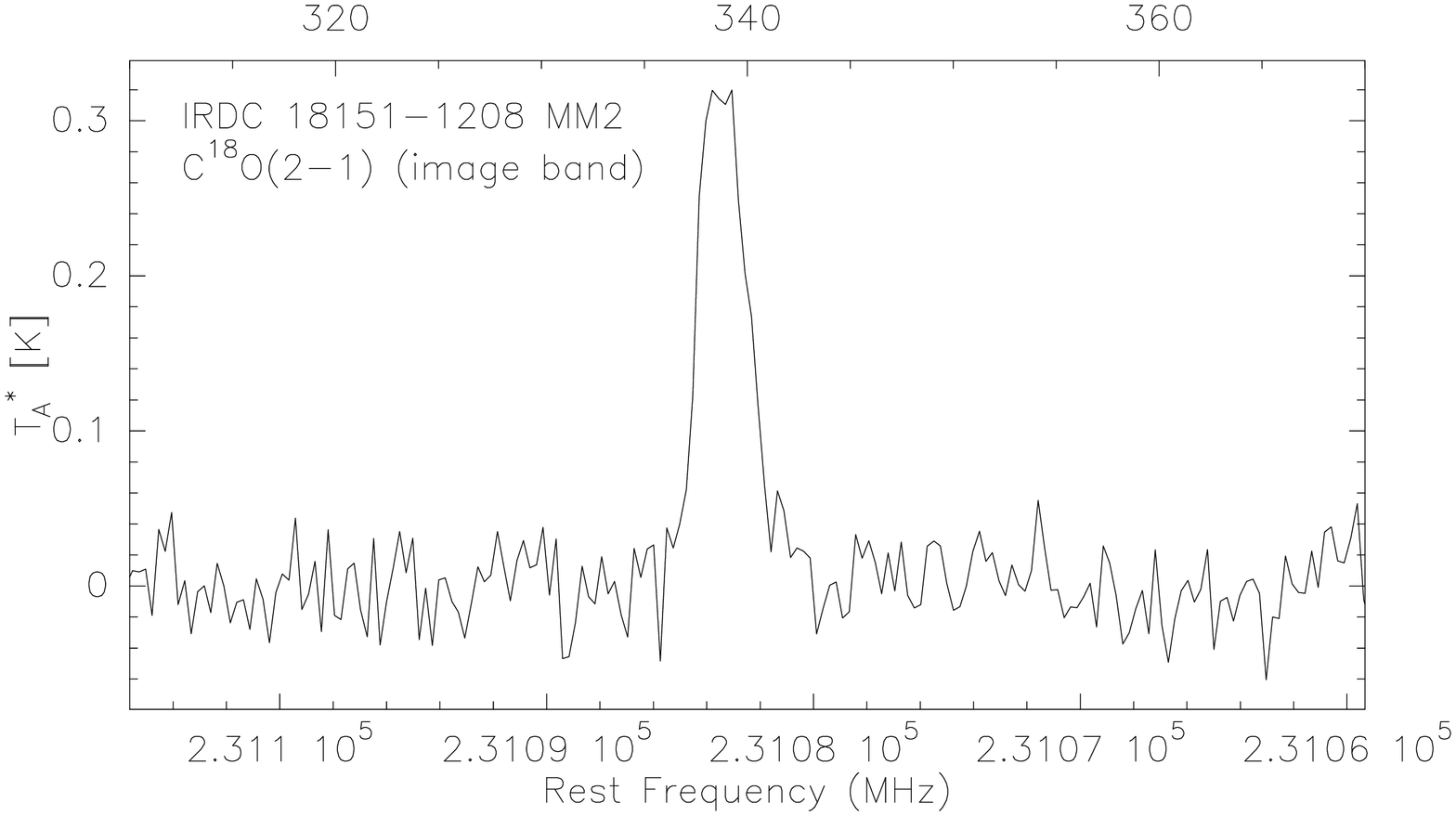}
\includegraphics[width=2.12cm,angle=-90]{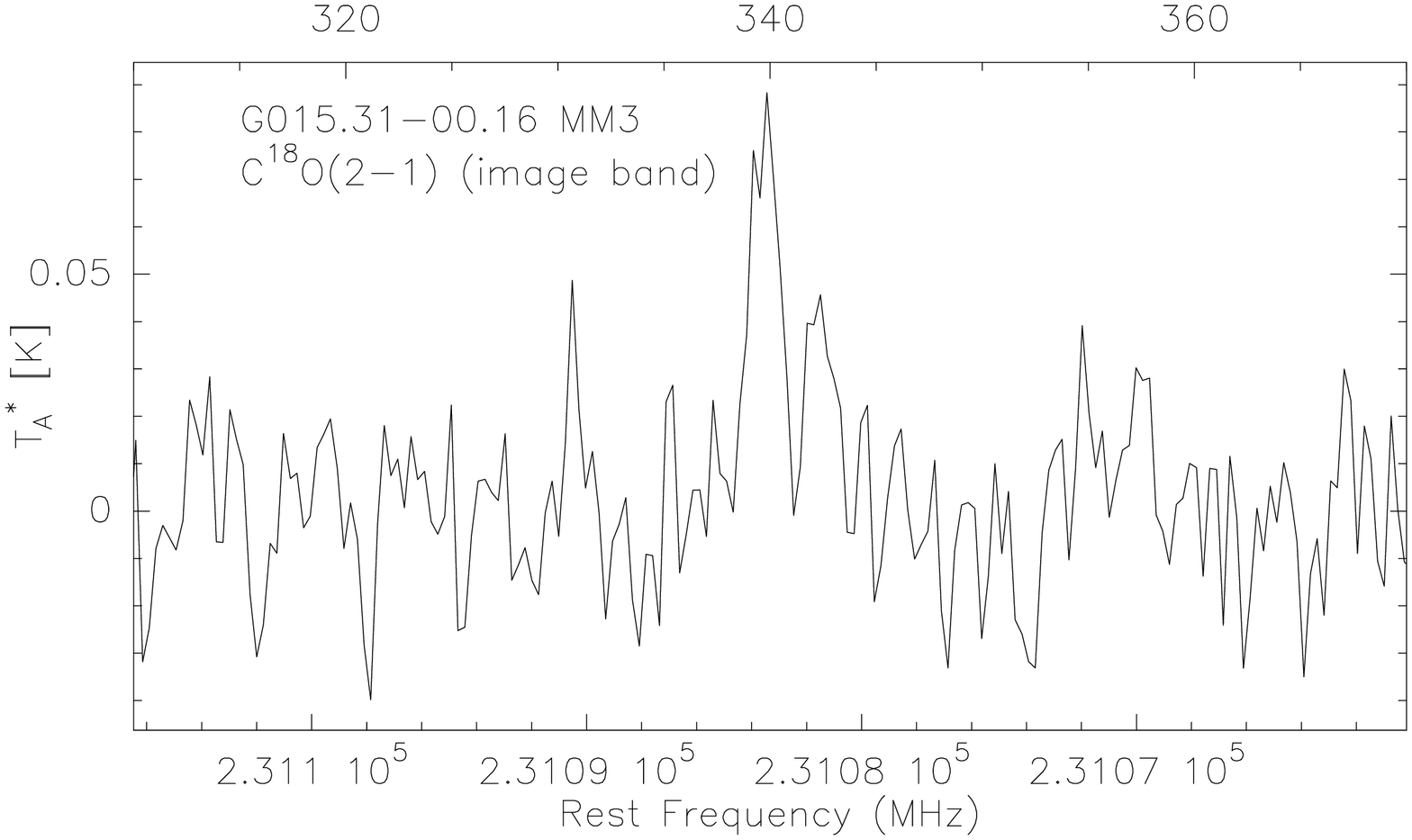}
\includegraphics[width=2.12cm,angle=-90]{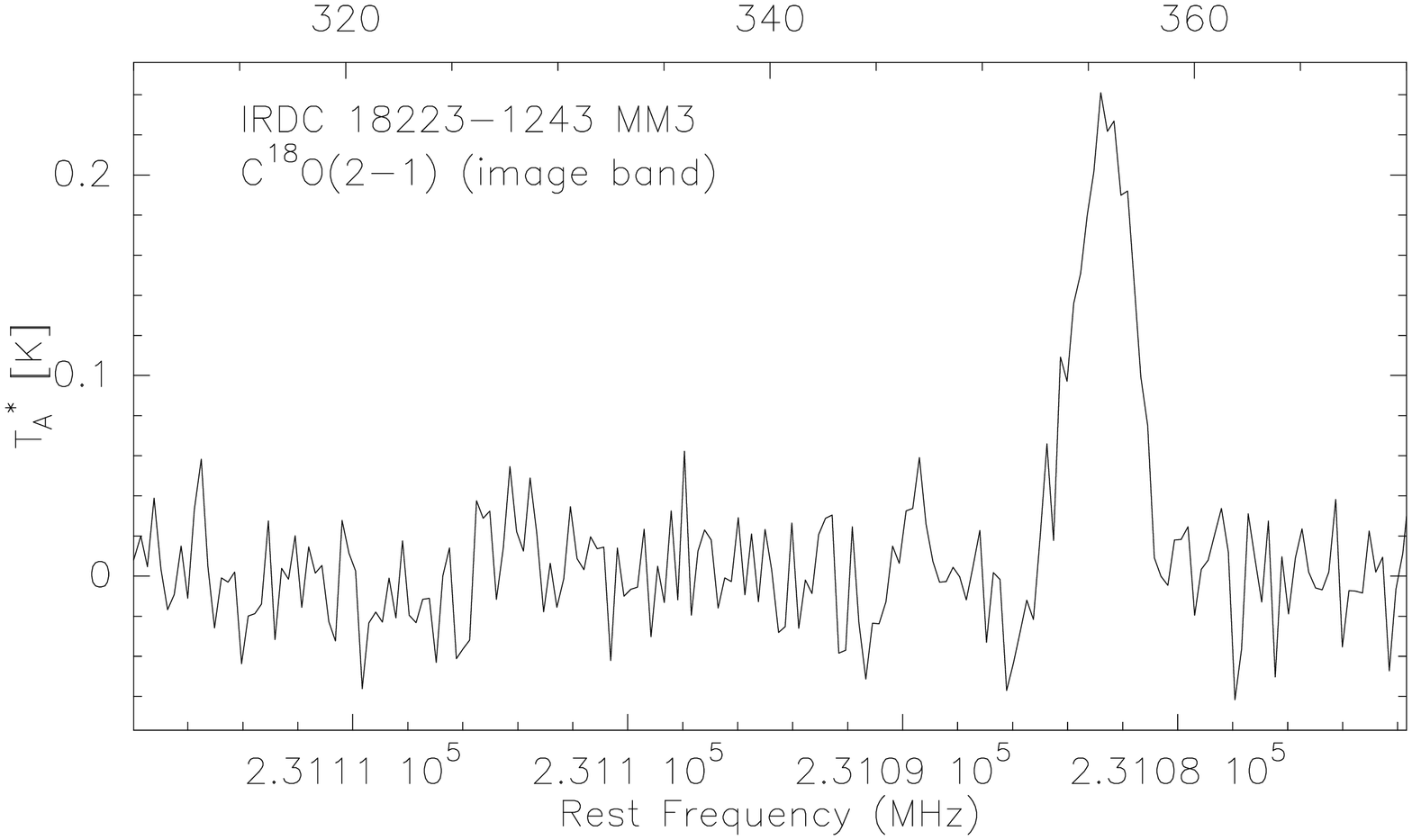}
\includegraphics[width=2.1cm,angle=-90]{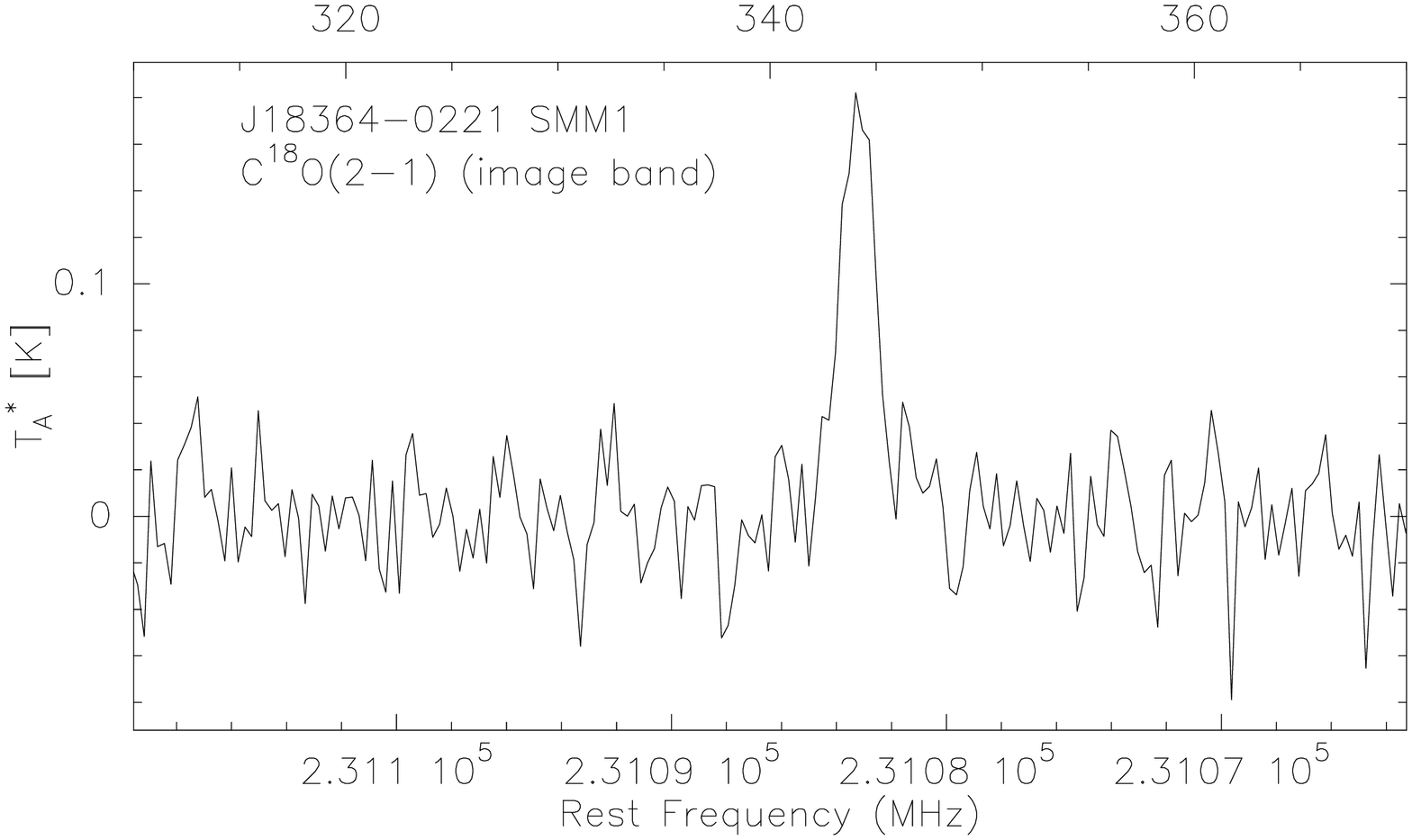}
\caption{C$^{18}$O$(2-1)$ lines at 219\,560.3568 MHz (JPL) arising from the 
image sideband.}
\label{figure:Ulines}
\end{center}
\end{figure*}

The $J_{K_a,K_c}=3_{3,0}-2_{2,1}$ transition of \textit{ortho}-c-C$_3$H$_2$ 
at $\sim216.3$ GHz in the LSB is also detected towards I18102 MM1, I18182 
MM2, and I18223 MM3 (Fig.~\ref{figure:C3H2}). Thermal (${\rm v}=0$) 
SiO$(6-5)$ line at $\sim260.5$ GHz in the USB is seen towards I18102 MM1, 
I18151 MM2, and I18223 MM3 (Fig.~\ref{figure:sio}). In particular, the 
SiO$(6-5)$ line of I18151 MM2 exhibits very broad wing emission. 
Finally, as in the cases of I18102 MM1 and I18182 MM2, we detected the $J=2-1$ 
transition of C$^{18}$O at $\sim219.56$ GHz in the image side band towards 
G015.05 MM1, I18151 MM2, G015.31 MM3, I18223 MM3, and J18364 SMM1 
(Fig.~\ref{figure:Ulines}). The intensities of these C$^{18}$O lines, 
$T_{\rm A}^*\sim0.09-0.31$ K, are mostly comparable to those possibly detected 
towards I18102 MM1 and I18182 MM2. Note that the lines ``leaking'' from the 
rejected image sideband are heavily attenuated by the sideband filter. Thus, 
we cannot establish the correct intensity scale for the image-band 
lines.

In Table~\ref{table:otherparameters} we list the observed extra 
transitions, their rest frequencies (observed frequency for the U-line), and 
upper-state energies. The rest frequencies and upper-state energies were 
assigned using JPL, CDMS, and 
Splatalogue\footnote{{\tt http://www.splatalogue.net/}} 
(\cite{remijan2007}) spectroscopic databases. The additional lines/species 
are discussed further in Appendix A.

\begin{table}
\caption{Other candidate species/transitions observed.}
{\scriptsize
\begin{minipage}{1\columnwidth}
\centering
\renewcommand{\footnoterule}{}
\label{table:otherparameters}
\begin{tabular}{c c c}
\hline\hline 
Species/ & $\nu$ & $E_{\rm u}/k_{\rm B}$\\  
transition\tablefootmark{a}  
        & [MHz] & [K]\\
\hline
\textit{ortho}-c-C$_3$H$_2(J_{K_a,K_c}=3_{3,0}-2_{2,1})$ ${\rm v}=0$ & 216\,278.7560 (JPL) & 19.5\\

C$_2$D$(N_{J,F}=3_{7/2,9/2}-2_{5/2,7/2})$\tablefootmark{b} & 216\,372.8300 (CDMS) & 20.8\\

CH$_2$CHC$^{15}$N$(J_{K_a,K_c}=25_{2,24}-25_{0,25})$\tablefootmark{c} & 216\,428.6683 (JPL) & 151.8\\

U & 216\,452.5110 & \ldots \\

CNCHO$(J_{K_a,K_c}=23_{0,23}-22_{0,22})$\tablefootmark{d} & 219\,531.6000 (CDMS) & 127.2\\

C$^{18}$O$(J=2-1)$\tablefootmark{d} & 219\,560.3600 (JPL) & 15.8\\

CH$_3$NH$_2$-E$(J_{K_a,K_c}=7_{2,5}-7_{1,5})$\tablefootmark{e} & 231\,060.6041 (JPL) & 75.6\\

OCS$(J=19-18)$ & 231\,060.9830 (JPL) & 110.9\\

CH$_3$COCH$_3$-EA$(J_{K_a,K_c}=15_{9,7}-14_{8,6})$ ${\rm v}=0$ & 231\,080.9647 (JPL) & 95.9\\

$^{13}$CS$(J=5-4)$ & 231\,220.9960 (JPL) & 33.3\\

SiO$(J=6-5)$ & 260\,518.0200 (JPL) & 43.8\\
\hline 
\end{tabular} 
\tablefoot{
\tablefoottext{a}{For asymmetric top molecules, $K_a$ and $K_c$ refer to 
the projection of the angular momentum along the $a$ and $c$ principal axes. 
The CH$_2$CHC$^{15}$N and CNCHO transitions are of type $a$ 
($\Delta K_a=0,\pm2,\ldots$ and $\Delta K_c=\pm1,\pm3,\ldots$), 
the \textit{o}-c-C$_3$H$_2$ and CH$_3$COCH$_3$-EA transitions are $b$-type 
($\Delta K_a=\Delta K_c=\pm1,\pm3,\ldots$), and the CH$_3$NH$_2$-E transition 
exbits $c$-type selection rules ($\Delta K_a=\pm1,\pm3,\ldots$ and 
$\Delta K_c=0,\pm2,\ldots$) (\cite{gordy1984}).} \tablefoottext{b}{This is the strongest hf component and it is blended with the $J,F=7/2,5/2-5/2,3/2$ and $J,F=7/2,7/2-5/2,5/2$ hf components.} \tablefoottext{c}{Blended with the C$_2$D$(3-2)$ hf group.} \tablefoottext{d}{Seen in the image band. The candidate lines of CNCHO, CH$_3$NH$_2$-E, OCS, and CH$_3$COCH$_3$-EA could be blended with C$^{18}$O$(2-1)$ coming from the image band.} \tablefoottext{e}{The corresponding vibrational level is $E_1$, $l=-1$. This line is blended with OCS$(19-18)$.}}
\end{minipage} }
\end{table}

\subsection{Spectral line parameters}

The spectral line parameters are given in Table~\ref{table:lineparameters}. 
In this table we give the radial velocity (${\rm v}_{\rm LSR}$), FWHM 
linewidth ($\Delta {\rm v}$), peak intensity ($T_{\rm A}^*$), integrated 
intensity ($\int T_{\rm A}^* {\rm dv}$), peak optical depth of the line
($\tau_0$), and excitation temperature ($T_{\rm ex}$). The values of 
${\rm v}_{\rm LSR}$ and $\Delta {\rm v}$ for C$^{17}$O, H$^{13}$CO$^+$, 
DCO$^+$, N$_2$H$^+$, N$_2$D$^+$, and C$_2$D were derived through fitting the 
hf structure. For the other lines these parameters were derived by 
fitting a single Gaussian to the line profile. The values of 
$T_{\rm A}^*$ were also determined by fitting the lines with a single Gaussian. 
The velocity-integrated intensities $\int T_{\rm A}^* {\rm dv}$ were calculated 
over the velocity range given in Col.~(6) of Table~\ref{table:lineparameters}.
 The uncertainties reported in ${\rm v}_{\rm LSR}$, $\Delta {\rm v}$, 
$T_{\rm A}^*$, and $\int T_{\rm A}^* {\rm dv}$ represent the formal $1\sigma$ 
errors determined by the fitting routine. We used 
{\tt RADEX}\footnote{{\tt http://www.strw.leidenuniv.nl/$\sim$moldata/radex.html}} (\cite{vandertak2007}) to determine the valu\-es of $\tau_0$ and 
$T_{\rm ex}$ for the lines of C$^{17}$O, H$^{13}$CO$^+$, DCO$^+$, N$_2$H$^+$, 
c-C$_3$H$_2$, OCS, $^{13}$CS, and SiO. {\tt RADEX} modelling is described in 
more detail in Sect.~4.2 where we discuss the determination of molecular 
column densities. For N$_2$D$^+(3-2)$ we used as $T_{\rm ex}$ the value 
derived for N$_2$H$^+(3-2)$ or, when not possible, we adopted 
$T_{\rm ex}[{\rm N_2H^+(1-0)}]$ from SSK08. We note that the $T_{\rm ex}$ 
values obtained for the $J=3-2$ transition of N$_2$H$^+$ are comparable to 
those of $J=1-0$ from SSK08.

\begin{table*}
\caption{Spectral line parameters.}
{\scriptsize
\begin{minipage}{2\columnwidth}
\centering
\renewcommand{\footnoterule}{}
\label{table:lineparameters}
\begin{tabular}{c c c c c c c c}
\hline\hline 
Source & Transition & ${\rm v}_{\rm LSR}$ & $\Delta {\rm v}$ & $T_{\rm A}^*$ & $\int T_{\rm A}^* {\rm dv}$\tablefootmark{a} & $\tau_0$\tablefootmark{b} & $T_{\rm ex}$\tablefootmark{c}\\
     & & [km~s$^{-1}$] & [km~s$^{-1}$] & [K] & [K~km~s$^{-1}$] & & [K]\\
\hline
IRDC 18102-1800 MM1 & C$^{17}$O$(2-1)$ & $21.5\pm0.02$ & $2.36\pm0.06$ & $0.85\pm0.06$ & $2.58\pm0.05$ [18.4,\, 25.0] & $0.08\pm0.01$ & $18.5\pm0.4$\\
                    & H$^{13}$CO$^+(3-2)$ & $21.7\pm0.08$ & $2.92\pm0.53$ & $0.37\pm0.05$ & $1.25\pm0.06$ [18.5,\, 24.2] & $0.99\pm0.45$ & $4.6\pm0.3$\\
		    & DCO$^+(3-2)$ & $22.0\pm0.2$ & $0.56\pm3.49$ & $0.04\pm0.01$ & $0.04\pm0.02$ [21.1,\, 22.8] & $0.05\pm0.01$ & $4.6\pm0.3$\\
                    & N$_2$H$^+(3-2)$ & $21.9\pm0.02$ & $3.56\pm0.02$ & $2.47\pm0.27$ & $17.62\pm0.12$ [15.6,\, 28.6] & $16.92\pm7.28$ & $8.2\pm0.1$\\
                    & N$_2$D$^+(3-2)$ & $22.0\pm0.1$ & $0.80\pm0.20$ & $0.07\pm0.01$ & $0.07\pm0.01$ [21.5,\, 23.0] & 0.06 & $8.2\pm0.1$\tablefootmark{d}\\
                    & c-C$_3$H$_2(3_{3,0}-2_{2,1})$ & $21.3\pm0.1$ & $1.10\pm0.26$ & $0.08\pm0.01$ & $0.07\pm0.02$ [20.6,\, 22.4] & $0.45\pm0.19$ & $3.4\pm0.1$\\
		    & CH$_3$NH$_2(7_{2,5}-7_{1,5})$\tablefootmark{e} & $21.1\pm0.2$ & $5.07\pm0.58$ & $0.05\pm0.01$ & $0.24\pm0.03$ [16.7,\, 25.1] & 0.003 & 50.4\\
                    & OCS$(19-18)$\tablefootmark{e} & $21.7\pm0.2$ & $-\left| \, \right|-$ & $-\left| \, \right|-$ & $-\left| \, \right|-$ [17.4,\, 25.4] & $0.01\pm0.0003$ & $13.2\pm0.4$\\
                    & $^{13}$CS$(5-4)$ & $22.3\pm0.2$ & $2.69\pm0.38$ & $0.05\pm0.01$ & $0.18\pm0.02$ [19.7,\, 24.7] & $0.07\pm0.01$ & $4.7\pm0.2$\\
                    & SiO$(6-5)$ & $21.5\pm0.4$ & $6.66\pm1.28$ & $0.11\pm0.04$ & $1.30\pm0.11$ [16.5,\, 25.3] & $0.42\pm0.27$ & $4.0\pm0.7$\\
G015.05+00.07 MM1 & C$^{17}$O$(2-1)$ & $24.6\pm0.03$ & $2.22\pm0.24$ & $0.74\pm0.04$ & $2.05\pm0.05$ [21.6,\,26.9] & $0.08\pm0.01$ & $16.4\pm1.7$\\
 		  & H$^{13}$CO$^+(3-2)$ & $24.5\pm0.1$ & $1.66\pm0.24$ & $0.08\pm0.01$ & $0.20\pm0.02$ [22.8,\, 26.8]& $0.11\pm0.01$ & $4.9\pm0.1$\\
		  & DCO$^+(3-2)$ & $24.1\pm0.1$ & $0.67\pm0.15$ & $0.08\pm0.003$ & $0.07\pm0.01$ [23.3,\, 25.2] & $0.09\pm0.01$ & $4.9\pm0.1$\\
                  & N$_2$H$^+(3-2)$ & $24.7\pm0.1$ & $4.58\pm0.29$ & $0.48\pm0.06$ & $2.47\pm0.09$ [19.9,\, 29.3] & $0.98\pm0.14$ & $5.2\pm0.1$\\
                  & N$_2$D$^+(3-2)$ & $24.1\pm0.1$ & $0.53\pm0.13$ & $0.06\pm0.01$ & $0.03\pm0.01$ [23.4,\, 25.4] & 0.12 & $5.2\pm0.1$\tablefootmark{d}\\ 
IRDC 18151-1208 MM2 & C$^{17}$O$(2-1)$ & $29.5\pm0.02$ & $2.75\pm0.15$ & $1.03\pm0.04$ & $3.33\pm0.05$ [26.5,\, 32.8] & $0.09\pm0.01$ & $20.8\pm1.9$\\
		    & H$^{13}$CO$^+(3-2)$ & $29.5\pm0.04$ & $2.47\pm0.11$ & $0.64\pm0.05$ & $1.63\pm0.06$ [27.4,\, 31.8] & $0.44\pm0.10$ & $6.8\pm0.4$\\
                    & DCO$^+(3-2)$ & $29.4\pm0.03$ & $1.43\pm0.10$ & $0.33\pm0.03$ & $0.75\pm0.03$ [27.3,\, 31.3] & $0.15\pm0.03$ & $7.2\pm0.6$\\
                    & c-C$_3$H$_2(3_{3,0}-2_{2,1})$ & $29.5\pm 0.2$ & $2.87\pm0.49$ & $0.07\pm0.02$ & $0.18\pm0.05$ [27.2,\, 31.5] & $0.08\pm0.02$ & $4.8\pm0.3$\\
                    & N$_2$D$^+(3-2)$ & $29.4\pm0.1$ & $1.03\pm0.18$ & $0.11\pm0.03$ & $0.27\pm0.03$ [28.3,\, 30.9] & 0.04 & $14.0\pm2.0$\tablefootmark{f}\\
                    & C$_2$D$(3-2)$\tablefootmark{g} & $29.6\pm0.3$ & $2.68\pm0.77$ & $0.07\pm0.02$ & $0.45\pm0.07$\tablefootmark{h} & 0.02 & 20.8\\
                    & CH$_2$CHC$^{15}$N$(25_{2,24}-25_{0,25})$\tablefootmark{g} & $30.1\pm0.5$ & $3.15\pm0.97$ & $0.04\pm0.02$ & $0.14\pm0.01$ [27.8,\, 33.3] & 0.001 & 101.2\\
                    & U216452.511 & $29.4\pm0.2$ & $2.06\pm0.49$ & $0.05\pm0.02$ & $0.13\pm0.02$ [28.0,\, 31.3] & \ldots & \ldots\\
                    & SiO$(6-5)$ & $30.5\pm1.0$ & $46.80\pm2.36$ & $0.09\pm0.03$ & $1.93\pm0.20$ [-6.1,\, 72.6] & $0.09\pm0.01$ & $5.7\pm0.1$\\
G015.31-00.16 MM3 & C$^{17}$O$(2-1)$ & $31.0\pm0.03$ & $0.80\pm0.09$ & $0.41\pm0.07$ & $0.80\pm0.04$ [28.9,\, 32.3] & $0.10\pm0.01$ & $10.2\pm0.6$\\
                  & H$^{13}$CO$^+(3-2)$ & $31.4\pm0.2$ & $0.89\pm0.14$ & $0.03\pm0.01$ & $0.05\pm0.02$ [30.2,\, 32.5] & $0.18\pm0.01$ & $3.5\pm0.1$\\
                  & DCO$^+(3-2)$ & $30.8\pm0.2$ & $0.87\pm0.15$ & $0.03\pm0.01$ & $0.06\pm0.01$ [29.8,\, 31.9] & $0.13\pm0.02$ & $3.4\pm0.1$\\
                  & N$_2$H$^+(3-2)$ & $30.9\pm0.04$ & $0.86\pm0.18$ & $0.26\pm0.02$ & $0.31\pm0.02$ [29.9,\, 31.9]\tablefootmark{i} & $4.64\pm1.30$ & $3.9\pm0.1$\\ 
IRDC 18182-1433 MM2 & C$^{17}$O$(2-1)$ & $41.0\pm0.02$ & $1.32\pm0.08$ & $0.74\pm0.08$ & $1.82\pm0.03$ [38.0,\, 43.0] & $0.13\pm0.01$ & $12.6\pm0.5$\\
                    & H$^{13}$CO$^+(3-2)$ & $40.7\pm0.1$ & $1.20\pm0.34$ & $0.11\pm0.02$ & $0.22\pm0.02$ [39.4,\, 42.5] & $0.50\pm0.03$ & $3.8\pm0.1$\\
                    & DCO$^+(3-2)$ & $41.1\pm0.3$ & $2.08\pm0.90$ & $0.09\pm0.01$ & $0.20\pm0.03$ [38.6,\, 43.3] & $0.28\pm0.02$ & $3.8\pm0.1$\\
                    & c-C$_3$H$_2(3_{3,0}-2_{2,1})$ & $40.9\pm0.1$ & $1.58\pm0.31$ & $0.07\pm0.01$ & $0.12\pm0.02$ [39.3,\, 42.4] & $1.29\pm0.13$ & $3.1\pm0.1$\\
                    & CH$_3$COCH$_3(15_{9,7}-14_{8,6})$ ?\tablefootmark{j} & $41.3\pm0.1$ & $1.80\pm0.14$ & $0.17\pm0.01$ & $0.35\pm0.03$ [39.5,\, 42.9] & 0.01 & 63.9\\
IRDC 18223-1243 MM3 & C$^{17}$O$(2-1)$ & $45.5\pm0.04$ & $3.09\pm0.10$ & $0.56\pm0.04$ & $1.97\pm0.05$ [42.1,\, 49.0] & $0.06\pm0.01$ & $18.2\pm1.1$\\
                    & H$^{13}$CO$^+(3-2)$ & $45.5\pm0.04$ & $1.81\pm0.26$ & $0.39\pm0.03$ & $0.83\pm0.04$ [43.4,\, 47.0] & $0.47\pm0.05$ & $5.5\pm0.1$\\
                    & DCO$^+(3-2)$ & $45.9\pm0.1$ & $1.83\pm0.18$ & $0.11\pm0.01$ & $0.23\pm0.02$ [43.4,\, 47.6] & $0.08\pm0.01$ & $5.6\pm0.2$\\
                    & N$_2$D$^+(3-2)$ & $45.3\pm0.3$ & $1.13\pm0.63$ & $0.08\pm0.02$ & $0.20\pm0.03$ [44.2,\, 47.7] & 0.05 & $9.4\pm0.3$\tablefootmark{f}\\ 
                    & c-C$_3$H$_2(3_{3,0}-2_{2,1})$ & $45.7\pm0.1$ & $1.84\pm0.15$ & $0.15\pm0.03$ & $0.22\pm0.02$ [43.6,\, 47.0] & $0.38\pm0.05$ & $4.0\pm0.1$\\
                 & SiO$(6-5)$ & $44.4\pm0.6$ & $8.22\pm1.26$ & $0.07\pm0.03$ & $0.84\pm0.08$ [38.0,\, 52.6] & $0.11\pm0.01$ & $4.9\pm0.1$\\
ISOSS J18364-0221 SMM1 & C$^{17}$O$(2-1)$ & $34.8\pm0.02$ & $1.60\pm0.05$ & $0.79\pm0.03$ & $1.69\pm0.03$ [32.5,\, 37.0] & $0.17\pm0.01$ & $10.9\pm0.2$\\
		       & H$^{13}$CO$^+(3-2)$ & $34.7\pm0.05$ & $0.86\pm0.14$ & $0.30\pm0.04$ & $0.44\pm0.04$ [33.5,\, 36.1] & $1.07\pm0.38$ & $4.3\pm0.2$\\
                       & DCO$^+(3-2)$ & $34.9\pm0.03$ & $1.14\pm0.10$ & $0.33\pm0.02$ & $0.50\pm0.02$ [33.5,\, 36.3] & $0.69\pm0.20$ & $4.4\pm0.2$\\
\hline 
\end{tabular} 
\tablefoot{
\tablefoottext{a}{Intensities are integrated over the velocity range given in square-brackets. In the cases of I18102 MM1, I18151 MM2, and I18223 MM3, 92.5\% of the total N$_2$D$^+(3-2)$ hf component's intensity lie within the quoted velocity range. For G015.05 MM1 the fraction is 81.7\%.} \tablefoottext{b}{For C$^{17}$O, H$^{13}$CO$^+$, DCO$^+$, N$_2$H$^+$, and N$_2$D$^+$ $\tau_0$ is the optical thickness in the centre of a hypothetical unsplit line. For C$_2$D $\tau_0$ is the sum of the peak optical thicknesses of all the hf components.} \tablefoottext{c}{See Sect.~4.2 for details on $T_{\rm ex}$ determination.} \tablefoottext{d}{$T_{\rm ex}$ is assumed to be the same as for N$_2$H$^+(3-2)$.} \tablefoottext{e}{Another candidate for this spectral line is the OCS$(19-18)$ transition. Alternatively, the two lines could be blended. It could also be due to C$^{18}$O$(2-1)$ seen in the image band.} \tablefoottext{f}{$T_{\rm ex}[{\rm N_2H^+(1-0)}]$ from SSK08.} \tablefoottext{g}{Another detected hf group of C$_2$D$(3-2)$ is blended with CH$_2$CHC$^{15}$N$(25_{2,24}-25_{0,25})$.} \tablefoottext{h}{Integrated intensity over the two detected hf groups between [27.7,\, 30.7] and [-49.5,\, -46.5] km~s$^{-1}$.} \tablefoottext{i}{92.6\% of the hf component's intensity lie within the quoted velocity range.}\tablefoottext{j}{This line may also be due to C$^{18}$O$(2-1)$ seen in the image band.}}
\end{minipage} }
\end{table*}

\section{Analysis and results}

\subsection{Revision of clump properties}

In this subsection, we present recalculations of several clump properties 
presented previously in the literature (see 
Tables~\ref{table:sources} and \ref{table:properties}). 

\subsubsection{Kinematic distances}

SSK08 calculated the near kinematic distance 
for all of our clumps except J18364 SMM1 (see their Table~4). 
They used the LSR velocity of N$_2$H$^+(1-0)$ and the rotation curve of 
Clemens (1985) with the standard rotation parameters 
($\Theta_0$, $R_0$)$=$(220 km~s$^{-1}$, 8.5 kpc), where $\Theta_0$ is the
circular orbital speed of the Sun, and $R_0$ is the galactocentric distance 
of the Sun. Because N$_2$H$^+$ is a tracer of high density gas, the radial 
velocity derived from N$_2$H$^+$ lines is suitable to 
determine the source kinematic distance. The hf structure of 
N$_2$H$^+(1-0)$ was also resolved (to some degree) towards many of the sources 
by SSK08, and thus the hf-structure fitting is expected 
to yield a reliable centroid velocity. Towards the three sources for which we 
detected N$_2$H$^+(3-2)$ lines, the obtained values of ${\rm v}_{\rm LSR}$ are 
comparable to those from N$_2$H$^+(1-0)$. We recalculated the distances from 
SSK08 using the recent rotation curve of Reid et al. (2009) 
which is based on measurements of trigonometric parallaxes and proper motions 
of masers in high-mass star-forming regions. The best-fit rotation parameters 
of Reid et al. (2009) are ($\Theta_0$, $R_0$)$=$(254 km~s$^{-1}$, 8.4 kpc).
The resulting galactocentric distances and near kinematic distances 
($R_{\rm GC}$ and $d$) are given in Cols.~(6) and (7) of 
Table~\ref{table:sources}. The revised distances differ at most by 
0.2 kpc from those reported by SSK08.

The source J18364 SMM1 is an exception. Its CO$(1-0)$ radial velocity of 
$\sim33$ km~s$^{-1}$ corresponds to a kinematic distance of $\sim2.2$ kpc 
according to the Brand \& Blitz (1993) rotation curve 
(see \cite{birkmann2006}). We recomputed this distance from the LSR velocity 
of C$^{17}$O$(2-1)$ (34.8 km~s$^{-1}$) because it is tracing higher density 
gas than the main CO isotopologue (si\-milar radial velocity was obtained for 
the other detected transitions). The revised distance is $\sim2.5$ kpc 
according to the Reid et al. (2009) rotation curve; the previous value 
of 2.2 kpc is within the errors.

We note that for the sources I18102 MM1, I18151 MM2, I18182 MM2, and 
I18223 MM3 there are kinematic distance estimates (2.6, 3.0, 4.5, and 3.7 
kpc, respectively) based on the CS$(2-1)$ velocity and the rotation curve of 
Brand \& Blitz (1993) (see \cite{sridharan2002}). For both G015.05 MM1 and 
G015.31 MM3, Rathborne et al. (2006) used the kinematic distance of 3.2 kpc 
derived from the $^{13}$CO$(1-0)$ velocity and the rotation curve of Clemens 
(1985). Our distances differ by 0.1--1.0 kpc from the above values.

\subsubsection{Gas and dust temperatures}

Sridharan et al. (2005) determined the NH$_3$ rotational temperature, 
$T_{\rm rot}({\rm NH_3})$, for three of our sources (I18151 MM2, I18182 MM2, 
and I18223 MM3) at the $40\arcsec$ angular resolution. SSK08 
have determined $T_{\rm rot}({\rm NH_3})$ for all of our sources except 
J18364 SMM1 (see their Table~10). The values of $T_{\rm rot}({\rm NH_3})$ 
from the above two studies are otherwise similar except in the case of 
I18223 MM3. Sridharan et al. reported the value 32.7 K for this source, 
whereas SSK08 obtained a factor of two lower tempe\-rature 
($16.2^{+1.0}_{-0.9}$ K) at about 1.8 times poorer angular resolution 
($73\arcsec$). The $T_{\rm rot}({\rm NH_3})$ value for J18364 SMM1 at 
the $40\arcsec$ resolution is 10.75 K (\cite{krause2003}). 
We converted $T_{\rm rot}({\rm NH_3})$ from Krause (2003) and SSK08 
into the gas kinetic temperature, $T_{\rm kin}$, using the relationship given 
by Tafalla et al. (2004; Appendix B therein). The uncertainty in 
$T_{\rm kin}$ was calculated by propagating the larger of the two 
$T_{\rm rot}$-errors given by SSK08.

For three of our sources, G015.05 MM1, I18223 MM3, and J18364 SMM1, dust 
temperature estimates have been made by Rathborne et al. (2010), 
Beuther et al. (2010), and Birkmann et al. (2006), respectively. 
The determined dust temperature va\-lues of G015.05 MM1, 11.0--36.0 K, 
bracket the $T_{\rm kin}$ value of 17.2 K. Dust temperature of I18223 MM3 was 
also derived by Beuther \& Steinacker (2007) from the source SED using the 
\textit{Spitzer}/MIPS data at 24 and 70 $\mu$m, and MAMBO 1.2 mm and 
PdBI 3.2 mm data. The value 15 K they obtained for the cold part of the 
spectrum is three kelvins lower than the value 18 K derived recently by 
Beuther et al. (2010). The latter also used the \textit{Herschel}/PACS 
(70, 100, and 160 $\mu$m) and SPIRE (250, 350, and 500 $\mu$m) data, 
and SCUBA 850 $\mu$m data to construct the source SED (but not the 3.2-mm 
flux density). The dust tempe\-rature of I18223 MM3 is comparable to its gas 
temperature of 18.7 K. By utilising the far-infrared and submm flux density 
ratios, Birkmann et al. (2006) deduced the value 
$T_{\rm dust}=16.5_{-3.0}^{+6.0}$ K for J18364 SMM1, which is 
higher than the gas temperature 11.4 K.

The gas and dust temperatures of the clumps are given in Cols.~(2) and (3) of 
Table~\ref{table:properties}. 

\subsubsection{Clump masses, radii, and H$_2$ column and number 
densities}

We calculated the masses and beam-averaged H$_2$ column densities of the 
clumps using the formulas 

\begin{equation}
\label{eq:mass}
M=\frac{S_{\nu}d^2}{B_{\nu}(T_{\rm dust})\kappa_{\nu}R_{\rm d}} \,,
\end{equation}

\begin{equation}
\label{eq:NH2}
N({\rm H_2})=\frac{I_{\nu}^{\rm dust}}{B_{\nu}(T_{\rm dust})\mu_{\rm H_2} m_{\rm H}\kappa_{\nu}R_{\rm d}} \,.
\end{equation}
In the above formulas, $I_{\nu}^{\rm dust}$ and $S_{\nu}$ are the peak surface 
brightness and integrated flux density of the (sub)mm dust emission; note that 
$I_{\nu}^{\rm dust}=S_{\nu}^{\rm peak}/\Omega_{\rm beam}$, where 
$S_{\nu}^{\rm peak}$ is the peak flux density and $\Omega_{\rm beam}$ is 
the solid angle of the telescope beam. $B_{\nu}(T_{\rm dust})$ is the 
Planck function for a dust temperature $T_{\rm dust}$, $\kappa_{\nu}$ is the 
dust opacity per unit dust mass, 
$R_{\rm d}\equiv \langle M_{\rm dust}/M_{\rm gas}\rangle$ is the average 
dust-to-gas mass ratio, $\mu_{\rm H_2}$ is the mean molecular weight per H$_2$ 
molecule (2.8 for the He/H abundance ratio of 0.1), and $m_{\rm H}$ is the mass 
of a hydrogen atom.

The $I_{\nu}^{\rm dust}$ values were determined from the (sub)mm maps 
shown as contours in Fig.~\ref{figure:maps}. The angular resolutions of the 
maps are different: the beam size of the SCUBA 850 $\mu$m, Bolocam 1.1 mm, 
and MAMBO 1.2 mm data are 14\farcs4, 31\arcsec, and 11\arcsec, respectively. 
The values of $S_{\nu}$ were taken from the literature as follows: for 
I18102 MM1, I18151 MM2, I18182 MM2, and I18223 MM3 we used the MAMBO 
1.2-mm flux densities from Beuther et al. (2002a); for G015.05 MM1 and G015.31 
MM3 we used the MAMBO-II 1.2-mm flux densities from Rathborne et al. (2006); 
for J18364 SMM1 we used the SCUBA 850-$\mu$m flux density from Birkmann et al. 
(2006). These flux densities are given in Col.~(4) of 
Table~\ref{table:properties}. 

As the dust temperature of the clumps, we used the gas kinetic temperatures 
listed in Col.~(2) of Table~\ref{table:properties}, and assumed that 
$T_{\rm dust}=T_{\rm kin}$.  We extrapolated the valu\-es 
of $\kappa_{\nu}$ from the Ossenkopf \& Henning (1994) model for dust grains 
with thin ice mantles, coagulated for $10^5$ yr at a gas density of 
$n_{\rm H}=n({\rm H})+2n({\rm H_2})\simeq 2n({\rm H_2})=10^6$ cm$^{-3}$ [their 
Table 1, Col.~(6)]. This is expected to be a reasonable dust model for the 
sources within cold and dense IRDCs. 
At the wavelengths considered in the present work, these values are 
$\kappa_{\rm 850\,\mu m}=0.197$ m$^2$~kg$^{-1}$, 
$\kappa_{\rm 1.1\,mm}=0.121$ m$^2$~kg$^{-1}$, and 
$\kappa_{\rm 1.2\,mm}=0.106$ m$^2$~kg$^{-1}$. The canonical value $1/100$ was 
adopted for $R_{\rm d}$. We note that the dust opacities 
are likely to be uncertain by a factor of $\gtrsim2$ (\cite{ossenkopf1994}; 
\cite{motte2001}). For comparison, Williams et al. (2004), Enoch et al. 
(2006), and Rygl et al. (2010) used the same dust model as here and 
interpolated the values of $\kappa_{\nu}$ at 850 $\mu$m, 1.1 mm, 
and 1.2 mm to be 0.154, 0.114, and 0.1 m$^2$~kg$^{-1}$, respectively. These 
are slightly smaller than our values, and the difference is likely to be caused
by different extrapolation methods (e.g., log-interpolation) and/or different 
dust emissivity index, $\beta$, used by the authors to determine the value of 
$\kappa_{\nu}\propto \nu^{\beta}$. 

In the papers by Beuther et al. (2002a) and Rathborne et al. (2006), 
the reported clump sizes refer to their FWHM sizes (diameters) resulting from 
two-dimensional Gaussian fits. However, the integrated 
flux densities used to calculate the masses refer to larger clump areas, $A$. 
Thus, in order to properly calculate the volume-average H$_2$ number density, 
$\langle n({\rm H_2}) \rangle$, one way is to use the so-called 
effective radius of the clump, $R_{\rm eff}=\sqrt{A/\pi}$ (see, e.g., 
\cite{miettinenharju2010}). An alternative way to calculate 
$\langle n({\rm H_2}) \rangle$ is to compute the amount of mass within the 
FWHM contour and use the FWHM radius. For sources with a Gaussian shape, the 
mass within the FWHM contour is a fraction $\ln~2\simeq0.693$ of the total 
mass (see \cite{kauffmann2010}). Following Kauffmann \& Pillai (2010), we 
reduce the clump masses by the above factor and use half the FWHM size as the 
effective radius of the clump. The density $\langle n({\rm H_2}) \rangle$ was 
then calculated using the formula

\begin{equation}
\label{eq:density}
\langle n({\rm H_2}) \rangle=\frac{\langle \rho \rangle}{\mu_{{\rm H_2}}m_{\rm H}}\,,
\end{equation}
where $\langle \rho \rangle=M/\left(4/3 \pi R_{\rm eff}^3\right)$ is 
the mass density, and $M$ refers to the reduced clump mass as described 
above. For the source J18364 SMM1, we scaled the effective radius 
$\approx 0.2$ pc reported by Birkmann et al. (2006) using the revised distance 
($R_{\rm eff}=0.23$ pc), and used the total mass within the clump area.

The results of the above calculations are presented in Cols.~(5)--(8) of 
Table~\ref{table:properties}. The uncertainties in the derived quantities 
were propagated from the uncertainties in $d$ and $T_{\rm kin}$, but 
we note that the uncertainty in (sub)mm dust opacity is the largest source of 
error in $M$ and $N({\rm H_2})$.

The derived values of $M$, $N({\rm H_2})$, and 
$\langle n({\rm H_2}) \rangle$ are mostly within a factor of $\lesssim2$ of 
those derived by Beuther et al. (2002a; where the revised equations by Beuther 
et al. (2005b) are employed), Rathborne et al. (2006), Birkmann et al. (2006), 
and Hennemann et al. (2009).
We note that besides the different source distances, the dust 
parameters used by the authors in the above re\-ference studies were different 
than here. Beuther et al. (2002a, 2005b) used $T_{\rm dust}$ values (35--50 K) 
resulting from the cold part of the source SEDs; the SEDs were constructed 
from the IRAS and mm data of the main core in the source region which explains 
the rather high dust temperatures. In addition, Beuther et al. (2002a, 2005b) 
used the grain radius, mass density, and emissivity index values of 
$a=0.1$ $\mu$m, $\rho_{\rm dust}=3$ g~cm$^{-3}$, and $\beta=2$, respectively, 
in the calculation of mass and column density. Following Hildebrand (1983), 
these values correspond to $\kappa_{\rm 1.2\,mm}\approx0.5$ m$^2$~kg$^{-1}$. 
Rathborne et al. (2006) used the values $T_{\rm dust}=15$ K and 
$\kappa_{\rm 1.2\,mm}=0.1$ m$^2$~kg$^{-1}$. For the source J18364 SMM1, Birkmann 
et al. (2006) used $T_{\rm dust}=16.5_{-3.0}^{+6.0}$ K instead of $T_{\rm kin}$, and 
they had $\kappa_{\rm 850\,\mu m}=0.180$ m$^2$~kg$^{-1}$ which is slightly smaller 
than ours even the adopted dust model was the same.

\begin{table*}
\caption{Physical properties of the sources.}
\begin{minipage}{2\columnwidth}
\centering
\renewcommand{\footnoterule}{}
\label{table:properties}
\begin{tabular}{c c c c c c c c}
\hline\hline
Source & $T_{\rm kin}$\tablefootmark{a} & $T_{\rm dust}$\tablefootmark{b} & $S_{\nu}$\tablefootmark{c} & $M$ & $N({\rm H_2})$ & $R_{\rm eff}$ & $\langle n({\rm H_2}) \rangle$\\
       &  [K] & [K] & [Jy] & [M$_{\sun}$] & [$10^{22}$ cm$^{-2}$] & [pc] & [$10^4$ cm$^{-3}$]\\
\hline
IRDC 18102-1800 MM1 & $21.3\pm1.6$ & \ldots & 3.3 & $357\pm163$ & $6.5\pm0.7$ & 0.34 & $2.9\pm1.3$\\
G015.05+00.07 MM1 & $17.2\pm2.1$ & 11.0-36.0 & 0.47 & $63\pm31$ & $1.0\pm0.2$ & 0.15 & $5.9\pm2.0$\\
IRDC 18151-1208 MM2 & $21.1\pm2.0$ & \ldots & 2.6 & $285\pm111$ & $12.9\pm1.6$ & 0.17 & $18.5\pm7.2$\\
G015.31-00.16 MM3 & $13.7\pm2.8$ & \ldots & 0.74 & $183\pm83$ & $0.8\pm0.3$ & 0.38 & $1.1\pm0.5$\\
IRDC 18182-1433 MM2 & $15.3\pm1.5$ & \ldots & 0.3 & $86\pm23$ & $5.2\pm0.8$ & 0.26 & $1.6\pm0.4$\\
IRDC 18223-1243 MM3 & $18.7\pm1.3$ & 15.0/18.0 & 0.8 & $173\pm43$ & $1.4\pm0.1$ & 0.18 & $9.5\pm2.3$\\
ISOSS J18364-0221 SMM1 & 11.4 & $16.5^{+6.0}_{-3.0}$ & 2.11 & $169\pm54$ & 7.6 & 0.23 & $6.4\pm2.0$\\
\hline 
\end{tabular} 
\tablefoot{
\tablefoottext{a}{Calculated from $T_{\rm rot}({\rm NH_3})$ from Krause (2003) and SSK08. Krause (2003) reported a slightly higher $T_{\rm kin}$ value of 11.8 K for J18364 SMM1.} \tablefoottext{b}{For G015.05 MM1, $T_{\rm dust}$ is from Rathborne et al. (2010). For I18223 MM3, the two $T_{\rm dust}$ values, 15 and 18 K, are from Beuther \& Steinacker (2007) and Beuther et al. (2010), respectively. The $T_{\rm dust}$ value for J18364 SMM1 is from Birkmann et al. (2006).} \tablefoottext{c}{For those sources whose name start with ``IRDC'', $S_{\nu}$ refers to MAMBO 1.2-mm flux density from Beuther et al. (2002a). For G015.05 MM1 and G015.31 MM3, the quoted value is the MAMBO-II 1.2-mm flux density (\cite{rathborne2006}). In the case of J18364 SMM1, we give the SCUBA 850-$\mu$m flux density from Birkmann et al. (2006).} }
\end{minipage}
\end{table*}

\subsection{Molecular column densities and fractional abundances}

The beam-averaged column densities of C$^{17}$O, H$^{13}$CO$^+$, DCO$^+$, 
N$_2$H$^+$, SiO, OCS, $^{13}$CS, and \textit{o}-c-C$_3$H$_2$ were derived 
using a one-dimensional spherically symmetric non-LTE radiative transfer code 
called {\tt RADEX} (see Sect.~3.2). {\tt RADEX} uses the method of mean escape
probability for an isothermal and homogeneous medium. The molecular data 
files (collisional rates) used in the {\tt RADEX} excitation analysis were 
taken from the LAMDA database (\cite{schoier2005}). The C$^{17}$O, 
H$^{13}$CO$^+$, DCO$^+$, and N$_2$H$^+$ transitions are treated as a 
hypothetical unsplit transition.
The input parameters in the off-line mode of {\tt RADEX} are the gas kinetic 
temperature, H$_2$ number density, and the width (FWHM) and intensity of the 
spectral line. We used the values of $T_{\rm kin}$ and 
$\langle n({\rm H_2}) \rangle$ listed in Table~\ref{table:properties}. 
However, we multiplied the densities by 1.2 
(He/H$_2=0.2$) to take the collisions with He into account 
(see Sect.~4.1 of the {\tt RADEX} 
manual\footnote{{\tt http://www.sron.rug.nl/$\sim$vdtak/radex/radex$_{-}$manual.pdf}}). As the input line intensity we used the main-beam brightness 
temperature $T_{\rm MB}=T_{\rm A}^*/ \eta_{\rm MB}$. 
When the source is resolved, $T_{\rm MB}$ is equal to the Rayleigh-Jeans 
equivalent radiation temperature, $T_{\rm R}$. The simulations aim to 
reproduce the observed line intensity and yield the values of the line peak 
optical thickness ($\tau_0$), excitation tempe\-rature ($T_{\rm ex}$), and the 
total co\-lumn density of the molecule ($N_{\rm tot}$). We varied $T_{\rm kin}$ 
and $\langle n({\rm H_2}) \rangle$ according to their errors to estimate the 
uncertainties associated with $\tau_0$, $T_{\rm ex}$, and $N_{\rm tot}$. 
The lines appear to be optically thin in most cases. The optical thickness 
is $\tau_0 \gtrsim 1$ for all the N$_2$H$^+$ lines, for H$^{13}$CO$^+$ 
towards I18102 MM1 and J18364 SMM1, and for c-C$_3$H$_2$ of 
I18182 MM2. The $T_{\rm ex}$ values for C$^{17}$O are close to 
$T_{\rm kin}$, indicating that the lines are nearly thermalised. 
Also, $T_{\rm ex}(\rm H^{13}CO^+)$ is found to be equal to 
$T_{\rm ex}(\rm DCO^+)$ within the errors. We note that 
the column density determination is in some cases hampered by self-absorbed 
line profiles.

\begin{figure}[!h]
\begin{center}
\resizebox{0.7\columnwidth}{!}{\includegraphics[angle=-90]{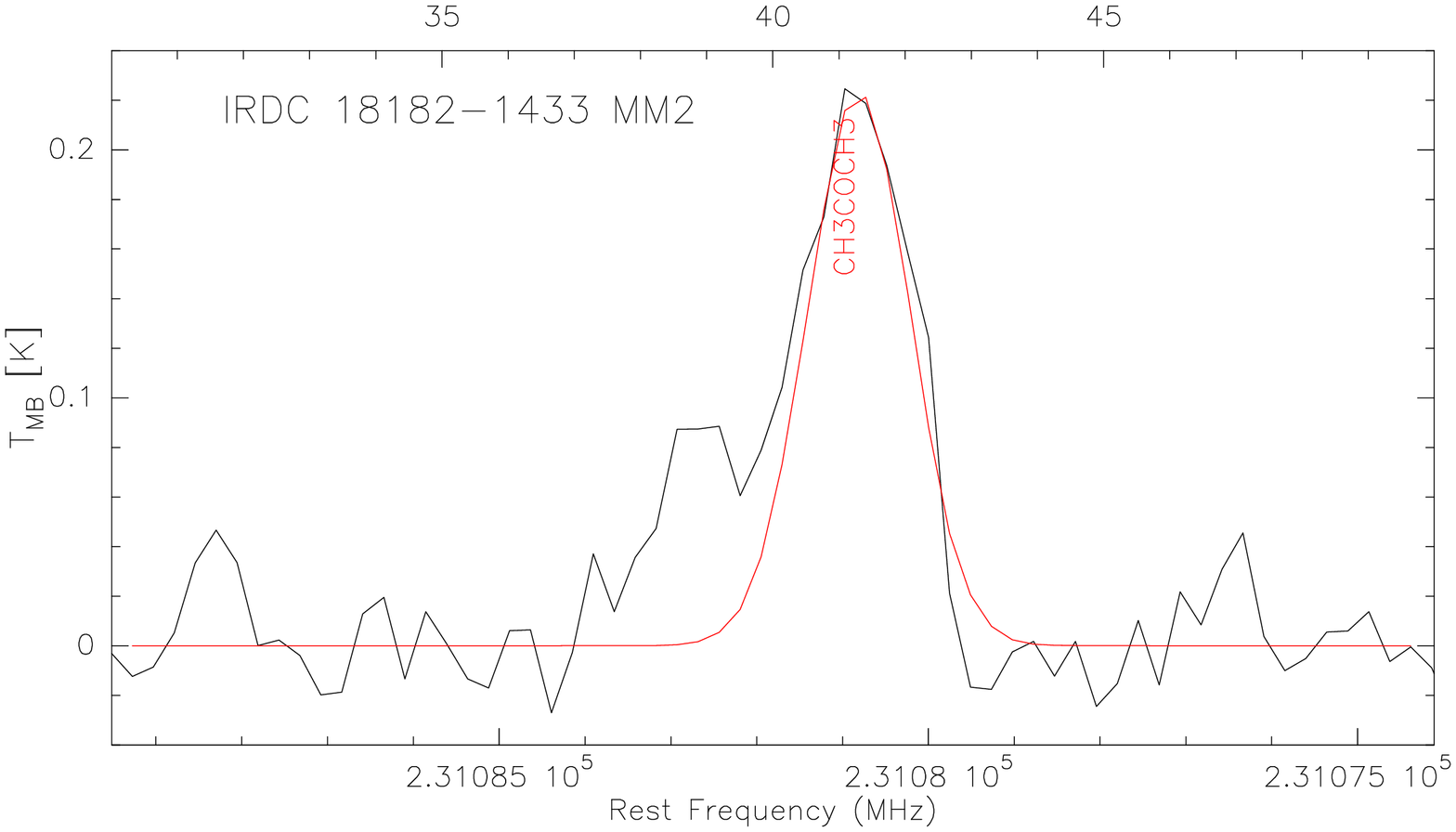}}
\end{center}
\caption{Example of the Weeds LTE modelling outlined in Sect.~4.2. The 
synthetic model spectrum is overlaid as a red line. Note that 
the intensity scale is $T_{\rm MB}$ for modelling purposes.}
\label{figure:model}
\end{figure}

For the rest of the observed molecules, there are no molecu\-lar data files 
available in the LAMDA database. The line optical thicknesses and total 
beam-averaged column densities of these molecules were determined through LTE 
modelling with CLASS/Weeds. The input parameters for a Weeds model are 
$N_{\rm tot}$, $T_{\rm ex}$, source size ($\theta_{\rm s}$), linewidth 
(FWHM), and offset from the reference-channel velocity. The linewidth 
is directly determined from the observed line profile, so there are 
basically three free parameters left ($N_{\rm tot}$, $T_{\rm ex}$, 
$\theta_{\rm s}$). Some of the model parameters may be degenerate, and cannot 
be determined independently (\cite{schilke2006}; \cite{maret2011}). 
The source size is degenerate with excitation temperature 
in the case of completely optically thick lines ($\tau \gg 1$), and with 
column density if the lines are completely optically thin ($\tau \ll 1$). 
We assumed that the source fills the telescope beam, i.e., that the source 
solid angle, $\Omega_{\rm s}\propto \theta_{\rm s}^2$, equals the beam solid 
angle, $\Omega_{\rm A}\propto \theta_{\rm HPBW}^2$, or 
$\theta_{\rm s}=\theta_{\rm HPBW}$. When the beam filling factor is unity, 
the line brightness temperature is $T_{\rm B} \simeq T_{\rm MB}$ 
[see Eqs.~(1) and (2) in Maret et al. (2011)]. 
For N$_2$D$^+$, we used as $T_{\rm ex}$ the values 
obtained for N$_2$H$^+$ from {\tt RADEX} simulations or from SSK08 
(Sect.~3.2). For the rest of the transitions we adopted 
the value $T_{\rm ex}=E_{\rm u}/k_{\rm B}$ for linear rotors, and 
$T_{\rm ex}=2/3\times E_{\rm u}/k_{\rm B}$ for asymmetric top rotors. 
The adopted rotational excitation temperatures give a lower limit to the 
total beam-averaged column densities (\cite{hatchell1998}). 
The input $N_{\rm tot}$ was then varied until a reasonable fit to the line was 
obtained (see Fig.~\ref{figure:model}). We note that for the 
complex organics we have detected only one transition per species. 
We cannot thus apply the method of rotation diagram to derive rotational 
temperatures and molecular column densities (e.g., \cite{goldsmith1999}).

We also determined the HCO$^+$ column density from the co\-lumn density of 
H$^{13}$CO$^+$. For this calculation, it was assumed that the carbon-isotope 
ratio $[^{12}{\rm C}]/[^{13}{\rm C}]$ depends on 
$R_{\rm GC}$ accor\-ding to the relationship given by Wilson \& Rood (1994):

\begin{equation}
\label{eq:C-ratio}
\frac{[^{12}{\rm C}]}{[^{13}{\rm C}]}=7.5\times R_{\rm GC}[{\rm kpc}]+7.6 \,.
\end{equation}
For the $R_{\rm GC}$ values considered here, $5.1-6.3$ kpc, the above ratio 
lies in the range $\sim46-55$.\footnote{Spectral lines of 
$^{12}$C-isotopologue of HCO$^+$ are likely to be optically thick. Therefore, 
the HCO$^+$ deuteration can be better investigated through the 
DCO$^+$/H$^{13}$CO$^+$ column density ratio. However, a caveat should be noted 
here. The HCO$^+$ molecules are produced directly from CO (see reactions 3 and 
7 in Table~\ref{table:reactions}). On the other hand, at low temperature, CO is 
susceptible to the exothermic isotopic charge exchange reaction 
${\rm ^{13}C^+}+{\rm ^{12}CO}\rightarrow {\rm ^{12}C^+}+{\rm ^{13}CO}+\Delta E$, where $\Delta E/k_{\rm B}=35$ K (\cite{watson1976}). This is expected to 
cause considerable $^{13}$C-fractionation in cold and dense gas, which 
complicates the deuteration analysis.} 

For I18151 MM2, I18182 MM2, and I18223 MM3 we do not have N$_2$H$^+$ data. 
For these sources we computed the N$_2$H$^+$ column density from the $J=1-0$ 
line parameters ($T_{\rm ex}$ and $\tau$) from SSK08 [see, e.g., Eq.~(10) in 
Miettinen et al. (2010)]. We obtain about 1.1--1.2 times higher N$_2$H$^+$ 
column densities compared to SSK08 who used the optically thin approximation 
(see Sect.~5.1 for a further discussion).

We calculated the fractional abundances of the molecules by dividing the 
molecular column density by the H$_2$ column density: 
$x({\rm mol})=N({\rm mol})/N({\rm H_2})$. For this purpose, the values of 
$N({\rm H_2})$ were derived from the (sub)mm dust continuum maps smoothed to 
the corresponding resolution of the line observations. With a 
resolution of $31\arcsec$, the Bolocam 1.1-mm data could not be smoothed to 
correspond the resolution of the line observations. In these cases we used 
the original Bolocam data; the $31\arcsec$ resolution is in most 
cases comparable to that of the line observations ($22\farcs3-28\farcs9$), 
and thus we do not expect this to be a significant source of error. 
The derived column densities and abundances are listed in 
Tables~\ref{table:column} and \ref{table:otherabundances}. 
The abundance errors were derived by propagating the errors in $N({\rm mol})$ 
and $N({\rm H_2})$.

\begin{table*}
\caption{Molecular column densities and fractional abundances with respect to 
H$_2$.}
\begin{minipage}{2\columnwidth}
\centering
\renewcommand{\footnoterule}{}
\label{table:column}
\begin{tabular}{c c c c c c c}
\hline\hline 
Source & $N({\rm C^{17}O})$ & $N({\rm N_2H^+})$ & $N({\rm N_2D^+})$ & $N({\rm H^{13}CO^+})$ & $N({\rm HCO^+})$ & $N({\rm DCO^+})$\\
        & [$10^{15}$ cm$^{-2}$] & [$10^{13}$ cm$^{-2}$] & [$10^{11}$ cm$^{-2}$] & [$10^{13}$ cm$^{-2}$] & [$10^{14}$ cm$^{-2}$] & [$10^{11}$ cm$^{-2}$]\\
\hline
IRDC 18102-1800 MM1 & $1.4\pm0.1$ & $30.1\pm13.3$ & 4.9 & $1.9\pm1.0$ & $9.7\pm5.1$ & $1.9\pm0.9$\\
G015.05+00.07 MM1 & $1.2\pm0.1$ & $3.3\pm0.6$ & 9.2 & $0.1\pm0.03$ & $0.5\pm0.2$ & $3.0\pm0.6$\\
IRDC 18151-1208 MM2 & $2.1\pm0.1$ & $5.0\pm1.1$\tablefootmark{a} & 5.2 & $0.5\pm0.1$ & $2.6\pm0.5$ & $8.2\pm1.9$\\
G015.31-00.16 MM3 & $0.3\pm0.1$ & $3.8\pm0.6$ & \ldots & $0.3\pm0.04$ & $1.5\pm0.2$ & $11.0\pm2.1$\\
IRDC 18182-1433 MM2 & $0.7\pm0.1$ & $2.1\pm0.4$\tablefootmark{a} & \ldots & $0.7\pm0.1$ & $3.2\pm0.5$ & $45.4\pm5.4$\\
IRDC 18223-1243 MM3 & $1.2\pm0.1$ & $5.1\pm0.5$\tablefootmark{a} & 6.7 & $0.4\pm0.1$ & $1.9\pm0.5$ & $6.6\pm0.9$\\
ISOSS J18364-0221 SMM1 & $1.1\pm0.1$ & \ldots & \ldots & $0.6\pm0.2$ & $3.3\pm1.1$ & $39.2\pm13.8$\\
\hline
& $x({\rm C^{17}O})$ & $x({\rm N_2H^+})$ & $x({\rm N_2D^+})$ & $x({\rm H^{13}CO^+})$ & $x({\rm HCO^+})$ & $x({\rm DCO^+})$\\
& [$10^{-7}$] & [$10^{-9}$] & [$10^{-11}$] & [$10^{-10}$] & [$10^{-8}$] & [$10^{-11}$]\\
\hline 
IRDC 18102-1800 MM1 & $1.5\pm0.2$ & $27.7\pm12.6$ & $5.0\pm0.5$ & $18.2\pm9.8$ & $9.3\pm5.0$ & $2.0\pm1.0$\\
G015.05+00.07 MM1 & $1.2\pm0.2$ & $3.4\pm0.9$ & $9.5\pm1.7$ & $1.0\pm0.4$ & $0.5\pm0.2$ & $3.1\pm0.8$\\
IRDC 18151-1208 MM2 & $0.6\pm0.1$ & $0.9\pm0.2$ & $1.3\pm0.2$ & $1.2\pm0.3$ & $0.6\pm0.1$ & $2.3\pm0.6$\\
G015.31-00.16 MM3 & $0.4\pm0.2$ & $4.5\pm1.6$ & \ldots & $3.5\pm1.2$ & $1.8\pm0.6$ & $13.0\pm4.8$\\
IRDC 18182-1433 MM2 & $1.1\pm0.2$ & $2.5\pm0.6$ & \ldots & $9.9\pm2.1$ & $4.5\pm1.0$ & $72.8\pm14.6$\\
IRDC 18223-1243 MM3 & $0.9\pm0.1$ & $3.7\pm0.5$\tablefootmark{b} & $4.9\pm0.5$ & $2.9\pm0.8$ & $1.4\pm0.4$ & $4.8\pm0.8$\\
ISOSS J18364-0221 SMM1 & $1.6\pm0.1$ & \ldots & \ldots & $7.3\pm2.4$ & $4.0\pm1.3$ & $58.7\pm20.7$\\
\hline 
\end{tabular} 
\tablefoot{
\tablefoottext{a}{Calculated from the SSK08 parameters ($18\arcsec$ resolution).} \tablefoottext{b}{The corresponding H$_2$ column density was estimated from the Bolocam map of $31\arcsec$ resolution.}}
\end{minipage} 
\end{table*}

\begin{table}
\caption{Column densities and fractional abundances of the other 
candidate species observed.}
{\scriptsize 
\begin{minipage}{1\columnwidth}
\centering
\renewcommand{\footnoterule}{}
\label{table:otherabundances}
\begin{tabular}{c c c c}
\hline\hline 
Source & Molecule & $N_{\rm tot}$ [cm$^{-2}$] & $x \equiv N_{\rm tot}/N({\rm H_2})$\\  
\hline
IRDC 18102-1800 MM1 & \textit{o}-c-C$_3$H$_2$ & $1.7\pm0.8\times10^{13}$ & $1.8\pm0.8\times10^{-9}$\\
                    & CH$_3$NH$_2$\tablefootmark{a} & $5.0\times10^{14}$\tablefootmark{b} & $5.1\pm0.6\times10^{-8}$\\
                    & OCS\tablefootmark{a} & $2.3\pm0.8\times10^{15}$ & $2.4\pm0.9\times10^{-7}$\\
                    & $^{13}$CS & $6.0\pm2.9\times10^{13}$ & $6.2\pm3.1\times10^{-9}$\\
                    & SiO & $1.2\pm0.8\times10^{15}$ & $1.1\pm0.8\times10^{-7}$\\ 
IRDC 18151-1208 MM2 & \textit{o}-c-C$_3$H$_2$ & $7.0\pm1.7\times10^{12}$ & $1.9\pm0.5\times10^{-10}$\\
                    & C$_2$D\tablefootmark{c} & $1.5\times10^{13}$\tablefootmark{d} & $4.2\pm0.5\times10^{-10}$\\
                    & CH$_2$CHC$^{15}$N & $2.5\times10^{16}$$^b$ & $6.9\pm0.9\times10^{-7}$\\
                    & SiO & $1.1\pm0.3\times10^{15}$ & $2.5\pm0.8\times10^{-8}$\\
IRDC 18182-1433 MM2  & \textit{o}-c-C$_3$H$_2$ & $7.7\pm0.8\times10^{13}$ & $1.2\pm0.2\times10^{-8}$\\
                     & CH$_3$COCH$_3$ & $2.0\times10^{15}$\tablefootmark{b} & $4.1\pm0.6\times10^{-7}$\\
IRDC 18223-1243 MM3 & \textit{o}-c-C$_3$H$_2$ & $2.2\pm0.3\times10^{13}$ & $1.6\pm0.3\times10^{-9}$\\
                    & SiO & $4.0\pm0.6\times10^{14}$ & $2.9\pm0.5\times10^{-8}$\\
\hline 
\end{tabular} 
\tablefoot{
\tablefoottext{a}{The corresponding line is blended with OCS.} \tablefoottext{b}{Calculated by assuming that $T_{\rm ex}=2/3\times E_{\rm u}/k_{\rm B}$.} \tablefoottext{c}{The other C$_2$D hf group detected is possibly blended with CH$_2$CHC$^{15}$N.} \tablefoottext{d}{Calculated by assuming that $T_{\rm ex}=E_{\rm u}/k_{\rm B}$.}}
\end{minipage}  }
\end{table}

\subsection{CO depletion and deuterium fractionation}

To estimate the amount of CO depletion in the clumps, we 
calculated the CO depletion factor, $f_{\rm D}$, following the analysis 
presented in the paper by Fontani et al. (2006). If $x({\rm CO})_{\rm can}$ is 
the ``canonical'' (undepleted) abundance, and $x({\rm CO})_{\rm obs}$ is the 
observed CO abundance, $f_{\rm D}$ is given by 

\begin{equation}
\label{eq:depletion}
f_{\rm D}=\frac{x({\rm CO})_{\rm can}}{x({\rm CO})_{\rm obs}}\, .
\end{equation}
The ``canonical'' CO abundance at the galactocentric distance $R_{\rm GC}$ 
was calculated using the relationship [Eq.~(7) in \cite{fontani2006}]

\begin{equation}
x({\rm CO})_{\rm can}=9.5\times10^{-5}{\rm e}^{1.105-0.13R_{\rm GC}[{\rm kpc}]}\, .
\end{equation}
Note that this relationship results from using the value $R_0=8.5$ kpc, 
whereas our $R_{\rm GC}$ values are computed using $R_0=8.4$ kpc. Such a small 
discrepancy is however negligible. At $R_{\rm GC}=8.5$ kpc, the above 
relationship gives the standard value $9.5\times10^{-5}$ for the 
abundance of the main CO isotopologue in the solar neighbourhood 
(\cite{frerking1982}). To calculate the ``canoni\-cal'' C$^{17}$O abundance we 
take into account that the oxygen-isotopic ratio 
$[^{16}{\rm O}]/[^{18}{\rm O}]$ depends on 
$R_{\rm GC}$ according to the relationship (\cite{wilson1994})

\begin{equation}
\label{eq:O-ratio}
\frac{[^{16}{\rm O}]}{[^{18}{\rm O}]}=58.8\times R_{\rm GC}[{\rm kpc}]+37.1 \,.
\end{equation}
For the $R_{\rm GC}$ values of our clumps, the above ratio range from about 
337 to 407.5. When the above relationship is combined with the 
$[^{18}{\rm O}]/[^{17}{\rm O}]$ ratio, for which we use 
the standard value 3.52 (\cite{frerking1982}), the value of 
$x({\rm C^{17}O})_{\rm can}$ can be calculated as 

\begin{eqnarray}
x({\rm C^{17}O})_{\rm can} &=& \frac{x({\rm CO})_{\rm can}}{[^{18}{\rm O}]/[^{17}{\rm O}]\times [^{16}{\rm O}]/[^{18}{\rm O}]}\nonumber \\
   &=& \frac{x({\rm CO})_{\rm can}}{3.52\times(58.8\times R_{\rm GC}[{\rm kpc}]+37.1)} \, .
\end{eqnarray}
The depletion factor $f_{\rm D}$ is then calculated from
$f_{\rm D}=x({\rm C^{17}O})_{\rm can}/x({\rm C^{17}O})_{\rm obs}$, and the 
results, $f_{\rm D}=0.6\pm0.1-2.7\pm1.3$, are listed in Col.~(2) of 
Table~\ref{table:depletion}. The $\pm$-error quoted was calculated by 
propagating the uncertainty in $x({\rm C^{17}O})_{\rm obs}$. The low 
va\-lues of $f_{\rm D}$ indicate that CO is not significantly depleted, if at 
all, in our clumps.

The degree of deuterium fractionation in HCO$^+$ and N$_2$H$^+$ was calculated 
by dividing the column density of the deuterated isotopologue by its normal 
hydrogen-bearing form as 
$R_{\rm D}({\rm HCO^+})\equiv N({\rm DCO^+})/N({\rm HCO^+})$ and 
$R_{\rm D}({\rm N_2H^+})\equiv N({\rm N_2D^+})/N({\rm N_2H^+})$. The 
values of $R_{\rm D}({\rm HCO^+})$ and $R_{\rm D}({\rm N_2H^+})$ are in the range 
$0.0002\pm0.0001-0.014\pm0.003$ and $0.002\pm0.001-0.028\pm0.005$, 
respectively; the error in $R_{\rm D}$ was derived from the errors in the 
corresponding column densities [see Cols.~(3) and (4) of 
Table~\ref{table:depletion}].

\subsection{Ionisation degree}

We have derived fractional abundances of different ionic species and their 
different isotopologues. In principle, a lower limit 
to the ionisation degree can be obtained by simply summing up these 
abundances (e.g., \cite{casellietal2002}; \cite{miettinen2009}):

\begin{equation}
\label{eq:ion1}
x({\rm e})>x({\rm HCO^+})+x({\rm H^{13}CO^+})+x({\rm DCO^+})+x({\rm N_2H^+})+x({\rm N_2D^+}) \, .
\end{equation}
This is based on the gas quasi-neutrality: the electron abundance equals the 
difference between the total abundances of the cations and the anions.
The resulting values, $x({\rm e})>\sum_{i} x({\rm ions})_{i}$, are 
listed in Col.~(5) of Table~\ref{table:depletion}. These lower limits are in 
the range $7\times10^{-9}-1.2\times10^{-7}$.

The fractional ionisation in dense molecular clouds can also be inferred by 
employing ion-molecule chemical reaction schemes with the observed 
molecular abundances. All the reactions considered here, and notes on the 
associated rate coefficients, are given in Table~\ref{table:reactions}.

At first, we derive another estimate for a lower limit to $x({\rm e})$ through 
the ionisation balance determined by H$_3^+$, HCO$^+$, N$_2$H$^+$, and 
electrons following Qi et al. (2003). The H$_3^+$ and N$_2$H$^+$ are mainly 
destroyed by CO; at steady state, reactions~3, 11, and 
15 in Table~\ref{table:reactions} lead to the following equation for the lower 
limit to $x({\rm e})$ [cf. Eq.~(6) in Qi et al. (2003)]:

\begin{equation}
\label{eq:ion2}
x({\rm e})\geq \frac{k_{15}x({\rm N_2H^+})x({\rm CO})}{\beta_{11}x({\rm HCO^+})}\, .
\end{equation}
The rate coefficients $k_{15}$ and $\beta_{11}$ were adopted from the UMIST
database\footnote{{\tt http://www.udfa.net/}} (\cite{woodall2007}). The 
derived values are listed in  Col.~(6) of Table~\ref{table:depletion}. For 
I18151 MM2 and I18223 MM2 the $x({\rm e})$ values derived from 
Eq.~(\ref{eq:ion2}) are similar to those computed from Eq.~(\ref{eq:ion1}). 
For I18102 MM1, G015.31 MM3, and I18182 MM2 the summed abundance of different 
ionic species is clearly higher (by factors $\sim3.5-16.3$) than the lower 
limit to $x({\rm e})$ resulting from Eq.~(\ref{eq:ion2}). In the case of 
G015.05 MM1, on the other hand, the summed ionic abundance is about six times 
lower. These discrepancies are not surprising because the chemical scheme 
behind Eq.~(\ref{eq:ion2}) is certainly oversimplified.

Next, we determine the degree of ionisation by utilising the abundance ratios 
$R_{\rm D}({\rm HCO^+})\equiv [{\rm DCO^+}]/[{\rm HCO^+}]$ and 
$R_{\rm H}\equiv [{\rm HCO^+}]/[{\rm CO}]$. The first studies of
fractional ionisation based on the above ratios were 
carried out over three decades ago (e.g., \cite{guelin1977}; \cite{watson1978}; 
\cite{wootten1979}). A similar analysis was subsequently applied in the papers 
by Caselli et al. (1998), Williams et al. (1998), Bergin et al. (1999), 
Anderson et al. (1999), and Caselli (2002).

We note that the following analysis includes several assumptions: \textit{i}) 
HCO$^+$ is mainly produced in the reaction bet\-ween H$_3^+$ and CO (reaction 3 
in Table~\ref{table:reactions}); \textit{ii}) all deuteration is due 
to the reaction between H$_3^+$ and HD (reaction 2); 
\textit{iii}) the pre\-sence of atomic deuterium, which could 
(slightly) increase the deuteration degree, is ignored; 
\textit{iv}) ionic species are destroyed mainly by electrons, the most 
important neutrals (CO and O), and negatively charged dust grains; 
\textit{v}) except CO and O, we neglect the contribution of some other neutral 
species, such as N$_2$, O$_2$, and H$_2$O, in the destruction of H$_3^+$ and 
H$_2$D$^+$; and \textit{vi}) we ignore the effects of refractory metals 
(\cite{anderson1999}; \cite{casellietal2002}). 
Concerning part \textit{v}), the abundances of neutrals, such as 
N$_2$ and H$_2$O, are poorly known and/or low. For instance, the first 
results from \textit{Herschel} have shown that H$_2$O abundance is relatively 
low in high-mass star-forming regions ($\sim10^{-10}-10^{-8}$; 
\cite{vandertak2010}; \cite{marseille2010a}; \cite{chavarria2010}). 
Bergin et al. (1999) varied the nitrogen abundance in their chemical model 
and found that it does not affect the electron abundance. Also, the 
destructive reaction with O$_2$ is very slow ($k=9.3\times10^{-10}$ 
cm$^3$~s$^{-1}$; UMIST). 

By writing steady-state equations for the abundances of H$_2$D$^+$ 
(reactions 2 and 7--10), DCO$^+$ (reactions 7 and 13--14), and HCO$^+$ 
(reactions 3, 7, and 11--12), it can be shown that 

\begin{equation}
\label{eq:RD}
R_{\rm D}({\rm HCO^+})=\frac{1}{3}\frac{k_{+}x({\rm HD})}{k_7 x({\rm CO})+\beta_8 x({\rm e})+k_9 x({\rm O})+k_{10}x({\rm g})} \, .
\end{equation}
Solving $x({\rm e})$ from the above formula yields

\begin{equation}
\label{eq:ion3}
x({\rm e})=\frac{1}{\beta_8}\left[\frac{k_{+}x({\rm HD})}{3R_{\rm D}({\rm HCO^+})}-k_7 x({\rm CO})-k_9 x({\rm O})-k_{10}x({\rm g})\right] \, .
\end{equation}
This represents the upper limit to $x({\rm e})$.
We have assumed that $k_7=1/3\times k_3$, because H$_2$D$^+$ can transfer a 
proton to CO producing HCO$^+$; the rate for this is two times higher than for 
the channel producing DCO$^+$.
For the reaction rate of the deuteration reaction H$_3^+ +{\rm HD}$ ($k_{+}$) 
we used the recent results by Hugo et al. (2009; their Table~VIII). 
For example, at 10 K the deuteration of H$_3^+$ proceeds about 4.4 times 
faster accor\-ding to Hugo et al. (2009) than suggested by laboratory 
measurements of Gerlich et al. (2002). Such a high rate was also used in the 
studies cited above, i.e., $k_{+}=1.5\times10^{-9}$ cm$^3$~s$^{-1}$ 
(e.g., \cite{caselli1998}). We take the HD abundance to be twice the elemental 
D/H-ratio, i.e., $x({\rm HD})=2\times[{\rm D}]/[{\rm H}]\sim 3\times10^{-5}$ 
(e.g., \cite{linsky2006}; \cite{prodanovic2010}). Because CO is not found to 
be depleted in our clumps, we assume that this is also the case for atomic 
oxygen and use the ``standard'' abundance relative to H$_2$ of 
$x({\rm O})=3.52\times10^{-4}$, i.e., comparable to $x({\rm CO})$ 
(see \cite{caselli1998}; \cite{caselli2002}). 
For the grain abundance, $x({\rm g})$, we use the value $2.64\times10^{-12}$, 
which is based on the va\-lues $a=0.1$ $\mu$m, $\rho_{\rm grain}=3$ g~cm$^{-3}$, 
and $R_{\rm d}=1/100$ [see, e.g., Eq.~(15) in Pagani et al. (2009a)]. The 
derived upper limits to $x({\rm e})$ are shown in Col.~(7) of 
Table~\ref{table:depletion}. The values lie in the range 
$\sim2\times10^{-6}-\sim3\times10^{-4}$, i.e., much larger than the lower 
limits estimated above. 

By deriving a steady-state equation for the H$_3^+$ abundance (see reactions 
1--6 in Table~\ref{table:reactions}), and applying it in the correspon\-ding 
equation for HCO$^+$, it can be shown that

\begin{equation}
\label{eq:RH}
R_{\rm H}=\frac{[\zeta_{\rm H_2}/n({\rm H_2})]k_3}{\left[\beta_4 x({\rm e}) + k_3 x({\rm CO})+k_5 x({\rm O})+k_6 x({\rm g})\right]\left [\beta_{11} x({\rm e})+k_{12}x({\rm g}) \right]} \, .
\end{equation}

After H$_2^+$ has formed via cosmic-ray ionisation of H$_2$ (reaction 1), 
it quickly reacts with H$_2$ to form H$_3^+$ (\cite{solomon1971}). 
Thus the H$_3^+$ abundance is governed by the rate $\zeta_{\rm H_2}$. 
When the fractional ionisation in the source is determined, Eq.~(\ref{eq:RH}) 
can be used to infer the cosmic-ray ionisation rate of H$_2$. To calculate 
$\zeta_{\rm H_2}$, for each source we adopted as $x({\rm e})$ the summed 
abundance of ionic species. Using the derived upper limits to 
$x({\rm e})$ would result in unrealistically high values of $\zeta_{\rm H_2}$. 
This suggests that the upper $x({\rm e})$ limits are clearly 
higher than the true values. The obtained results are shown in the last 
column of Table~\ref{table:depletion}. In most cases, the 
$\zeta_{\rm H_2}$ values lie in the range 
$\sim1\times10^{-17}-5\times10^{-16}$ s$^{-1}$. However, towards I18102 MM1 we 
obtain a very high rate of $\sim1\times10^{-15}$ s$^{-1}$. Because 
I18102 MM1 is associated with a bright MIR point-source, and shows no signs of 
CO depletion, it is probably a rapidly evolving source. Therefore, a steady 
state assumption used in the above analysis may be invalid for I18102 MM1.

\begin{table}
\caption{Ion-molecule reactions included in the analysis of ionisation degree.}
{\scriptsize 
\begin{minipage}{1\columnwidth}
\centering
\renewcommand{\footnoterule}{}
\label{table:reactions}
\begin{tabular}{c c c}
\hline\hline 
No. & Reaction & Note on \\
    &          & rate coefficient\tablefootmark{a} 
 \\
\hline
1 & ${\rm H_2} + {\rm crp}\Arrow^{\zeta_{\rm H_2}}{\rm H_2^+}+{\rm e^-}$ & $\zeta_{\rm H_2}$ [s$^{-1}$]\\
  & ${\rm H_2^+}+{\rm H_2}\Arrow{\rm H_3^+} + {\rm H}$ &\\
2 & ${\rm H_3^+}+{\rm HD}\Harpoons^{k_+}_{k_{-}}{\rm H_2D^+}+{\rm H_2}$ & $k_+$ from Hugo et al. (2009)\\
3 & ${\rm H_3^+}+{\rm CO}\Arrow^{k_3}{\rm HCO^+}+{\rm H_2}$ & $k_3=1.7\times10^{-9}$ cm$^3$~s$^{-1}$\\
4 & ${\rm H_3^+} + {\rm e^-}\Arrow^{\beta_4}{\rm H_2+H}$ & $\beta_4$ from Pagani et al. (2009a)\\
  & \hspace{1.69cm} ${\rm H+H+H}$ & \\
5 & ${\rm H_3^+}+{\rm O}\Arrow^{k_5}{\rm H_2O^+}+{\rm H}$ & $k_5=1.2\times10^{-9}$ cm$^3$~s$^{-1}$\\
  & \hspace{1.36cm} ${\rm OH^++H_2}$ & \\
6\tablefootmark{b} & ${\rm H_3^+}+{\rm g^-}\Arrow^{k_6}{\rm g^0}+{\rm H_2}+{\rm H}$ & $k_6$ from Pagani et al. (2009a)\\
  & \hspace{1.6cm} ${\rm g^0+H+H+H}$& \\
7 & ${\rm H_2D^+}+{\rm CO}\Arrow^{k_7}{\rm HCO^+}+{\rm HD}$ ($\frac{2}{3}$) & $k_7=1/3 \times k_3$\\
  & \hspace{1.72cm} ${\rm DCO^+}+{\rm H_2}$ ($\frac{1}{3}$) &\\
8 & ${\rm H_2D^+} + {\rm e^-}\Arrow^{\beta_8}{\rm H+H+D}$ & $\beta_8$ from Pagani et al. (2009a)\\
  & \hspace{1.3cm} ${\rm HD+H}$ & \\  
  & \hspace{1.3cm} ${\rm H_2+D}$ & \\
9 & ${\rm H_2D^+} + {\rm O}\Arrow^{k_9}{\rm OH^++HD}$ & $k_9=k_5$\\
  & \hspace{1.45cm} ${\rm OD^+ +H_2}$ & \\
10 & ${\rm H_2D^+} + {\rm g^-}\Arrow^{k_{10}}{\rm g^0}+{\rm H_2}+{\rm D}$ & $k_{10}$ from Pagani et al. (2009a)\\
  & \hspace{1.72cm} ${\rm g^0+HD+H}$ & \\ 
  & \hspace{1.9cm} ${\rm g^0+D+H+H}$ & \\
11 & ${\rm HCO^+}+{\rm e^-}\Arrow^{\beta_{11}}{\rm CO}+{\rm H}$ & $\beta_{11}=2.4\times10^{-7}\left(\frac{T_{\rm kin}}{300\,{\rm K}}\right)^{-0.69}$\\
12 & ${\rm HCO^+}+{\rm g^-}\Arrow^{k_{12}}{\rm g^0+CO+H}$ & $k_{12}$ from Pagani et al. (2009a)\\
13 & ${\rm DCO^+}+{\rm e^-}\Arrow^{\beta_{13}}{\rm CO}+{\rm D}$ & $\beta_{13}=\beta_{11}$\\
14 & ${\rm DCO^+}+{\rm g^-}\Arrow^{k_{14}}{\rm g^0}+{\rm CO}+{\rm D}$ & $k_{14}=k_{12}$\\
15 & ${\rm N_2H^+}+{\rm CO}\Arrow^{k_{15}}{\rm HCO^+}+{\rm N_2}$ & $k_{15}=8.8\times10^{-10}$ cm$^3$~s$^{-1}$\\
\hline 
\end{tabular} 
\tablefoot{
\tablefoottext{a}{The rate coefficients are taken from the UMIST database 
unless otherwise stated. The temperature-dependent rates were 
calculated by using the $T_{\rm kin}$ values listed in Col.~(2) of 
Table~\ref{table:properties}.}\tablefoottext{b}{The label {\rm g} refers to 
the dust grains.}}
\end{minipage} }
\end{table}

\begin{table*}
\caption{Depletion-, deuteration-, and ionisation parameters of the clumps.}
\begin{minipage}{2\columnwidth}
\centering
\renewcommand{\footnoterule}{}
\label{table:depletion}
\begin{tabular}{c c c c c c c c}
\hline\hline 
Source & $f_{\rm D}$ & $R_{\rm D}({\rm HCO^+})$ & $R_{\rm D}({\rm N_2H^+})$ & $\sum_{i} x({\rm ions})_{i}$ & $x({\rm e})_l$ & $x({\rm e})_u$ & $\zeta_{\rm H_2}$ \\
       &            &              &         & [$10^{-8}$] & [$10^{-8}$] & [$10^{-5}$] & [$10^{-17}$ s$^{-1}$]\\
\hline
IRDC 18102-1800 MM1 & $0.7\pm0.1$ & $0.0002\pm0.0001$ & $0.002\pm0.001$ & $12.3\pm6.4$ & 3.5 & 29.2 & 115\\
G015.05+00.07 MM1 & $0.8\pm0.1$ & $0.006\pm0.003$ & $0.028\pm0.005$ & $0.9\pm0.3$ & 5.6 & 0.8 & 1.2\\
IRDC 18151-1208 MM2 & $1.6\pm0.3$ & $0.003\pm0.001$ & $0.010\pm0.002$\tablefootmark{a} & $0.7\pm0.1$ & 0.7 & 1.8 & 5.0\\
G015.31-00.16 MM3 & $2.7\pm1.3$ & $0.007\pm0.002$ & \ldots & $2.3\pm0.8$ & 0.6 & 0.7 & 5.5\\
IRDC 18182-1433 MM2 & $1.1\pm0.2$ & $0.014\pm0.003$ & \ldots & $4.9\pm1.1$ & 0.3 & 0.2 & 19.7\\
IRDC 18223-1243 MM3 & $1.3\pm0.1$ & $0.003\pm0.001$ & $0.013\pm0.001^a$ & $1.8\pm0.5$ & 1.6 & 1.8 & 13.1\\
ISOSS J18364-0221 SMM1 & $0.6\pm0.1$ & $0.012\pm0.006$ & \ldots & $4.1\pm1.3$ & \ldots & 0.3 & 51.1\\
\hline 
\end{tabular} 
\tablefoot{
\tablefoottext{a}{Calculated by utilising $N({\rm N_2H^+})$ from SSK08 (see Sect.~4.2 and 5.1).}}
\end{minipage}
\end{table*}

\section{Discussion}

In the following subsections, we discuss our results, and compare them with 
the results from previous studies.

\subsection{Molecular column densities and abundances}

Some of the molecular column densities and fractional abundances derived 
here have been determined in previous studies. Assuming LTE conditions 
and optically thin emission, SSK08 estimated N$_2$H$^+$ column 
densities from the $J=1-0$ transition for I18102 MM1, G015.05 MM1, and 
G015.31 MM3 which are clearly lower than our values (by factors $\sim2-6$).
SSK08 assumed that $T_{\rm ex}({\rm N_2H^+})=T_{\rm rot}({\rm NH_3})$ even though 
the $T_{\rm ex}({\rm N_2H^+})$ values are considerably lower than 
$T_{\rm rot}({\rm NH_3})$ (their Table~4). 
On the other hand, $T_{\rm ex}[{\rm N_2H^+}(1-0)]$ from SSK08 are very 
similar to those from our {\tt RADEX} simulations for the $J=3-2$ transition. 
Also, the line optical thicknesses SSK08 list in their Table~4 indicate 
optically thick emission ($\tau \sim2-11$). This can explain the 
discrepancy in the derived column densities.

Marseille et al. (2008) modelled the N$_2$H$^+$, HCO$^+$, and 
H$^{13}$CO$^+$ emission of I18151 MM2, and obtained abundances which are 
comparable within the errors to those we have derived. This strengthens the 
reliability of our HCO$^+$ abundances derived from H$^{13}$CO$^+$ by utilising 
the [$^{12}$C]/[$^{13}$C] abundance ratio.

Sakai et al. (2010; hereafter SSH10) derived H$^{13}$CO$^+$ co\-lumn densities 
from the $J=1-0$ transition for I18102 MM1, I18151 MM2, 
and I18223 MM3. Their results are otherwise similar to ours except that for 
I18102 MM1 the column densities differ by a factor of $\sim4$. SSH10 assumed 
optically thin emission and that 
$T_{\rm ex}({\rm H^{13}CO^+})=T_{\rm rot}({\rm CH_3OH})\pm5$ 
K$=16.7\pm5$ K$\sim3.6\times T_{\rm ex}[{\rm H^{13}CO^+}(3-2)]$. 
SSH10 also derived SiO column densities from the $J=2-1$ transition 
for the above three sources. We have obtained 36--86 times higher 
column densities for these sources from the $J=6-5$ transition 
through {\tt RADEX} calculations. The spatial resolution of SiO 
observations by SSH10 was 18\arcsec, and as $T_{\rm ex}({\rm SiO})$ 
they used $T_{\rm rot}({\rm CH_3OH})\pm5$ K, or assumed that 
$T_{\rm ex}({\rm SiO})=20$ K, which are over four times higher than 
$T_{\rm ex}[{\rm SiO}(6-5)]$ derived here. For comparison, we performed the 
LTE modelling of SiO$(6-5)$ lines in Weeds, and found that column densities 
comparable to those obtained from {\tt RADEX} are needed to explain the 
observed line intensities. Of course, similarly to SSH10, our 
{\tt RADEX} and Weeds analyses of a single SiO transition include several 
(uncertain) assumptions. The large discrepancy in the derived $N({\rm SiO})$ 
values could also be due to the fact that the $J=2-1$ and $J=6-5$ lines, which 
have completely different upper-state energies (6.3 K and 43.8 K, 
res\-pectively), originate in different parts within a clump.

\begin{figure*}
\begin{center}
\includegraphics[width=0.98\textwidth]{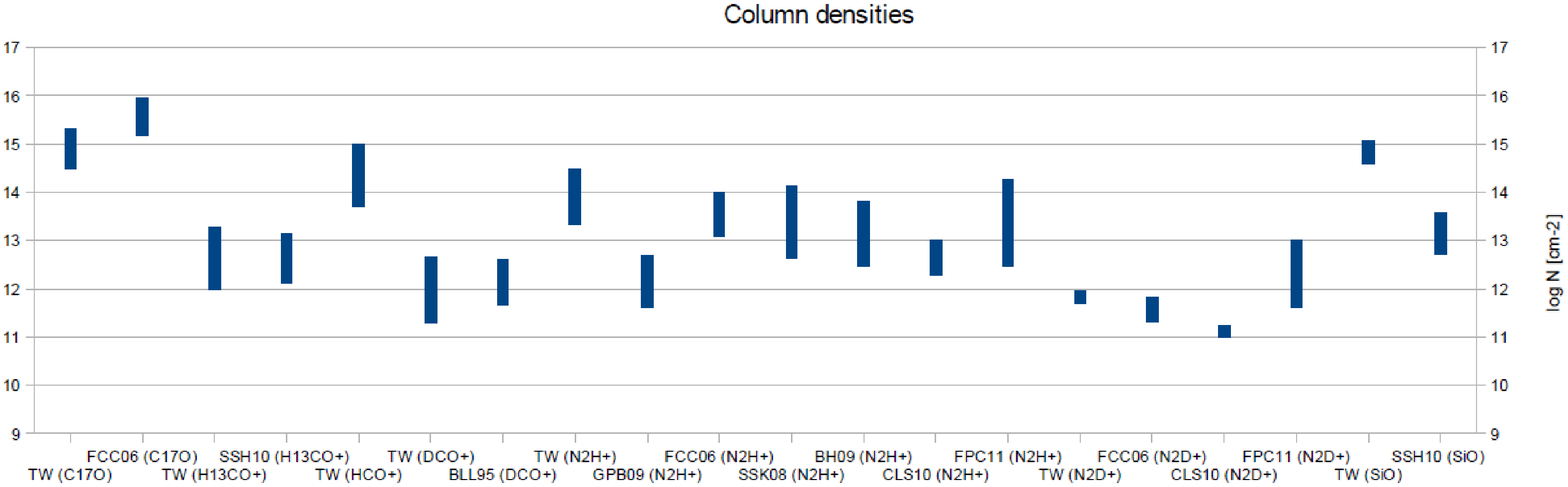}
\includegraphics[width=0.98\textwidth]{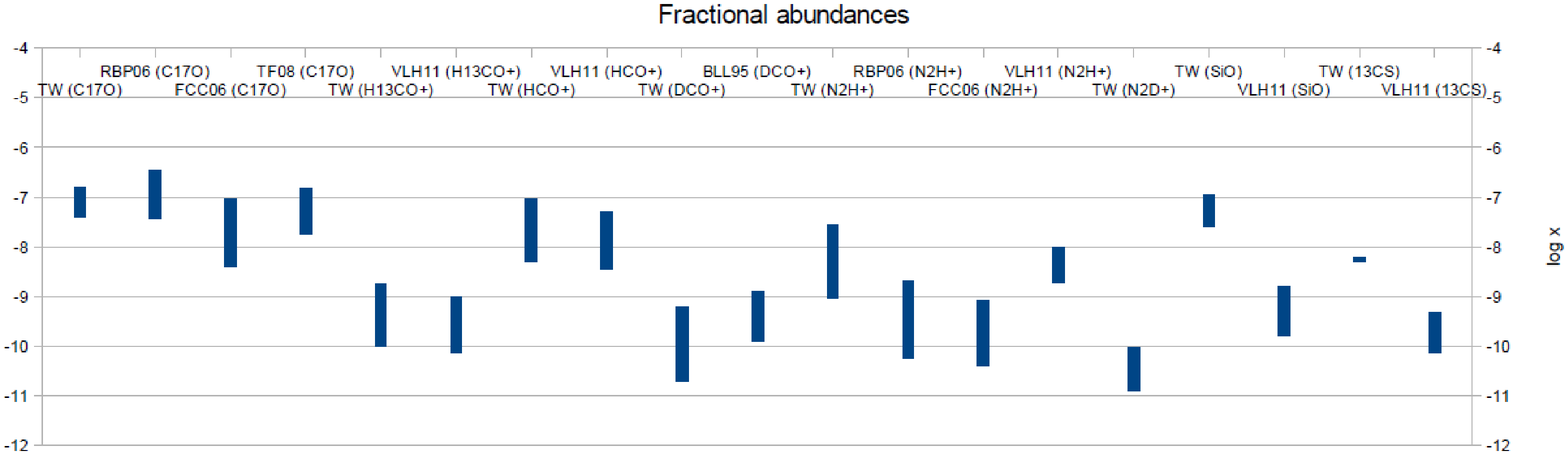}
\caption{The range of molecular column densities 
(\textit{upper panel}) and fractional abundances (\textit{lower panel}) 
for our sources as compared with similar samples of other studies. 
The y-axes are shown on a logarithmic scale. The 
abbreviation TW refers to 'this work'. The references are coded as follows: 
FCC06$=$Fontani et al. (2006), SSH10$=$Sakai et al. (2010), BLL95$=$Butner et 
al. (1995), GPB09$=$Gibson et al. 2009, SSK08$=$Sakai et al. (2008), 
BH09$=$Beuther \& Henning (2009), CLS10$=$Chen et al. (2010), 
FPC11$=$Fontani et al. (2011), RBP06$=$Ragan et al. (2006), 
TF08$=$Thomas \& Fuller (2008), VLH11$=$Vasyunina et al. (2011). 
The $^{13}$CS abundance of $\sim6.2\times10^{-9}$ we derived towards I18102 
MM1 is shown by a slightly stretch stock for a better illustration.}
\label{figure:CD}
\end{center}
\end{figure*}

In Fig.~\ref{figure:CD}, we show stock charts of the derived molecular 
column densities and fractional abundances. For compa\-rison, we plot column 
densities and abundances from several previous studies. The quoted reference 
studies deal with IRDCs (\cite{ragan2006}; SSK08; \cite{gibson2009}; 
\cite{beuther2009}; \cite{chen2010}; SSH10; \cite{vasyunina2011}), 
high-mass young stellar objects (HMYSOs; \cite{fontani2006}; 
\cite{thomas2008}), and low-mass starless and protostellar cores 
(\cite{butner1995}). The latter study was included because we found no other 
reported DCO$^+$ surveys towards IRDCs or HMYSOs. We also plot values from 
Fontani et al. (2011) who studied a sample of 27 high-mass cores (including 
starless cores, HMYSOs, and UC H{\scriptsize II} regions). The C$^{18}$O 
abundances from Ragan et al. (2006) were converted into $x({\rm C^{17}O})$ by 
utilising the abundance ratio [$^{18}$O]/[$^{17}$O]$=3.52$.

As can be seen in Fig.~\ref{figure:CD}, the column densities and 
abundances we have derived are generally comparable to those observed in other 
sources. The most evident difference is the high SiO and $^{13}$CS column 
densities and abundances we have obtained compared to the IRDCs studied 
by SSH10 and Vasyunina et al. (2011).

SSH10 found that, in general, the $N({\rm SiO})/N({\rm H^{13}CO^+})$ 
ratio is higher for the MSX dark sources than for the sources associated with 
MIR emission, and suggested that this ratio represents the fraction of the 
shocked gas. For the sources I18102 MM1, I18151 MM2, and I18223 MM3, SSH10 
found the ratios $N({\rm SiO})/N({\rm H^{13}CO^+})=2.9_{-1.2}^{+1.9}$, 
$2.4_{-0.9}^{+1.5}$, and $1.6_{-0.6}^{+1.0}$, respectively. Using our column 
densities these ratios become very different: $57.1\pm46.8$, $220\pm74.4$, and 
$80\pm20$, respectively. The latter values are in better agreement 
with the general trend found by SSH10 for a sample of 20 sources. 
SSH10 suggested that the SiO emission 
from the MIR dark objects originates in newly formed shocks, whereas the 
SiO emission from more evolved, MIR bright objects could originate in gas 
shocked earlier in time. This is supported by the fact that we see the 
narrowest SiO line towards I18102 MM1, pre\-sumably the most evolved source 
in our sample with detected SiO emission (see Sect.~B.1 of Appendix 
B). Follow-up studies of larger source samples are needed to test 
the statistical significance of the hypothesis by SSH10.
 
\subsection{Depletion and deuteration}

The CO depletion factors derived in this work are small, only in the range 
$\sim0.6-2.7$. Thus, CO molecules do not appear to be significantly 
depleted, if at all, in the studied clumps. This conforms to the fact that 
the clumps show signs of star formation activity, such as outflows, which 
presumably release CO from the dust grains back into the gas phase 
(Appendix B). The presence of outflow-shocked gas is revealed by the 
SiO line wings (particularly towards I18151 MM2) and very high SiO abundances 
found in the present work. Also, the temperature in the envelopes of I18102 
MM1 and I18151 MM2 is, likely due to heating by embedded YSOs, slightly 
higher than the CO sublimation temperature of 20 K (e.g., \cite{aikawa2008}). 
Moreover, relatively high cosmic-ray ionisation rates towards the clumps were 
found. Thus, the cosmic-ray impulsive hea\-ting could also play a role in 
returning CO into the gas phase (\cite{hasegawa1993}).

However, the derived CO depletion factors represent 
only the average values along the line of sight within the 27\farcs8 beam. 
Because the studied clumps lie at kiloparsec distances from us, the 
observations could be ``contaminated'' by non-depleted gas along the line of 
sight (\cite{fontani2006}). It is also uncertain whether the ``canonical'' CO 
abundance used to calculate $f_{\rm D}$ is exactly correct. 
Its value is known to vary by a factor of $\sim2$ bet\-ween different 
star-forming regions (\cite{lacy1994}). Finally, the H$_2$ column densities 
derived from dust emission suffer from the uncertainty of dust opacity, which 
may be a factor of 2--3. These uncertainties are likely to explain the unsual 
values of $f_{\rm D}<1$.

For comparison, Fontani et al. (2006) determined the amount of 
CO depletion in HMYSOs. They found CO depletion 
factors in the range 0.4--35.8 with a median value of 3.2.
Thomas \& Fuller (2008) found depletion factors $\lesssim10$ for their sample 
of HMYSOs and deduced that the sources are young, a few times $10^5$ 
yr. 

At high angular resolution, it is possible to identify more CO depleted 
regions within the clumps. Zhang et al. (2009) studied two massive clumps in 
the IRDC G28.34+0.06 at $\sim1\arcsec$ resolution with the SMA. They found CO 
depletion factors up to $\sim10^2-10^3$ towards the two clumps, where 
the largest values were found in the cores within one of the clumps. Similarly, 
with the $\sim3\arcsec$ resolution PdBI observations, Beuther \& Henning 
(2009) found a large CS depletion factor of $\sim100$ in IRDC 19175-4.

\begin{figure}[!h]
\begin{center}
\resizebox{0.8\hsize}{!}{\includegraphics{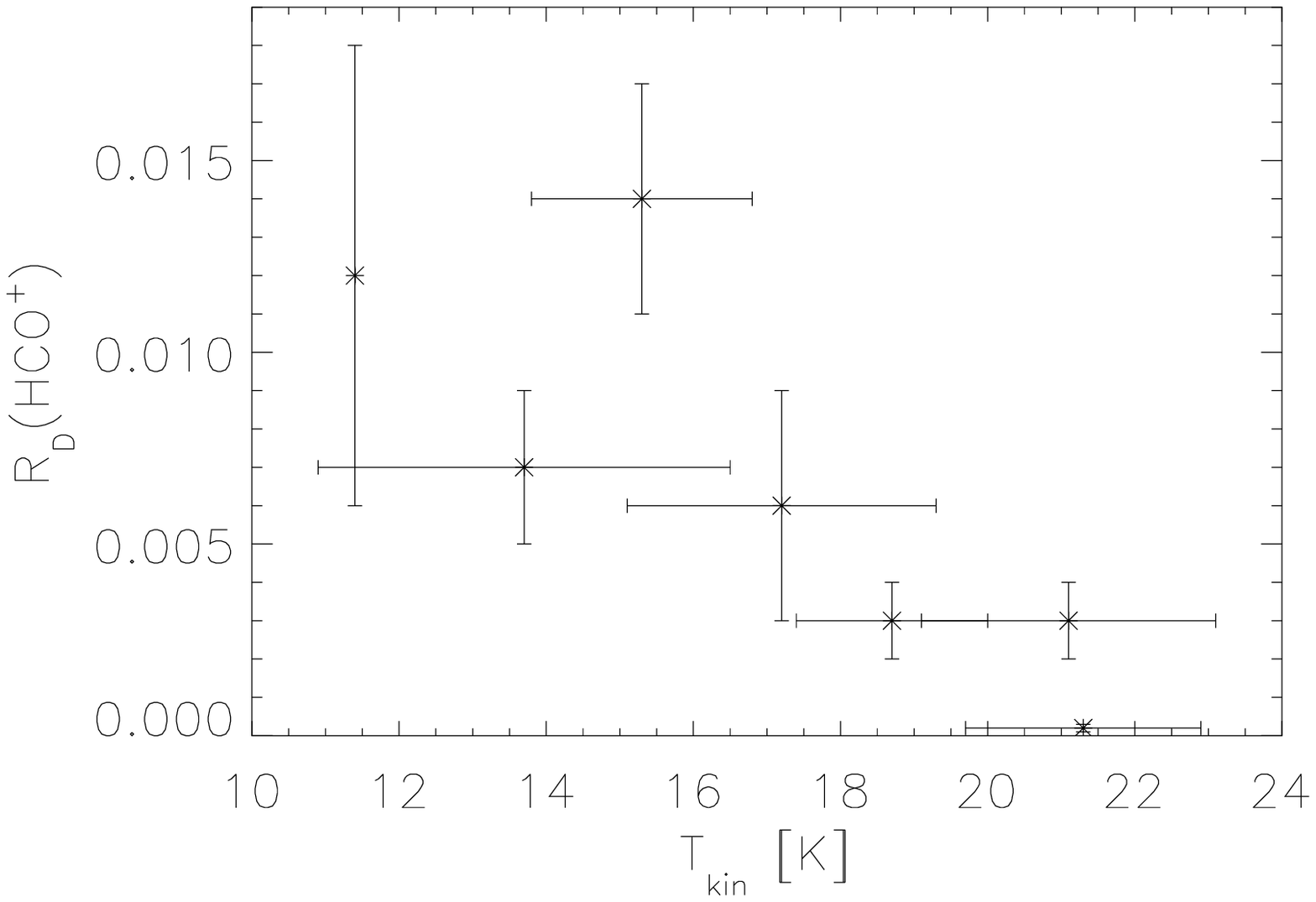}}
\end{center}
\caption{$N({\rm DCO^+})/N({\rm HCO^+})$ ratio versus gas kinetic temperature. 
As $T_{\rm kin}$ increases, the degree of deuterium fractionation decreases.}
\label{figure:correlation}
\end{figure}

The degree of deuteration in our clumps are $\sim0.0002-0.014$ in HCO$^+$ 
and $\sim0.002-0.028$ in N$_2$H$^+$. Compared to the ave\-rage cosmic D/H 
ratio of $\sim1.5\times10^{-5}$ (Sect.~4.4), the derived values 
are $\sim10^1-10^3$ times larger. Relatively low deuteration degrees in some 
of our sources are consistent with the oberved low CO depletion 
factors. CO is the main destroyer of H$_3^+$ and H$_2$D$^+$, and thus its 
presence can lower the deuterium fractio\-nation (e.g., \cite{caselli2008}). 
For example, Swift (2009) estimated the fractional abundance of 
\textit{ortho}-H$_2$D$^+$ in two IRDCs of only $\sim3-5\times10^{-13}$, 
which is up to three orders of magnitude lower than found in low-mass cores 
(e.g., \cite{caselli2008}; \cite{friesen2010}).
From a theoretical point of view, if there are no depletion gradients in the 
source, one would expect to see the equality 
$R_{\rm D}({\rm HCO^+})=R_{\rm D}({\rm N_2H^+})$ (\cite{rodgers2001b}). 
The fact that we found somewhat lower values for $R_{\rm D}({\rm HCO^+})$ than 
$R_{\rm D}({\rm N_2H^+})$ could be related to differential depletion of 
molecules, and radial density gradients (\cite{casellietal2002}). 
For high-mass star-forming clumps, the density profile is found to be of the 
form $n(r)\propto r^{-(1.6\pm0.5)}$ (e.g., \cite{beuther2002a}). 
CO is the parent species of HCO$^+$, but it destroys the N$_2$H$^+$ molecules. 
Deuteration proceeds most efficiently in regions where CO is mostly frozen 
onto dust grains. In the warmer envelope layers where CO is not depleted, 
HCO$^+$ can have a relatively high abundance, i.e., a lower value of 
$R_{\rm D}({\rm HCO^+})$ compared to that of N$_2$H$^+$ 
(\cite{emprechtinger2009}). In general, the amount of molecular deuteration 
is expected to decrease during protostellar evolution because of internal 
heating of the surrounding envelope (e.g., \cite{emprechtinger2009}). 
Indeed, there is a decreasing trend in $R_{\rm D}({\rm HCO^+})$ with 
the gas kinetic temperature as shown in Fig.~\ref{figure:correlation}. 
When the temperature becomes $\gtrsim20$ K, reaction 2 in 
Table~\ref{table:reactions} can proceed in both directions, and thus the 
destruction rate of H$_2$D$^+$ increases. Consequently, other deuterated 
molecules that form via reactions with H$_2$D$^+$ start to decrease in 
abundance. 

For comparison, Fontani et al. (2006) determined $R_{\rm D}({\rm N_2H^+})$ in 
HMYSOs and found the values in the range $\lesssim0.004-0.02$, with 
an average value of $\sim0.015$. Roberts \& Millar (2007) derived the upper 
limits to $R_{\rm D}({\rm N_2H^+})$ of $<0.002$ and 
$<0.006$ for the hot cores G34.26 and G75.78, respectively. 
Our source I18102 MM1 appears similar in this regard. Chen et al. (2010) 
found that $R_{\rm D}({\rm N_2H^+})=0.017-0.052$ in three cores within the 
IRDC G28.34+0.06. Moreover, they found that $R_{\rm D}({\rm N_2H^+})$ is lower 
in the more evolved stages of protostellar evolution. Recently, Fontani et al. 
(2011) found the values of $R_{\rm D}({\rm N_2H^+})=0.012-0.7$, 
0.017--$\leq0.4$, and 0.017--0.08 for their sample of high-mass starless cores, 
HMYSOs, and UC H{\scriptsize II} regions, respectively. The above results are 
quite similar to our findings. Another study of deuteration in IRDCs is that 
by Pillai et al. (2007) who derived NH$_2$D/NH$_3$ column density ratios in 
the range 0.005--0.386 for their sample of IRDC clumps. More recently, 
Pillai et al. (2011) found $\left[{\rm NH_2D}/{\rm NH_3}\right]$ ratios in the 
range 0.06--0.37 towards IRDCs. The $\left[{\rm NH_2D}/{\rm NH_3}\right]$ 
ratios obtained by Pillai et al. are even higher than those observed in 
low-mass dense cores (see \cite{pillai2011} and references therein).

Again, high resolution interferometric observations could reveal substructures 
within the clumps with higher degree of deuteration. For example, Fontani et 
al. (2006) found an ave\-rage value $R_{\rm D}({\rm N_2H^+})\sim0.01$ towards 
the high-mass star-forming region IRAS 05345+3157, but at high angular 
resolution ($\sim3\arcsec$) they resolved it into two N$_2$D$^+$ condensations 
each with ten times higher deuteration degree [$R_{\rm D}({\rm N_2H^+})=0.11$; 
\cite{fontani2008}]. Such high values are comparable to those found in low-mass
starless cores [\cite{crapsi2005}; $R_{\rm D}({\rm N_2H^+})<0.02-0.44$] and 
Class 0 protostellar envelopes [\cite{emprechtinger2009}; 
$R_{\rm D}({\rm N_2H^+})<0.029-0.271$]. For their sample of low-mass dense 
cores, Caselli et al. (1998) found $R_{\rm D}({\rm HCO^+})$ to lie in the 
range 0.025--0.07. For a sample of more massive cores, Bergin et al. (1999) 
found that most sources have values of $R_{\rm D}({\rm HCO^+})<0.035$. This 
trend conforms to our results towards more massive clumps.
It is possible that the cold phase during which the 
deuterium fractionation takes place is so short for HMYSOs that very 
high deuteration degrees are not reached (\cite{rodgers2007}). On the 
other hand, the high $\left[{\rm NH_2D}/{\rm NH_3}\right]$ ratios found by 
Pillai et al. (2007, 2011) could indicate a different production mechanism of 
deuterated ammonia compared to those of N$_2$H$^+$ and HCO$^+$ which are both 
purely produced in the gas phase.

One caveat to the deuteration analysis is that the FWHM linewidths of 
the detected N$_2$H$^+$ transitions are significantly larger than those of 
N$_2$D$^+$ [Table~\ref{table:lineparameters}, Col.~(4)]. Similarly, the 
linewidths of H$^{13}$CO$^+(3-2)$ and DCO$^+(3-2)$ are quite different compared 
to each other towards some of the sources. The above transition pairs have 
similar critical densities, and the lines were observed at comparable spatial 
resolutions. Therefore, the difference between the linewidths suggests that 
the corresponding transitions may not be tracing the same gas component. 
In this picture, the calculated deuteration degrees should only be interpreted 
as an average values along the line of sight.

\subsection{Ionisation}

We have calculated the average ionisation degree along the line of sight 
towards the clumps by utilising three different me\-thods. The summed 
abundances of the ionic species yield lower limits to $x({\rm e})$ in the range 
$\sim0.7-12.3\times10^{-8}$. A very simple chemical scheme including HCO$^+$ 
and N$_2$H$^+$ yielded another estimates for the lower li\-mits, 
about $0.3-5.6\times10^{-8}$. Thirdly, we used the 
$R_{\rm D}({\rm HCO^+})$ values to derive an upper limit to fractional 
ioni\-sation, and found the range $x({\rm e})=0.2-29.2\times10^{-5}$. 
To our knowledge, these are the first reported estimates of the 
ionisation degree in IRDCs made so far. 
As discussed below, the derived lower limits are likely to be more realistic 
estimates of $x({\rm e})$, whereas the upper limits appear to be very high. 
The $x({\rm e})$ values should be taken as rough estimates because of the 
simplicity of the analytical model we have utilised. For example, our analysis 
is based on chemical equilibrium which may not be valid 
(\cite{caselli1998}; \cite{lintott2006}). Other factors that cause 
uncertainties are the possible variation of CO depletion and fractional 
molecular abundances along the line of sight. 
Electron abundance depends strongly on the depletion 
of CO because CO destroys ionic species such as H$_3^+$ and N$_2$H$^+$. 
In this regard, one could think that the lower limit to $x({\rm e})$ derived 
by summing up the abundances of ionic species is the most reliable one. 
However, there is also a possibility that the line emission from 
different molecules originate in different layers of the clump. 
Indeed, the different linewidths between N$_2$H$^+(3-2)$ and N$_2$D$^+(3-2)$ 
point towards such a possibility (Sect.~5.2). In this case, the physical 
meaning of the summed abundance of ionic species is unclear. 
Furthermore, the fact that we have no information about the abundances of some 
important molecular ions, such as H$_3^+$ and H$_3$O$^+$, nor the metal ions, 
hampers the determination of $x({\rm e})$. Metal ions play an important role 
in the ionisation balance in the outer layers of dense clouds 
(\cite{caselli2002}). The fact that the sources show low degrees of 
deuteration conforms to the fairly high ionisation levels of gas. However, 
detailed chemical mo\-delling is needed to perform a more accurate analysis 
of the ionisation degree.

The cosmic-ray ionisation rates of H$_2$ we found are in the range 
$\zeta_{\rm H_2}\sim1\times10^{-17}-1\times10^{-15}$ s$^{-1}$. 
Six of the seven sources have estimated values between 
$\sim1\times10^{-17}$ and $\sim5\times10^{-16}$ s$^{-1}$, but the highest rate 
of $\sim10^{-15}$ s$^{-1}$ is seen towards I18102 MM1. 
It should be emphasised that the $\zeta_{\rm H_2}$ values were 
derived from the lower limits to $x({\rm e})$. However, as discussed above, 
we believe that these estimates are more accurate than those 
uti\-lising the upper limits to $x({\rm e})$. Also, the 
applicability of the volume-averaged H$_2$ number densities used to compute 
$\zeta_{\rm H_2}$ is uncertain because all the detected molecular-ion 
transitions have a much higher critical density (\cite{hezareh2008}). 
The standard relation between the electron abundance and the H$_2$ number 
density is $x({\rm e})\sim1.3\times10^{-5}n({\rm H_2})^{-1/2}$ 
(\cite{mckee1989}). This is based on the pure cosmic-ray ionisation with the 
rate $1.3\times10^{-17}$ s$^{-1}$ and includes no depletion of heavy elements. 
By utilising the densities derived in the present work, $1.1-18.5\times10^4$ 
cm$^{-3}$, the standard relation yields the values 
$x({\rm e})\sim3-12\times10^{-8}$. These are comparable to the estimated lower 
limits to $x({\rm e})$. 

How do our results compare to those found in other sources ? 
Caselli et al. (1998) determined $x({\rm e})$ in a sample of 24 low-mass 
cores consisting of both starless and protostellar objects. 
Their analysis was based on observations of CO, HCO$^+$, and DCO$^+$, and the 
resulting values were in the range 
$10^{-8}$ to $10^{-6}$. They argued that the variation in 
$x({\rm e})$ among the sources is due to variations in metal abundance and 
$\zeta_{\rm H_2}$. Williams et al. (1998) used observations of C$^{18}$O, 
H$^{13}$CO$^+$, and DCO$^+$ to determine the values 
$10^{-7.5}\lesssim x({\rm e}) \lesssim 10^{-6.5}$ in a similar sample of 
low-mass cores as Caselli et al. (1998), but they used a slightly different 
analysis. Applying the same analysis as Williams et al. (1998), 
Bergin et al. (1999) found the ionisation levels of 
$10^{-7.3}\lesssim x({\rm e})\lesssim 10^{-6.9}$ towards more massive cores 
(in Orion) than to those studied by Williams et al.
The most massive sources were found to have the lowest electron 
abundances, $x({\rm e})<10^{-8}$. This conforms to the results by de 
Boisanger et al. (1996) who found that $x({\rm e})\sim10^{-8}$ in the 
massive star-forming regions NGC 2264 IRS1 and W3 IRS5. Hezareh et al. (2008) 
studied the high-mass star-forming region DR21(OH) in Cygnus X, and found 
that $x({\rm e})=3.2\times10^{-8}$. These are consistent with our lower limits 
to $x({\rm e})$ in massive clumps within IRDCs. The very high upper limits 
to $x({\rm e})$ we have obtained, particularly that in I18102 MM1 
($\sim10^{-4}$), resemble those found in PDRs, where carbon can provide most 
of the charge and $x({\rm e})\simeq x({\rm C^+})$ (\cite{goicoechea2009}).

Observational results presented in the literature suggest re\-latively large 
environmental variations in $\zeta_{\rm H_2}$\footnote{In the following, we 
have used the relation $2.3\zeta_{\rm H}=1.5\zeta_{\rm H_2}$ to convert the 
ionisation rate per H atom, $\zeta_{\rm H}$, into $\zeta_{\rm H_2}$.}. 
van der Tak \& van Dishoeck (2000) used H$_3^+$ and H$^{13}$CO$^+$ observations 
to constrain the ionisation rate towards HMYSO envelopes, and obtained 
the best-fit value $\zeta_{\rm H_2}=(2.6\pm1.8)\times10^{-17}$ s$^{-1}$. 
We found comparable rates towards three of our sources.
Caselli et al. (1998) inferred the values of $\zeta_{\rm H_2}$ spanning a 
range of two orders of magnitude bet\-ween $10^{-18}-10^{-16}$ s$^{-1}$ in 
low-mass cores. Some of this variation could be due to different cosmic-ray 
flux in the source regions. Wlliams et al. (1998) deduced 
a mean value of $\zeta_{\rm H_2}=5\times10^{-17}$ s$^{-1}$. On the other 
hand, H$_3^+$ observations towards the diffure ISM suggest higher values in 
the range $\zeta_{\rm H_2}\sim0.5-1.2\times10^{-15}$ s$^{-1}$ 
(\cite{mccall2003}; \cite{indriolo2007}). Also, towards the 
Galactic Centre clouds, the ionisation rate is found to be very high, 
$\zeta_{\rm H_2}>10^{-15}$ s$^{-1}$ up to $\sim10^{-13}$ s$^{-1}$ (e.g., 
\cite{oka2005}; \cite{yusef2007}; \cite{goto2008}). However, such high rates 
can explain the high temperatures of $\sim70-100$ K in the Galactic Centre 
clouds, whereas the dense YSO envelopes are typically much colder.
van der Tak et al. (2006) estimated that 
$\zeta_{\rm H_2}\sim4\times10^{-16}$ s$^{-1}$ in the Sgr B2 envelope, 
whereas Hezareh et al. (2008) found a lower value of 
$\zeta_{\rm H_2}=3.1\times10^{-18}$ s$^{-1}$ towards DR21(OH). Observations 
of the Horsehead Nebula by Goicoechea et al. (2009) could only be reproduced 
with $\zeta_{\rm H_2}=7.7\pm4.6\times10^{-17}$ s$^{-1}$. Despite all 
the uncertainty factors in $\zeta_{\rm H_2}$, it seems possible that some 
of our clumps could be exposed to somewhat higher cosmic-ray flux, but 
the observed clump temperatures, $\simeq20$ K, suggest that 
$\zeta_{\rm H_2}$ is $\sim10^{-17}-10^{-16}$ s$^{-1}$ (see \cite{bergin1999}). 

Because (most of) our sources are associated with high-mass star 
formation, which could increase the local UV radiation field, the role played 
by photoionisation should also be considered. 
Photoionisation is expected to become negligible when the extinction 
is $A_{\rm V} \gtrsim 4$ mag (\cite{mckee1989}). In units of $N({\rm H_2})$, 
this corresponds to about $\gtrsim 4.9\times10^{21}$ cm$^{-2}$ 
(\cite{vuong2003}). Therefore, the ionisation in the studied clumps is very 
likely dominated by cosmic ray particles and the ambient UV radiation 
field is not important in this regard. If the clumps are inhomogeneous, 
however, external UV radiation could penetrate more efficiently into them 
and enhance the ionisation level (\cite{boisse1990}). 
Our sources also show high HCO$^+$ 
abundances ($\sim5\times10^{-9}-9\times10^{-8}$), whereas in the presence of 
signifi\-cant UV radiation one would expect to observe lower values of 
$x({\rm HCO^+})$ due to photoproduced electrons raising the HCO$^+$ 
recombination rate (\cite{jansen1995}). On the other hand, the HCO$^+$ 
abundance could be enhanced by the presence of protostellar outflows where 
shock-generated UV radiation is present (\cite{rawlings2000}). This could 
explain the high HCO$^+$ abundance in I18102 MM1 (Sect.~B.1).

\subsection{Coupling between the ions and neutrals}

The knowledge of the gas ionisation degree enables to estimate 
the level of coupling between the neutral gas component and the pervasive 
magnetic field. The coupling strength can be quantified through the 
wave-coupling number, which is defined as the ratio of the clump radius 
to the MHD cutoff wavelength, $W\equiv R/\lambda_{\min}$ (\cite{myers1995}). The 
wavelength $\lambda_{\min}$ represents the minimum wavelength for propagation 
of MHD waves, and is given by 
$\lambda_{\min} \equiv \pi {\rm v}_{\rm A}\tau_{\rm in}$, where ${\rm v}_{\rm A}$ 
is the Alfv\'en speed, and $\tau_{\rm in}$ is the ion-neutral collision time. 
When $\lambda<\lambda_{\min}$, the frequency of the wave is 
higher than the collision frequency between the ions and neutrals 
($\nu \gg \nu_{\rm in}$). In this case, the neutral fluid component is 
decoupled from the magnetic-field dynamics. If the clump is much 
larger than $\lambda_{\min}$, i.e., $W \gg 1$, the magnetic field is strongly 
coupled to the neutral component, and MHD waves can stabilise the clump. 

We calculated the $W$ number by utilising Eq.~(6) in Williams et al. (1998):

\begin{equation}
W=\frac{x({\rm e}) n({\rm H_2})^{1/2} \langle \sigma_{\rm in} {\rm v} \rangle}{\pi} \left\{\frac{5(1+p)}{4\pi \mu m_{\rm H}G-15\sigma_{\rm T}^2/ \left[n({\rm H_2})R^2 \right]} \right\}^{1/2} \,,
\end{equation}
where $\langle \sigma_{\rm in} {\rm v} \rangle$ is the ion-neutral 
collision rate, $p \in [0,\,1]$ is a parameter describing the level of 
turbulence (0 for a minimum amount of turbulent motions, 1 for a maximum level 
of turbulence), $\mu$ is the mean molecular weight per particle (taken to be 
2.33), $G$ is the gravitational constant, and 
$\sigma_{\rm T}=\sqrt{k_{\rm B}T_{\rm kin}/\mu m_{\rm H}}$ is the 
one-dimensional thermal velocity dispersion. Following Williams et al. (1998), 
we take $\langle \sigma_{\rm in} {\rm v} \rangle=1.5\times10^{-9}$ 
cm$^3$~s$^{-1}$. We used as $x({\rm e})$ the summed abundance of ionic 
species, and thus the values of $W$ should be taken as lower limits. 
We assumed that the clumps are turbulent ($p=1$), and clump radii were taken 
to be the effective radii listed in Col.~(7) of Table~\ref{table:properties}. 
Note that the latter were also used to derive the H$_2$ number densities. 
The computed values of $W$ and $\lambda_{\min}$ 
are given in Table~\ref{table:W}. The derived $W$ values are in the range 
$\sim2-19$, whereas the minimum wavelengths lie between 0.018 and 0.174 pc. 
Note that the clump radii range from 0.15 to 0.34 pc, and are mostly 
comparable to each other. Thus, the large variation of the $W$ values 
cannot (solely) be explained by different source sizes. Because $W$ represents 
the ratio of maximum to minimum wavelength, the maximum wavelength corresponds 
to the clump radius. For comparison, Bergin et al. (1999) deduced that massive 
cores in Orion (comparable in size to our clumps) are characterised by the 
value $W=20$. The MHD wave power transmission at the lowest frequency of wave 
propagation (i.e., the longest wavelength) rapidly drops below unity when 
$W\lesssim100$ (\cite{myers1998}; their Fig.~1). Therefore, the values $W<100$ 
imply that the coupling is ``marginal'', and that MHD waves have a quite 
narrow band of wavelengths to propagate above $\lambda_{\min}$. 
However, for I18102 MM1 and J18364 SMM1 the allowed wavelengths of MHD waves 
could extend from the effective radii $\sim0.2-0.3$ pc down to 
$\lambda_{\min}\sim 0.02-0.04$ pc. This is expected to lead to a clumpy 
structure of a medium (\cite{myers1995}), just like is found to be the case 
in J18364 SMM1 which is fragmented into two subcores (see Sect.~B.6).

Another approach to investigate the coupling of the neutral gas and the 
magnetic field is to determine the ratio between the ambipolar diffusion (AD) 
timescale and free-fall timescale, $\tau_{\rm AD}/\tau_{\rm ff}$. In the process 
of AD, neutrals drift quasi-statically (under their own self-gravity) relative 
to ions and magnetic field (\cite{mestel1956}). It can be shown, that the 
ratio $\tau_{\rm AD}/\tau_{\rm ff}$ is approximately given by 
$W\sim \tau_{\rm ff}/\tau_{\rm in}$ (see \cite{mouschovias1991}; 
\cite{williams1998}). The free-fall timescales of our clumps, 
$\tau_{\rm ff}=\sqrt{3\pi/32G \langle \rho \rangle}\sim8\times10^4-3\times10^5$ 
yr, imply AD timescales in the range 
$\tau_{\rm AD}\sim W \times \tau_{\rm ff} \sim2\times10^5-4\times10^6$ yr 
[see the first values in Col.~(4) of Table~\ref{table:W}]. We note that 
$\tau_{\rm AD}$ depends strongly on the ionic composition, and can be even much 
longer than the above rough estimates. Assuming that HCO$^+$ is the dominant 
ionic species, the AD timescale is given by 
$\tau_{\rm AD}\sim1.3\times10^{14}x({\rm e})$ yr
[see Eq.~(5) in Walmsley et al. (2004)]. Using again the lower limits to 
$x({\rm e})$, we obtain timescales in the range 
$\tau_{\rm AD}\sim1-16\times10^6$ yr which are about four times longer than 
the above estimates [see the second values in Col.~(4) of Table~\ref{table:W}]. 
Therefore, on the scale probed by our single-dish observations, AD is 
not expected to be a main driver of clump evolution unless it occurs on 
timescales much larger than $10^6$ yr. However, the situation may be different 
at smaller scales, where density is higher, and therefore $x({\rm e})$ is 
lower and $\tau_{\rm AD}$ is shorter.

\begin{table}
\caption{The field-neutral coupling parameter ($W$), MHD cutoff wavelength 
($\lambda_{\min}$), and ambipolar diffusion timescale ($\tau_{\rm AD}$).}
\begin{minipage}{1\columnwidth}
\centering
\renewcommand{\footnoterule}{}
\label{table:W}
\begin{tabular}{c c c c}
\hline\hline 
Source & $W$ & $\lambda_{\min}$ & $\tau_{\rm AD}$\tablefootmark{a}\\  
       &     & [pc] & [$10^5$ yr]\\
\hline
IRDC 18102-1800 MM1 & 18.5 & 0.018 & 37/162\\ 
G015.05+00.07 MM1 & 2.1 & 0.072 & 3/12\\ 
IRDC 18151-1208 MM2 & 2.6 & 0.065 & 2/9\\ 
G015.31-00.16 MM3 & 2.2 & 0.174 & 7/30\\ 
IRDC 18182-1433 MM2 & 6.0 & 0.044 & 16/65\\ 
IRDC 18223-1243 MM3 & 4.9 & 0.037 & 5/24\\ 
ISOSS J18364-0221 SMM1 & 8.9 & 0.036 & 12/54\\ 
\hline 
\end{tabular} 
\tablefoot{\tablefoottext{a}{The first $\tau_{\rm AD}$ value was estimated from 
the computed value of $W$, and the second one was calculated by assuming that 
HCO$^+$ is the dominant ionic species (see text).}}
\end{minipage} 
\end{table}

\section{Summary and conclusions}

We have carried out a molecular-line study of seven massive clumps associated 
with IRDCs. We used APEX to observe transitions of C$^{17}$O, H$^{13}$CO$^+$, 
DCO$^+$, N$_2$H$^+$, and N$_2$D$^+$. The principal aim of this study was to 
investigate the depletion, deuterium fractionation, and the degree of 
ionisation in the sources. Our main results are summarised as follows:

1. The CO molecules do not appear to be significantly depleted, if at all, 
in the observed sources. The largest CO depletion factor, $\sim2.7$, is found 
towards G015.31 MM3 which is dark at MIR. In many sources, CO is likely to 
be released from dust grains into the gas phase due to shock evaporation.

2. The deuteration degree in HCO$^+$ was found to range from 0.0002 
to 0.014, whereas that in N$_2$H$^+$ lies in the range 0.002--0.028. 
These are smaller values than found in low-mass starless cores and 
protostellar envelopes, but still significantly larger than the cosmic 
D/H-ratio $\sim10^{-5}$. The degree of deuteration in HCO$^+$ appears to 
decrease with increasing gas temperature, as expected theoretically. 
This is likely to reflect the evolutio\-nary stage of the clump: at early 
stages the source is cold and as the evolutionary stage progresses the 
temperature rises. In the course of evolution the amount of deuterium 
fractio\-nation decreases.

3. For the first time, we have estimated the level of ionisation in IRDCs.
The ionisation degree is difficult to determine accurately, especially from 
the analytical basis like in the present work. However, a lower limit to 
electron abundance appears to be $x({\rm e})\gtrsim10^{-8}-10^{-7}$, which 
is comparable to those seen in other star-forming regions. The cosmic-ray 
ionisation rate of H$_2$ were estimated to vary from $\sim10^{-17}$ s$^{-1}$ 
up to $\sim10^{-15}$ s$^{-1}$ in one target position. Similar 
estimates have been found in a variety of Galactic environments.

4. Additional molecular species were detected towards some of the sources. 
In particular, \textit{ortho}-c-C$_3$H$_2$ was detected in four sources, and 
SiO towards three sources. This is the first reported observation of 
c-C$_3$H$_2$ towards IRDCs. The SiO lines showed large linewidths indicative 
of outflows. Also for the first time towards IRDCs, the deuterated 
ethynyl, C$_2$D, was detected in I18151-1208 MM2. We estimate the 
C$_2$D/C$_2$H column density ratio to be about 0.07, in good 
agreement with the results from other sources (Appendix A). 

5. The ambipolar-diffusion timescale implied by the lower limits to 
$x({\rm e})$ is typically $\sim$several Myr. Therefore, on the scale 
probed by our observations, clump evolution cannot be mainly driven by 
ambipolar diffusion unless it occurs on timescales much larger than $10^6$ yr.

The present single-dish study cannot follow the chemical structure of the 
sources down to small scales where the evolution of the massive young stellar 
object(s) is taking place (e.g., the case of J18364 SMM1). Follow-up 
high-resolution studies are needed \textit{i)} to resolve whether the clumps are
fragmented into smaller subunits (as is known to be the case in J18364 SMM1); 
\textit{ii)} to examine the kinematics (infall, outflow, relative motions) of 
the subfragments, and, importantly, their chemical partition (such as 
depletion and deuteration); and \textit{iii)} to locate the detected complex 
molecules (envelope or hot core ?). In this regard, 
follow-up observations with ALMA seem particularly worthwhile. Finally, 
detailed chemical models are needed to better understand the observed 
molecular abundances, and how these are related 
to the timescale of the high-mass star formation process.

\begin{acknowledgements}

We are indebted to the referee for his/her valuable comments which helped 
significantly in making the paper more readable. We also thank the editor, 
Malcolm Walmsley, for his useful comments. We thank the staff at the 
APEX telescope for performing the service mode observations presented in 
this paper, and the ESO archive team for their help with the data products. 
O.~M. acknowledges support from the Academy of Finland grant 
132291. Many thanks to Henrik Beuther for providing us the 
MAMBO 1.2-mm dust continuum map of IRDC 18151-1208. We would also like to 
thank the people who maintain the CDMS and JPL molecular spectroscopy 
databases, and the \textit{Splatalogue} Database for Astronomical Spectroscopy. 

The present study makes use of the JCMT Archive project, which is a 
collaboration between the Canadian Astronomy Data Centre (CADC), Victoria, and 
the James Clerk Maxwell Telescope (JCMT), Hilo. This work also makes use of 
observations made with the Spitzer Space Telescope, which is operated by the 
Jet Propulsion Laboratory (JPL), California Institute of Technology under a 
contract with NASA. Moreover, this research has made use of NASA's 
Astrophysics Data System and the NASA/IPAC Infrared Science Archive, which is 
operated by the JPL, California Institute of Technology, under contract with 
the NASA.

\end{acknowledgements}

\Online

\begin{appendix}

\section{Other detected transitions}

Besides the molecular species discussed in Sect.~5.1, the sources I18102 MM1, 
I18151 MM2, I18182 MM2, and I18223 MM3 show emission lines also from other 
species, some of which are complex organics. These include 
\textit{o}-c-C$_3$H$_2$, C$_2$D, C$^{18}$O (seen in the image band), and 
possibly CNCHO (in the image band), CH$_3$NH$_2$ (blended with OCS), 
CH$_2$CHC$^{15}$N, and CH$_3$COCH$_3$. In this Appendix, we discuss 
each of these species (except C$^{18}$O) and their derived column densities and 
abundances. 

\textit{Cyclopropenylidene (c-C$_3$H$_2$)}. \,\, We have detected the 
\textit{ortho} form of cyclic-C$_3$H$_2$ in four sources, and found the column 
densities and abundances in the range $\sim7.0\times10^{12}-7.7\times10^{13}$ 
cm$^{-2}$ and $\sim1.9\times10^{-10}-1.2\times10^{-8}$, respectively. The 
c-C$_3$H$_2$ molecule has been detected in several different Galactic sources 
but to our knowledge, this is the first reported detection of this molecule 
towards massive clumps associated with IRDCs. 
For comparison, the beam-averaged c-C$_3$H$_2$ column densities and abundances 
of $\sim4-9\times10^{13}$ cm$^{-2}$ and $3-7\times10^{-11}$ have been found in 
Sgr B2 (\cite{turner1991}; \cite{nummelin2000}). These column densities are 
comparable to those in our clumps, but the abundances appear lower in Sgr B2.

The major formation channel of c-C$_3$H$_2$ is 
${\rm C_3H^+}+{\rm H_2}\rightarrow {\rm c-C_3H_3^+}$; ${\rm c-C_3H_3^+}+{\rm e^-}\rightarrow{\rm c-C_3H_2}$ (e.g., \cite{park2006}). The very low rotational 
excitation temperatures we have derived, 3.1--4.8 K, suggest that 
c-C$_3$H$_2$ emission mainly comes from the cool and relatively low-density 
envelope (\cite{turner1991}). Also, higher electron abundance in the outer 
layers can promote the formation of c-C$_3$H$_2$. This is consistent with the 
derived high fractional abundances.

\textit{Cyanoformaldehyde or formyl cyanide (CNCHO)}. \,\, The CNCHO 
transition is possibly seen in the image sideband towards I18102 MM1. 
Because of attenuated intensity, we cannot determine the column density of 
this molecule. Remijan et al. (2008) determined the values 
$N({\rm CNCHO})=1-17\times10^{14}$ cm$^{-2}$ and 
$x({\rm CNCHO})=0.7-11\times10^{-9}$ towards Sgr B2(N). They suggested that 
CNCHO is likely formed in a neutral-radical reaction of formaldehyde (H$_2$CO) 
and the cyanide (CN) radical.

\textit{Methylamine (CH$_3$NH$_2$)}. \,\, Rotational levels of CH$_3$NH$_2$ 
undergo the so-called $A$- and $E$-type transitions due to the internal 
rotation of the CH$_3$-group.
We have a possible detection of the $E$-type CH$_3$NH$_2$ transition, and we  
derived the values $N({\rm CH_3NH_2})=5\times10^{14}$ cm$^{-2}$ and 
$x({\rm CH_3NH_2})=5.1\pm0.6\times10^{-8}$. However, our 
CH$_3$NH$_2$ line is blended with OCS line, and thus these values should be 
taken with caution. Beam-averaged co\-lumn densities 
of $N({\rm CH_3NH_2})\sim1\times10^{14}-8\times10^{15}$ cm$^{-2}$ have been 
found in Sgr B2 (\cite{turner1991}; \cite{nummelin2000}). 
The origin of CH$_3$NH$_2$, and nitrogen-bearing organics in general, 
could be in the evaporation of ice mantles, indicative of a hot-core 
chemistry (e.g. \cite{rodgers2001a}). 

\textit{Carbonyl sulfide (OCS)}. \,\, The high-$J$ transition of OCS we have 
possibly detected (blended with CH$_3$NH$_2$) implies a co\-lumn density and 
fractional abundance of $2.3\pm0.8\times10^{15}$ cm$^{-2}$ and 
$2.4\pm0.9\times10^{-7}$, respectively. For example, Qin et al. (2010) detected 
OCS in the high-mass star-forming region G19.61-0.23. Assuming 
$T_{\rm rot}=E_{\rm u}/k_{\rm B}$, they derived the beam-averaged column density 
and fractional abundance of $2.2\pm0.1\times10^{16}$ cm$^{-2}$ and 
$2.7\pm0.1\times10^{-8}$, respectively. Note that Qin et al. achieved a much 
higher angular resolution with their SMA observations, and thus their large 
column density value could partly be due to filtering out the extended envelope.
Our coarser angular resolution probably causes a significant beam dilution. 
SSH10 derived OCS column densities of $1.2-5.5\times10^{14}$ 
cm$^{-2}$ for their sample of 20 massive clumps associated with IRDCs, i.e.,  
lower than found here towards I18102 MM1.

The OCS molecules form on grain surfaces through the addition 
of S atom to CO, or via the O atom addition to CS (e.g., \cite{charnley2004}). 
Solid-state OCS has been detected towards high-mass star-forming regions by, 
e.g, Gibb et al. (2004). SSH10 suggested that OCS is released 
from the grains into the gas phase through protostellar shocks. This could 
well be the case in I18102 MM1 (Sect.~B.1). 

\textit{Deuterated ethynyl (C$_2$D)}. \,\, We have made the first detection 
of C$_2$D towards IRDCs. The column density and abundance estimated from the 
$N=3-2$ transition are $\sim1.5\times10^{13}$ cm$^{-2}$ and 
$4.2\pm0.5\times10^{-10}$, respectively. SSH10 detected the normal isotopologue
C$_2$H$(N=1-0)$ towards I18151 MM2. The C$_2$H column density they derived, 
$\sim2.2\times10^{14}$ cm$^{-2}$, together with $N({\rm C_2D})$ derived by us, 
suggest a deuteration degree of $\sim0.07$ in C$_2$H. 
Vrtilek et al. (1985) detected the C$_2$D$(N=2-1)$ transition near the 
Orion-KL position, and obtained the co\-lumn density $\sim1.8\times10^{13}$ 
cm$^{-2}$. Moreover, they derived the $N({\rm C_2D})/N({\rm C_2H})$ ratio of 
0.05. These are very similar to what we have found. Quite similarly, 
Parise et al. (2009) found upper limits of $N({\rm C_2D})<2.5\times10^{13}$ 
cm$^{-2} $ and $x({\rm C_2D})<2\times10^{-10}$ in a clump associated with the 
Orion Bar.

The formation of C$_2$D is believed to take place in the gas phase through 
the route ${\rm CH_3^+}\rightarrow {\rm CH_2D^+}\rightarrow {\rm C_2D}$ 
(see, e.g., \cite{parise2009}). The $N({\rm C_2D})/N({\rm C_2H})$ ratio we 
have obtained is comparable with those predicted by the low-metal abundance 
model by Roueff et al. (2007) at temperature around 30--40 K. We note that the 
C$_2$H abundance, $\sim6\times10^{-9}$, calculated from the observed 
deuteration degree (0.07) is comparable to the va\-lues 
$x({\rm C_2H})=2.5\times10^{-9}-5.3\times10^{-8}$ recently found by Vasyunina 
et al. (2011) towards IRDCs. 

\textit{Acrylonitrile or vinyl cyanide, $^{15}$N isotopologue 
(CH$_2$CHC$^{15}$N)}. \,\, The Weeds modelling suggests that there is 
a CH$_2$CHC$^{15}$N transition blended with the hf group of the 
detected C$_2$D line. Thus, reliable column density estimate cannot be 
performed. Assuming that the detected line is 
completely due to CH$_2$CHC$^{15}$N emission, we derive very high values of 
$2.5\times10^{16}$ cm$^{-2}$ and $\sim7\times10^{-7}$ for the column density 
and fractional abundance. For comparison, a column-density upper limit of 
$3\times10^{15}$ cm$^{-2}$ towards Sgr B2(N) was derived by M{\"u}ller et al. 
(2008).

If present, this molecule would indicate a hot-core che\-mistry. The main 
isotopologue C$_2$H$_3$CN is expected to form through gas-phase reactions 
after the ethyl cyanide (C$_2$H$_5$CN), forming on dust grains, 
evaporates into the gas phase (\cite{caselli1993}). 

\textit{Acetone (CH$_3$COCH$_3$)}. \,\, The first detection of CH$_3$COCH$_3$ 
in the interstellar medium was made by Combes et al. (1987) towards Sgr B2. 
They found the column density and fractional abundance of this molecule to be 
$5\times10^{13}$ cm$^{-2}$ and $5\times10^{-11}$. These are significantly lower 
than we have estimated towards I18182 MM2 ($2\times10^{15}$ cm$^{-2}$ and 
$\sim4\times10^{-7}$). Combes et al. (1987) found that the CH$_3$COCH$_3$ 
abundance is about 1/15 of its precursor molecule CH$_3$CHO 
(acetaldehyde), and suggested the formation route of acetone to be 
${\rm CH_3^+}+{\rm CH_3CHO}\rightarrow({\rm CH_3})_2{\rm CHO^+}+h\nu; \,\, ({\rm CH_3})_2{\rm CHO^+}+{\rm e^-}\rightarrow{\rm CH_3COCH_3}+{\rm H}$. However, 
Herbst et al. (1990) showed that this radiative 
association reaction is likely too slow to be consistent with the observed 
abundance in Sgr B2. The chemistry behind the formation of acetone is not 
clear. It could be due to some other gas-phase ion-molecule reactions, or due 
to grain chemistry (\cite{herbst1990}). 

Friedel et al. (2005) found that acetone in Orion BN/KL is concentrated 
towards the hot core. They derived the beam-averaged column densities 
of $\sim2-8\times10^{16}$ cm$^{-2}$. More recently, Goddi et al. (2009) found 
the column density $N({\rm CH_3COCH_3})=5.5\times10^{16}$ cm$^{-2}$ in Orion 
BN/KL, in agreement with the Friedel et al. results.
The very high column densities and abundances of CH$_3$COCH$_3$ found in 
Orion BN/KL and in I18182 MM2 in the present work require some other reaction 
pathway(s) than the above radiative association reaction to be efficient. 
Grain surface and hot-core gas-phase chemistry may both play critical roles 
(\cite{garrod2008}).

The detection of complex molecules in clumps associated with IRDCs, indicating 
the presence of hot cores, supports the scenario that high-mass star formation 
can take place in these objects. For example, Rathborne et al. (2007, 2008) 
found that the clumps G024.33+00.11 MM1 and G034.43+00.24 MM1 are both 
likely to contain a hot molecular core. On the other hand, complex organics 
could also be ejected from grain mantles through shocks, and cosmic rays 
heating up the dust can also have some effect (e.g., \cite{requena2008}, and 
re\-ferences therein).

\section{Discussion on individual sources}

\subsection{IRDC 18102-1800 MM1}

The clump I18102 MM1 is the warmest (21.3 K) and most massive 
($\sim360$ M$_{\sun}$) source of our sample. It is associated with 
\textit{Spitzer} point sources at 8 and 24 $\mu$m, and high-mass star 
formation is taking place within it as indicated by the presence of the 
6.7 GHz Class {\footnotesize II} CH$_3$OH maser (\cite{beuther2002b}). 
Among our sample, the lowest degree of deuteration in both HCO$^+$ and 
N$_2$H$^+$ is found for I18102 MM1. On the other hand, the source shows 
the highest degree of ionisation.

Fuller et al. (2005) detected central dips and red asymmetries in 
the spectral lines HCO$^+(1-0)$, HCO$^+(4-3)$, N$_2$H$^+(1-0)$, and 
H$_2$CO$(2_{1,2}-1_{1,1})$. A similar line profile is seen in the $J=3-2$ 
transition of N$_2$H$^+$ in the present study.
These profiles indicate the presence of expanding or outflowing gas 
(see, e.g., \cite{park2000} and references therein). 
Beuther \& Sridharan (2007) detected 
very broad (42.2 km~s$^{-1}$ down to zero intensity) SiO$(2-1)$ wings towards 
I18102 MM1, indicative of bipolar outflows. This source showed the second 
broadest SiO line in the sample of Beuther \& Sridharan (2007). 
Also, the SiO$(2-1)$ line detected by SSH10 towards I18102 MM1 was very wide, 
$\Delta {\rm v}=13\pm1$ km~s$^{-1}$. The SiO$(6-5)$ line detected in the 
present study has $\Delta {\rm v}\simeq 4.7$ km~s$^{-1}$, and the width down 
to zero intensity is $\sim10$ km~s$^{-1}$. We have also detected the $^{13}$C 
isotopologue of CS in this source. SSH10 suggested that CS could 
originate in the shock evaporation of grain mantles or radiative heating.

Beuther \& Sridharan (2007) detected CH$_3$CN$(6_K-5_K)$ and 
CH$_3$OH$(5_K-4_K)$ lines towards I18102 MM1, and derived the abundances of 
$3\times10^{-11}$ and $4\times10^{-11}$ for CH$_3$CN and CH$_3$OH, 
respectively. SSK08 detected CH$_3$OH$(7_K-6_K)$ and 
HC$_3$N$(5-4)$ lines in this source, and SSH10 detected 
C$_2$H$(N=1-0)$ and a hint of CH$_3$OH$(2_1-1_1)$ A$^-$. We note that 
SSK08 and SSH10 did not detected the lines of CCS$(4_3-3_2)$, 
SO$(2_2-1_1)$, or OCS$(8-7)$ in their surveys. The upper limit they derived 
for the OCS column density, $\lesssim2.7\times10^{14}$ cm$^{-2}$, is about 8.5 
times less than the $N(\rm OCS)$ value we have obtained. The 
observational results cumulated so far indicate hot-core chemistry in 
I18102 MM1. This conforms to the fact that this clump is giving birth to a 
high-mass star(s), and possibly through disk accretion as indirectly 
suggested by the outflow signatures. The associated hot core is likely to be in 
its later stages of evolution because it is associated with a methanol 
maser (cf. \cite{rathborne2008}).

\subsection{G015.05+00.07 MM1 and G015.31-00.16 MM3}

G015.05 MM1 is the lowest mass (63 M$_{\sun}$) and G015.31 MM3 is the 
second coldest (13.7 K) clump of our sample. Both clumps are dark in 
the \textit{Spitzer} 8 and 24 $\mu$m images. The highest deuteration degree 
in N$_2$H$^+$ (0.028) was found towards G015.05 MM1, whereas G015.31 MM3 shows 
the largest CO depletion factor (2.7) in our sample. G015.05 MM1 is associated 
with H$_2$O maser (\cite{wang2006}), indicative of star formation activity. 
Rathborne et al. (2010) derived the following properties for G015.05 MM1 
from the broadband SED: $T_{\rm dust}=11-36$ K, $L=15.5-362$ L$_{\sun}$, 
and $M=35-158$ M$_{\sun}$ (scaled to the revised distance 2.6 kpc). The values 
$T_{\rm kin}=17.2$ K and $M=63$ M$_{\sun}$ derived in the present study lie at 
the low end of the above values of temperature and mass [at the derived clump 
densities, it is expected that $T_{\rm kin}\simeq T_{\rm dust}$ 
(\cite{goldsmith1978})]. 

SSK08 barely detected the HC$_3$N$(5-4)$, CH$_3$OH$(7_K-6_K)$, 
and CCS$(4_3-3_2)$ lines towards G015.05 MM1 and G015.31 MM3. Actually, 
all the CH$_3$OH detected objects in the survey by SSK08 
are associated with the \textit{Spitzer} 24-$\mu$m sources.

Both G015.05 MM1 and G015.31 MM3 are likely to be in a very early stage of 
evolution. Moreover, both sources are massive enough to allow high-mass star 
formation. G015.31 MM3 could represent or host the so-called high-mass 
prestellar core, whereas G015.05 MM1 could be slightly more 
evolved with H$_2$O maser emission but still lacking IR emission at 8 and 
24 $\mu$m. The small wave-coupling number of $W\sim2$ for G015.31 MM3 
suggests that ``magnetic turbulence'' is not able to fragment the clump into 
smaller pieces, strengthening the possibility that it is a massive 
prestellar ``core''.  

\subsection{IRDC 18151-1208 MM2}

The I18151 MM2 clump is dark in the \textit{Spitzer} 8-$\mu$m image 
(24 $\mu$m not available). It is associated with H$_2$O (\cite{beuther2002b})
and Class I CH$_3$OH masers (\cite{marseille2010b}). This clump has the 
highest volume-averaged H$_2$ number density ($\sim1.9\times10^5$ cm$^{-3}$) 
among our sources.

Beuther \& Sridharan (2007) detected the broadest SiO$(2-1)$ wing emission 
(65 km~s$^{-1}$ down to zero intensity) towards I18151 MM2 in their sample. 
Also, the $J=2-1$ and $3-2$ SiO lines detected recently by 
L{\'o}pez-Sepulcre et al. (2011) are very broad, i.e., 
FWZP$=84.3$ and 103.1 km~s$^{-1}$, respectively. 
The SiO$(6-5)$ line we detected is also very broad with the FWHM 46.8 
km~s$^{-1}$. The outflow activity within the clump was confirmed 
by Marseille et al. (2008) who, for the first time, found that I18151 MM2 
is driving a CO outflow and hosts a mid-IR-quiet, possibly a Class 0-like 
HMYSO (cf. \cite{motte2007}). 
Marseille et al. (2008) modelled the dust continuum 
emission (SED) of I18151 MM2 and found that the bolometric luminosity, 
mass, and the mean temperature of the source are $L_{\rm bol}=2190$ 
L$_{\sun}$, $M_{\rm gas}=373\pm81$ M$_{\sun}$, and 
$\langle T \rangle=19.4\pm0.2$ K (scaled to the revised distance 2.7 kpc). 
Within the errors, these mass and temperature values are comparable to the 
values derived in the present paper. Besides the Class I methanol maser 
tracing the molecular outflow, Marseille et al. (2010b) detected blue 
asymmetry in CH$_3$OH$(5_{-1,5}-4_{0,4})$ E, indicating infall motions. 
The 22 GHz H$_2$O maser is probably excited in the outflow shocked gas 
(cf. \cite{furuya2011}).

Beuther \& Sridharan (2007) also detected CH$_3$CN$(6_K-5_K)$ and 
CH$_3$OH$(5_K-4_K)$ lines towards I18151 MM2, which is a sign of hot-core 
chemistry, and they derived the abundances of $8\times10^{-11}$ and 
$6\times10^{-10}$ for CH$_3$CN and CH$_3$OH, respectively. 
SSK08 detected CH$_3$OH$(7_K-6_K)$ lines towards I18151 MM2, but not 
CCS$(4_3-3_2)$ or HC$_3$N$(5-4)$. In their line survey, SSH10 detected 
C$_2$H$(N=1-0)$, but not SO$(2_2-1_1)$, OCS$(8-7)$, 
or CH$_3$OH$(2_1-1_1)$ A$^-$ lines. The C$_2$H column density they derived, 
$\sim2.2\times10^{14}$ cm$^{-2}$, together with the value 
$N({\rm C_2D})=1.5\times10^{13}$ cm$^{-2}$ derived by us, suggest 
a deuteration degree of $\sim0.07$ in C$_2$H. Marseille et al. (2008) 
concluded from their modelling of CS transitions that CS is depleted in 
I18151 MM2. We derived only a small CO depletion factor of $\sim1.6$ for this 
clump, and low degrees of deuteration in HCO$^+$ and N$_2$H$^+$, namely 0.3\% 
and 1\%, respectively. Also, this source shows the lowest lower limit to 
ionisation degree, only $7\times10^{-9}$. 

\subsection{IRDC 18182-1433 MM2}

The filamentary clump I18182 MM2 is associated with \textit{Spitzer} 8 and 
24-$\mu$m sources. This source shows the highest degree of deuteration in 
HCO$^+$ (0.014). 

Beuther \& Sridharan (2007) detected CH$_3$OH$(5_K-4_K)$ lines towards 
I18182 MM2, and derived the CH$_3$OH abundance of $2.1\times10^{-10}$. 
SSK08 did not detect CCS$(4_3-3_2)$, CH$_3$OH$(7_K-6_K)$, or 
HC$_3$N$(5-4)$ lines towards this source in their survey. Our tentative 
detection of the O-bearing species CH$_3$COCH$_3$ indicates 
hot-core chemistry within the clump. Moreover, O-bearing species are sign of 
the early stage of che\-mical evolution (e.g., \cite{shiao2010}). This conforms 
to the presumable young age of the clump as it is associated with IRDC.
Interestingly, the nearby clump I18182 MM1, which is associated with the 
HMYSO IRAS 18182-1433, also appears to contain a hot core 
(\cite{beuther2006}).

\subsection{IRDC 18223-1243 MM3}

I18223 MM3, part of a long filamentary IRDC, is a high-mass clump 
harbouring an embedded accreting low- to intermediate-mass protostar which 
could evolve to a high-mass star at some point in the future 
(see \cite{beuther2007b}; \cite{beuther2010}). Fallscheer 
et al. (2009) detected a mole\-cular outflow in this source and found evidence 
for a large rota\-ting structure, or toroid, perpendicular to the outflow. 
Evidence for outflow activity in this source was already found by Beuther et 
al. (2005c) and Beuther \& Sridharan (2007) who found that there are 4.5 
$\mu$m emission features at the clump edge and that the spectral lines of CO, 
CS, and SiO show broad wing emission. The line profile of SiO$(6-5)$ 
detected in the present study also indicates outflowing gas.

The clump shows 24 $\mu$m emission but is dark at the \textit{Spitzer} 
IRAC wavelengths (3.6, 4.5, 5.8, and 8.0 $\mu$m). 
Beuther et al. (2010) derived an SED for this source between 24 $\mu$m and 
1.2 mm, including the recent \textit{Herschel} PACS and SPIRE data, and 
obtained the total luminosity of 539 L$_{\sun}$ (scaled to $d=3.5$ kpc). 
They also found that the ratio between the total and submm luminosity 
(integrated longward of 400 $\mu$m) is only 11, suggesting that the source 
is very young. 

Beuther \& Sridharan (2007) detected CH$_3$CN$(6_K-5_K)$ and 
CH$_3$OH$(5_K-4_K)$ lines towards I18223 MM3, indicative of hot-core 
chemistry, and derived the abundances of $9\times10^{-11}$ and 
$6.1\times10^{-10}$ for CH$_3$CN and CH$_3$OH, respectively. 
SSK08 detected HC$_3$N$(5-4)$, only weak CH$_3$OH$(7_K-6_K)$ lines, 
and no CCS$(4_3-3_2)$ lines towards this 
source. SSH10 did not detect the lines SO$(2_2-1_1)$, 
OCS$(8-7)$, or CH$_3$OH$(2_1-1_1)$ A$^-$ in their survey; however, they 
detected the C$_2$H$(N=1-0)$ line. Based on the column 
densities of different species (e.g., SiO and H$^{13}$CO$^+$), 
SSH10 suggested that I18223 MM3 is in early stage of evolution. 
This conforms to the fact that the second largest value of 
$R_{\rm D}({\rm N_2H^+})$ in our sample (0.013) is found towards I18223 MM3.

\subsection{ISOSS J18364-0221 SMM1}

At the distance of $\sim2.5$ kpc, the clump J18364 SMM1 is the nearest 
source in our sample. It is also the coldest (11.4 K) clump of our 
sample. The clump is dark at 8 $\mu$m but it is associated with the 
24-$\mu$m point source. The second largest value of $R_{\rm D}({\rm HCO^+})$ in 
our sample, 0.012, is found towards this source. It also 
shows high cosmic-ray ionisation rate of H$_2$ ($\sim5\times10^{-16}$ s$^{-1}$).
Birkmann et al. (2006) studied this clump through $J=3-2$ 
transitions of HCO$^+$ and H$^{13}$CO$^+$ (beam size $\sim9\farcs5$). 
The former line showed blue asymmetric profile, indicating infall. 
They also found significant CO$(2-1)$ line wings, indicating the presence of 
outflows. The H$^{13}$CO$^+(3-2)$ line we have observed shows an 
asymmetric profile with a central dip and slightly stronger red peak, 
contrary to that observed by Birkmann et al. (2006) in HCO$^+(3-2)$. 
This difference is probably due to the larger beam size of our observations 
($24\arcsec$), yielding a signature of outflowing gas motions.

More recently,  J18364 SMM1 was studied in detail by Hennemann et al. (2009). 
They resolved this contracting clump in the interferometric mm continuum 
into two compact cores, named SMM1 North and South separated by $9\farcs5$ 
(0.12 pc). Their positions are indicated in Fig.~\ref{figure:maps}. 
The peak H$_2$ column densities and dust temperatures were found to be 
$2.7\times10^{23}$ cm$^{-2}$ and 15 K for SMM1 North and, 
$2.4\times10^{23}$ cm$^{-2}$ and 22 K for SMM1 South. Using the revised 
distance 2.5 kpc, the radius, mass, and luminosity of the northern core 
are 0.06 pc, 19 M$_{\sun}$, and 26 L$_{\sun}$, whereas for the southern 
core these values are 0.05 pc, 13 M$_{\sun}$ and 230 L$_{\sun}$. 
Thus, the cores within the clump have comparab\-le sizes and masses but the 
southern one is associated with the \textit{Spitzer} 24 and 70 $\mu$m sources 
and is more luminous. Hennemann et al. (2009) found that SMM1 South 
drives an energetic mole\-cular outflow and that the core centre is 
supersonically turbulent. 
On the other hand, the IR-dark core SMM1 North shows lower levels of 
turbulence, but it also drives an outflow. Both the outflows from SMM1 
North and South are rather collimated and their estimated ages are 
$<10^4$ yr. The HCN$(1-0)$ modelling results by Hennemann et al. 
(2009) showed that the spectrum of SMM1 South can be explained with a 
collapse of the core. They obtained an infall velocity of 0.14 km~s$^{-1}$ 
and an estimated mass infall rate of $\sim3.4\times10^{-5}$ 
M$_{\sun}$~yr$^{-1}$ (scaled to the revised distance).

In summary, the clump J18364 SMM1 is fragmented into two cores which 
both harbour protostellar seeds, possibly evol\-ving into intermediate- to 
high-mass stars. As discussed in Sect.~5.4, MHD wave propagation 
could have played a role in fragmenting the parent clump. As the southern core 
harbours a 24 $\mu$m source, is highly turbulent in the central region, and 
shows jet features at large distance from the driving source, it appears to 
be more evolved than the northern core.

\end{appendix}

\end{document}